\documentclass[fleqn,usenatbib]{mnras}

\usepackage{newtxtext,newtxmath}
\usepackage[T1]{fontenc}
\usepackage{ae,aecompl}
\usepackage{comment}
\usepackage{tabularx}
\newcolumntype{C}{>{\centering\arraybackslash}X}
\usepackage{listings}
\usepackage{array}
\newcolumntype{P}[1]{>{\centering\arraybackslash}p{#1}}
\usepackage{multirow}
\usepackage{graphicx}
\usepackage{amsmath}
\usepackage{amssymb}
\usepackage{float}
\usepackage[caption=false]{subfig}

\title[Detectability of protoplanets from HD simulations]{Detectability of embedded protoplanets from hydrodynamical simulations}

\author[Sanchis et al.]{
E. Sanchis,$^{1,2}$\thanks{E-mail: esanchis@eso.org}
G. Picogna,$^{2}$
B. Ercolano$^{2,3}$
L. Testi$^{1,3,4}$
and G. Rosotti$^{5,6}$
\\
$^{1}$European Southern Observatory, Karl-Schwarzschild-Strasse 2, D-85748 Garching bei M{\"u}nchen, Germany\\
$^{2}$Universit{\"a}ts-Sternwarte, Ludwig-Maximilians-Universit{\"a}t M{\"u}nchen, Scheinerstrasse 1, D-81679 M{\"u}nchen, Germany\\
$^{3}$Excellence Cluster Origins, Boltzmannstrasse 2, D-85748 Garching bei M{\"u}nchen, Germany\\
$^{4}$INAF/Osservatorio Astrofisico di Arcetri, Largo E. Fermi 5, I-50125 Firenze, Italy\\
$^{5}$Institute of Astronomy, University of Cambridge, Madingley Road, Cambridge, UK\\
$^{6}$Leiden Observatory, Leiden University, P.O. Box $9531$, NL-$2300$ RA Leiden, the Netherlands
}

\date{Accepted XXX. Received YYY; in original form ZZZ}

\pubyear{2020}

\begin{document}
\label{firstpage}
\pagerange{\pageref{firstpage}--\pageref{lastpage}}
\maketitle

\begin{abstract}
We predict magnitudes for young planets embedded in transition discs, still affected by extinction due to material in the disc. We focus on Jupiter-size planets at a late stage of their formation, when the planet has carved a deep gap in the gas and dust distributions and the disc starts being transparent to the planet flux in the infrared (IR). Column densities are estimated by means of three-dimensional hydrodynamical models, performed for several planet masses. Expected magnitudes are obtained by using typical extinction properties of the disc material and evolutionary models of giant planets. For the simulated cases located at $5.2$ $\mathrm{AU}$ in a disc with local unperturbed surface density of $127$ $\mathrm{g} \cdot \mathrm{cm}^{-2}$, a $1$ $M_\mathrm{J}$ planet is highly extincted in $J$-, $H$- and $K$-bands, with predicted absolute magnitudes $\ge$ $50$ $\mathrm{mag}$. In $L$- and $M$-bands extinction decreases, with planet magnitudes between $25$ and $35$ $\mathrm{mag}$. In the $N$-band, due to the silicate feature on the dust opacities, the expected magnitude increases to $\sim 40$ $\mathrm{mag}$. For a $2$ $M_\mathrm{J}$ planet, the magnitudes in $J$-, $H$- and $K$-bands are above $22$ $\mathrm{mag}$, while for $L$-, $M$- and $N$-bands the planet magnitudes are between $15$ and $20$ $\mathrm{mag}$. For the $5$ $M_\mathrm{J}$ planet, extinction does not play a role in any IR band, due to its ability to open deep gaps. Contrast curves are derived for the transition discs in CQ~Tau, PDS~$70$, HL~Tau, TW~Hya and HD~$163296$. Planet mass upper-limits are estimated for the known gaps in the last two systems.
\end{abstract}



\section{Introduction}\label{sec:intro}
For the last two decades a large number of exoplanet detections have remarkably expanded and shaped the prevailing planet formation theories. Most of these discoveries have been accomplished using indirect techniques, although in a few cases detections were achieved using direct imaging \citep{quanz2015, reggiani2017, keppler+2018}. Direct observations are only possible for systems close enough to us, with planets far from their host stars \citep{rameau+2015}. When searching for young planets of only a few $\mathrm{Myr}$ an additional problem arises: while a young planet might itself be bright enough to be detected, it will still be embedded in the primordial disc, and thus hidden behind large columns of dust and gas.

Studying the interaction between the disc material and the young planet during its formation is crucial to understand the ongoing processes of planet formation and its further evolution. After the initial stages of planet formation, the protoplanet is likely accreting material and still surrounded by gas and dust. Several processes --mainly internal and/or external photo-evaporation, accretion onto the central star and planet, magneto-hydrodynamical winds \citep[see review by][]{ercolanopascucci2017} continuously reduce the disc material until its complete dispersal. The typical lifetimes for discs can vary considerably and are uncertain, in general the inner disc (within a fraction of AU from the central star) is expected to disperse within a few $\mathrm{Myr}$ \citep{hernandez+2008}, even though there is potential evidence for replenishment of the inner disc over longer timescales \citep[e.g.][]{2010ApJ...720.1108B,2014A&A...566L...3S}. The dissipation timescales of the outer disc, which is more relevant for the direct detectability of young protoplanets, is much more uncertain. Initial Atacama Large Millimeter/submillimeter Array (ALMA) surveys suggest that the depletion of the outer disc may also proceed on a similar timescale, following a simple estimate of disc masses based on (sub)-millimetre continuum emission from large dust grains \citep{2017AJ....153..240A}. More detailed studies seem to imply that gas and dust depletion may be substantial even at young ages  \citep[][]{miotello+2017,2018A&A...618L...3M}. As the disc keeps losing material, the extinction is reduced proportionally. If dust extinction has decreased enough, a detection of an embedded planet may be possible with state-of-the-art facilities observing at IR wavelengths, where the planet spectrum peaks.

Consequently, detections of protoplanets embedded in discs depend on the properties of the planet, its immediate surroundings, and also on the upper atmospheric layers of the disc. The search for indirect detections in (sub)-millimetre observations has been pursued during the past years \citep[e.g.,][]{2006A&A...460L..43P,2008ApJ...675L.109B}, more intensively once ALMA started operating \citep{alma2015}. Substructure like cavities, gaps and spirals could be first observed at these wavelengths, suggesting planet-disc interaction as a plausible cause of such features \citep[e.g.][]{2006A&A...453.1129P,2006MNRAS.373.1619R}. The DSHARP large program \citep{andrews+2018} has confirmed that substructure is ubiquitous in large discs when observed with enough resolution, although these disc features do not necessarily confirm the presence of planets. The first indirect detection of planets at these wavelengths was achieved from detailed analyses of the gas kinematics in the HD~$163296$ disc \citep{teague2018, pinte2018}.

While these indirect detections with ALMA and other (sub)-millimetre facilities tantalise evidence for young, and in some cases, massive planets in discs, the interpretation is not unique. Direct detection of young planet candidates is required to confirm their presence in discs. Additionally, direct detection from IR and spectroscopy is crucial to further characterise the planet properties (e.g. its atmosphere). Several attempts have been made to detect directly young planet candidates in discs, by means of near and mid-IR high contrast imaging, but the vast majority of these efforts resulted in no detections. \cite{testi2015} searched for planets in HL~Tau in $L$-band, without any point-sources detected but setting upper limits of planet masses at the rings location. Much more stringent upper limits were set for the TW~Hya system by \cite{ruane2017} with the Keck/NIRC2 instrument, and for the HD~$163296$ system by \citet{guidi2018}. In only a few occasions, point-like sources have been claimed as detections in other protoplanetary discs. \citet{2014ApJ...792L..23R} and \citet{quanz2015} announced direct evidence (as a point-sources) of protoplanets embedded in the HD~169142 and HD~100546 at $L$- and $M'$-bands respectively, using the NaCo instrument at the Paranal Observatory of the European Southern Observatory. While apparently convincing, both detections have been disputed and further confirmation is still pending. More recently, an additional candidate, still requiring confirmation, has been detected by \cite{reggiani2017}  in the spiral arm of MWC~758 in $L$-band. To date, the most convincing direct detection of a young planet in a protoplanetary disc, is the multi-wavelength detection of PDS~$70$b, confirmed using multiple epoch data \citep{keppler+2018}; the system also includes an additional planet, PDS~$70$c \citep{haffert+2019}.

Hydrodynamical (HD) simulations of planet-disc interactions have greatly improved our theoretical understanding of these systems. \cite{kley1999} and \cite{bryden1999} simulated a planet embedded in a disc in $2$D, showing the formation of gaps from the planet-disc interaction and deriving crucial properties of protoplanetary discs as mass accretion and viscosity. Many other studies followed, focusing on the characterisation of geometrical properties and the derivation of important disc evolution parameters \citep[like the viscosity, e.g.][]{szulagyi2014}. The implementation of nested grids alleviated the resolution limitation and computation times, allowing for the first $3$D simulations to be performed \citep[see][]{dangelo2003}.

The detectability of a protoplanet at different wavelengths can also been inferred from HD-simulations. Indeed, \cite{wolf2005} performed mock observations to infer planet signatures that could be detected using ALMA, while \cite{zhu2015} focused on the detectability in IR-bands of the circumplanetary disc (CPD) around highly accreting planets. Other indirect observable signatures due to the presence of planets have been also investigated: gap opening effects at various wavelengths \citep[e.g.][]{dong+2015a,rosotti2016,2017MNRAS.469.1932D,dong+2017,JangCondell2017}, disc inclination effects \citep{JangCondell+2013}, spiral arms in scattered light \citep[][]{dong+2015b,fung+2015,juhasz2015,juhasz_rosotti_2018}, planet shadowing \citep{JangCondell2009}, and even the effect of migrating planets \citep{meru2019,nazari2019}.

This study focuses on the early evolutionary stages, when the accretion of disc material onto the planet might still have an incidence on the total planet brightness. We performed 3D HD simulations, using high resolution nested grids in the planet surroundings. Column densities and extinction coefficients at the planet location are derived from the simulations, in order to infer planet magnitudes in $J$-, $H$-, $K$-, $L$-, $M$- and $N$-bands. This model can be used to guide future direct imaging observations of young planets embedded in protoplanetary discs.

The work is organised as follows: in Section~\ref{sec:methods} we describe the simulations set-up and the model. The results of the different simulated systems are presented in Section~\ref{sec:results}. The application of the model to known protoplanetary discs is discussed in Section~\ref{sec:applicationresults}, and the main implications of this work are summarised in Section~\ref{sec:conclusions}.


\section{Set-up and Model description}\label{sec:methods}
The HD simulations were performed with the PLUTO code \citep{mignone2010, mignone2012}, a 3-dimensional grid-code designed for astrophysical fluid dynamics. The details of the simulations set-up, the description of the planet flux model used (\ref{subsec:fluxes}) and the derivation of the expected planet magnitudes (\ref{subsec:model}) are presented in this section.

\subsection{Simulation Set-up}\label{subsec:setup}
We modelled a gaseous disc in hydrostatic equilibrium around a central star with a protoplanet on a fixed orbit. The disc evolution is determined by the $\alpha$-viscosity prescription \citep{shakurasunyaev}. The disc is considered to be locally isothermal, for which the Equation-of-State (EoS) is described as:
\begin{equation}\label{eq:eos}
P = n \cdot k_{B} \cdot T = \frac{\rho}{m_u \cdot \mu} \cdot k_{B} \cdot T
\end{equation}
where $P$ is the pressure, $n$ the total particle number density, $k_B$ the Boltzmann constant, $T$ the temperature, $\rho$ the gas density, $m_u$ the atomic mass unit and $\mu$ the mean molecular weight. The temperature in the disc varies only radially:
\begin{equation}\label{eq:temperatureprofile}
T(R) = T_0  \bigg( \frac{R}{R_0} \bigg) ^{q}
\end{equation}
with $R$ being the radius in cylindrical coordinates, $R_0 = 5.2$ AU, $T_0 = 121$ $K$ and $q$ the temperature exponent factor, set to $-1$. The adopted density distribution of the protoplanetary disc \citep[as in][]{nelson2013} is described as:
\begin{equation}\label{eq:radialstructure}
\rho (r, \theta) = \rho_0 {\bigg( \frac{R}{R_0} \bigg)}^{p} \exp \bigg[ \frac{G M_{\star}}{c_{\mathrm{iso}}^{2}} \cdot \Big( \frac{1}{r} - \frac{1}{R} \Big) \bigg]
\end{equation}
where $r$ refers to the radial distance from the centre in spherical coordinates, $\theta$ the polar angle (thus $R = r \cdot \cos(\theta)$), $\rho_0$ the density at the planet location, $p$ is the density exponent factor, in our case with value $-1.5$, $G$ is the universal gravitational constant, $M_\mathrm{\star}$ the mass of the central star, and $c_{\mathrm{iso}}$ the isothermal speed of sound. From this equation, the surface density $\Sigma$ scales radially as:
\begin{equation}\label{eq:surfacedensity}
\Sigma (R) = \Sigma_0 \cdot {\bigg( \frac{R}{R_0} \bigg)}^{-p'}
\end{equation}
with a power law with index $p'=-0,5$. The planet is included as a modification in the gravitational potential of the central star in the vicinity of the planet location, which is kept fixed. The gravitational potential $\phi$ considered in the simulations is:
\begin{equation}\label{eq:potential}
\phi = \phi_{\star} + \phi_{\mathrm{pl}} + \phi_{\mathrm{ind}}
\end{equation}
$\phi_{\star}$ is the term due to the star, $\phi_{\mathrm{pl}}$ the planet potential and $\phi_{\mathrm{ind}}$ accounts for the effect of the planet potential onto the central star. In the cells closest to the planet (cells at a distance to the planet lower that the smoothing length $d_{\mathrm{rsm}}$), $\phi_\mathrm{pl}$ is introduced with a cubic expansion as in \cite{klahrkley} to avoid singularities at the planet location:
\begin{equation}\label{eq:potentialexpansion}
\phi_\mathrm{pl} (d<d_{\mathrm{rsm}}) = - \frac{G M_{\mathrm{pl}}}{d} \bigg[ {\Big( \frac{d}{d_{\mathrm{rsm}}} \Big)}^{4} -2 {\Big( \frac{d}{d_{\mathrm{rsm}}} \Big)}^{3} +2 {\Big( \frac{d}{d_{\mathrm{rsm}}} \Big)} \bigg]
\end{equation}
with $d$ referring to the distance between cell and planet. We set $d_\mathrm{rsm}$ to be $0.1$ $R_\mathrm{Hill}$ in the first 4 simulations. In two additional runs we decreased this value and doubled the grid resolution in order to investigate the behaviour of the density in the cells close to the planet surface. In Table~\ref{tab:userparameters} all the values of $d_\mathrm{rsm}$ used in the simulations are shown.

The gas rotational speed ($\Omega$) is sub-keplerian; the additional terms arise from the force balance equations in radial and vertical directions \citep[see e.g.][]{nelson2013}. It is described by:
\begin{equation}\label{eq:subkeplerianspeed}
\Omega (R, z) = \Omega_K {\bigg[ (p + q) {\Big( \frac{h}{R} \Big)}^{2} + (1+q) - \frac{q R}{\sqrt{R^2 + z^2}} \bigg]}^{1/2}
\end{equation}
with $z$ being the vertical coordinate, $\Omega_k$ the keplerian orbital speed, and $h$ the vertical scale-height of the disc.

The simulated protoplanetary discs had a stellar mass of $1.6$ $M_{\odot}$ and a surface density at $5.2$ AU of 127 $\mathrm{g} \cdot \mathrm{cm}^{-2}$, consistent with estimates of the Minimum Mass Solar Nebula \citep[][]{1981PThPS..70...35H} and the densest discs in the star forming regions in the Solar neighbourhood. \citep[][]{2015PASP..127..961A,2017A&A...606A..88T}. The model can be adapted to systems with different characteristics by re-scaling our results to different surface densities and other parameters, as shown in section \ref{sec:applicationresults}.

We modelled three different planetary masses ($1$, $2$ and $5$ $M_\mathrm{J}$) embedded in a viscous disc with $\alpha = 0.003$. A fourth inviscid run with a $1$ $M_\mathrm{J}$ was performed in order to study the effect of viscosity on planet detectability, since recent studies with non-ideal magneto-hydrodynamical effects have shown that the disc can be laminar at the mid-plane \citep{paardekooper+2017}. Besides, two additional runs with doubled grid resolution were done for the $1$ and $5$ $M_\mathrm{J}$ cases, in order to improve our understanding of the disc-planet interaction at the upper atmospheric layers of the planet. In total 6 simulations were carried out, summarised in Table~\ref{tab:userparameters}. During the first $20$ orbits the planetary mass was raised until its final value ($1$, $2$ or $5$ $M_\mathrm{J}$) following a sinusoidal function, in order to prevent strong disturbances in the disc. The simulations ran for $200$ orbits, once a steady-state of the system is reached.
\begin{table*}
	\centering
    \caption[Properties of doubled resolution runs.]{Set-up parameters of the simulations and inferred column mass densities $\sigma$. The disc aspect ratio $H$, defined as $H = h/R$ was set to $0.05$ in all the simulations. The columns refer to: the $\alpha$-viscosity parameter for a viscously evolving disc; cell size at planet vicinity, given in Hill radii ($R_\mathrm{Hill}$) and in planet radii ($R_\mathrm{pl}$); smoothing and accretion radii; and predicted column mass densities in $\mathrm{g} \cdot \mathrm{cm}^{-2}$. $\sigma$ is obtained integrating over every cell above the planet except the cells within the $d_{\mathrm{rsm}}$. The uncertainty is computed from the dispersion of the $\sigma$ value in the last $10$ orbits. Planet radii are taken from the evolutionary models of giant planets of \cite{spiegel2012}.}
    \label{tab:userparameters}
	\begin{tabular}{lccccccr}
		\hline
Run & $\alpha$ & Cell size [$R_\mathrm{Hill}$] & Cell size [$R_\mathrm{pl}$] & $d_\mathrm{rsm}$ [$R_\mathrm{Hill}$] & $r_{\mathrm{sink}}$ [$R_\mathrm{Hill}$] &  & $\sigma$  \\
\hline
$1M_\mathrm{J}$ & $0.003$ & $0.02$ & $7.3$ & $0.10$ & $0.07$ &  & $2.8\pm0.9$ \\

$2M_\mathrm{J}$ & $0.003$ & $0.02$ & $9.5$ & $0.10$ & $0.07$ &  & $0.8\pm0.6$ \\

$5M_\mathrm{J}$ & $0.003$ & $0.02$ & $11.8$ & $0.10$ & $0.07$ &  & $0.011\pm0.007$ \\

$1M_\mathrm{J}$, inviscid disc & $\sim 0$ & $0.02$ & $7.3$ & $0.10$ & $0.07$ &  & $1.2\pm0.3$ \\

$1M_\mathrm{J}$, doubled resol. & $0.003$ & $0.01$ & $3.7$ & $0.03$ & $0.03$ &  & $2.7\pm1.2$ \\

$5M_\mathrm{J}$, doubled resol. & $0.003$ & $0.01$ & $5.9$ & $0.04$-$0.05$ & $0.03$ &  & $0.01\pm0.01$ \\ 
		\hline
	\end{tabular}
\end{table*}

\begin{figure}
  \centering
  \includegraphics[width=1.0\linewidth]{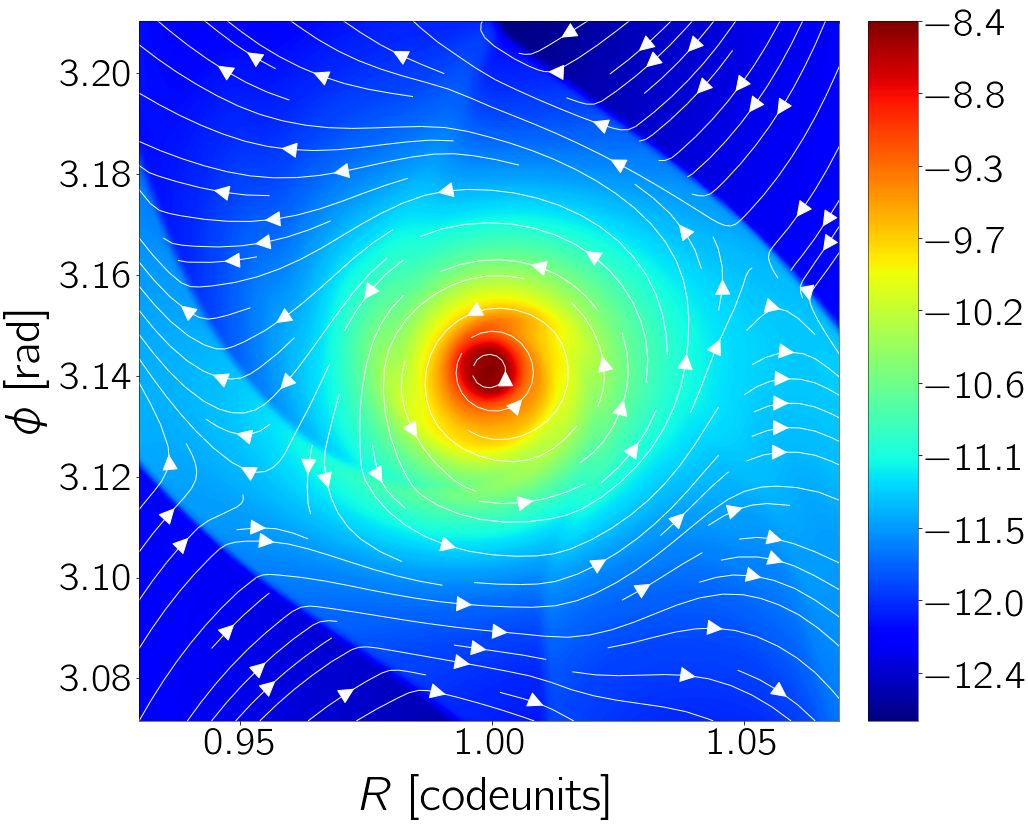}
  \caption[Gas stream lines near the planet location.]{Density map at the mid-plane near a $1$ $M_J$ planet. The gas streamlines are plotted on top, showing the circular motion of the gas around the planet. A value of 1 in radial code units is equivalent to $5.2$ $\mathrm{AU}$. The colour scale shows the logarithm of the density, in $\mathrm{g} \cdot \mathrm{cm}^{-3}$.}
  \label{fig:streamlinesXY}
\end{figure}

The resolution at the vicinity of the planet is crucial for our study in order to obtain realistic infrared (IR) optical depths due to the disc material close to the planet surface. To fulfil this, we used a 3-level nested grid with a maximum resolution of $0.02$ $R_\mathrm{Hill}$. For the $1$, $2$ and $5$ $M_\mathrm{J}$ planets, and considering planetary radii of $1.74$, $1.69$ and $1.87$ $R_J$ from the 1 Myr old \textit{hot-start} models of \cite{spiegel2012} (details of these models in Section~\ref{subsec:fluxes}), the maximum resolution corresponds to $7.3$, $9.5$ and $11.8$ planetary radii respectively. In the 2 additional runs with doubled resolution over the entire grid, the smallest cell-size was set to $0.01$ $R_\mathrm{Hill}$. To test the accuracy of our simulations, we inspected the gas streamlines in the vicinity of the planet (Section~\ref{subsec:streamlinesresults}). The grid is centred at the star location, thus close to the planet the grid appears to be Cartesian. As suggested by \cite{ormelkuiper, ormelshi}, the grid describes the CPD correctly if the gas streamlines form enclosed circular orbits around the planet location, which is confirmed in our simulations (Figure~\ref{fig:streamlinesXY}).

To save computational time, we assumed the disc to be symmetric with respect to the mid-plane. The simulated range for $\theta$ goes from the mid-plane up to $7^\circ$, adequate for a proper representation of the disc dynamics and the column density derivation. At the disc upper layers, density has decreased $\geq2$-$3$ orders of magnitude, and the contribution to the column densities of these upper layers is negligible. The details for the 3 grid regions are summarised in Table~\ref{tab:grid}.
\begin{table*}
	\centering
    \caption[Geometry of the resolution regions used in the simulations.]{Resolution and extension (as $\#$ of cells) for each coordinate (radial distance $R$, polar $\theta$ and azimuthal $\phi$ angles) of our 3-levels grid. There is no low-resolution level for the $\theta$ coordinate.}
    \label{tab:grid}
	\begin{tabular}{lccccr}
		\hline
 \multicolumn{1}{c}{ } & \multicolumn{2}{ c }{Level 1 (\textit{hi-res})} & \multicolumn{2}{ c }{Level 2 (\textit{mid-res})}  & \multicolumn{1}{ c }{Level 3 (\textit{low-res})}\\
Coord. & Resolution & \#cells & Resolution & \#cells & \#cells \\
		\hline
R & $0.02 R_\mathrm{Hill}$ & $128$ & $0.08 R_\mathrm{Hill}$ & $64$ & $128$ \\

$\theta$ & $0.02 R_\mathrm{Hill}$ & $64$  & $0.08 R_\mathrm{Hill}$ & $32$ & - \\

$\phi$ & $0.02 R_\mathrm{Hill}$ & $128$ & $0.08 R_\mathrm{Hill}$ & $128$ & $128$  \\

		\hline
	\end{tabular}
\end{table*}

\subsection{Intrinsic, accretion and total planet fluxes}\label{subsec:fluxes}
The total planet emission is considered as a combination of its intrinsic and accretion flux components. At the wavelengths studied, the intrinsic flux of the planet is expected to dominate, except in those cases with very high accretion rates onto the planet.

The intrinsic component of the planet flux is derived from the evolutionary models of \cite{spiegel2012}. These models provide the absolute magnitudes in $J$-, $H$-, $K$-, $L$-, $M$-, $N$- IR bands for a range of planet masses and as a function of age, up to $100Myr$. Following their nomenclature we refer to \textit{hot-start} and \textit{cold-start} models to the cases of a planet fully formed via disc instability or via core accretion respectively. In a more realistic scenario a planet would be characterised by an intermediate solution.

The total accretion luminosity is given by \citep{Frank1985}:
\begin{equation}\label{eq:laccr}
L_\mathrm{acc} \simeq \frac{G  M_\mathrm{pl} \dot
{M}_\mathrm{acc}}{R_\mathrm{acc}}\,
\end{equation}
where $M_\mathrm{pl}$ is the planetary mass, $\dot{M}_\mathrm{acc}$ is the mass accretion rate onto the planet and $R_\mathrm{acc}$ the accretion radius, the distance to the planet at which accretion shocks occur \citep{hartmann1998}. $R_\mathrm{acc}$ is typically $2$-$4$ times the planetary radius, we have chosen $R_\mathrm{acc} \equiv 4 R_\mathrm{pl}$ for the estimation of the accretion luminosity. The accretion shocks are not resolved in the simulations, this is out of the scope of this work. Nevertheless, the computation of $\dot{M}_\mathrm{acc}$ from the simulations (explained in last paragraph of this section) are independent of $R_\mathrm{acc}$. The accretion flux considered in this model accounts only for the accretion shocks' irradiation. Flux irradiated by the CPD is not taken into account.

The continuum emission of the accretion shocks is at temperatures ($\equiv T_{\mathrm{acc}}$) of the order of $\sim 10^4$ $K$ \citep{hartmann+2016}. To obtain the accretion flux in each band, we approximate the shocks' emission to a black body \citep{mendigutia+2011} that emits at a radius $R_{\mathrm{acc}}$. The surface area covered by the shocks is a fraction (defined as $b_{\mathrm{acc}}$) of the spherical surface with same radius. In this model, the accretion flux in a given band ($F_\mathrm{acc}^{\mathrm{band}}$) is computed as a fraction $b_{\mathrm{acc}}$ of the flux within the same band of a spherical black body ($F_\mathrm{bb}^{\mathrm{band}}$) with temperature $T_{\mathrm{acc}}$ and radius $R_{\mathrm{acc}}$:
\begin{equation}\label{eq:faccrband}
F_\mathrm{acc}^{\mathrm{band}} = b_\mathrm{acc} \cdot F_\mathrm{bb}^\mathrm{band}
\end{equation}

The factor $b_{\mathrm{acc}}$ can be estimated as the fraction between the total accretion luminosity $L_{\mathrm{acc}}$ (from Eq.~\ref{eq:laccr}) and the bolometric luminosity of the same black body:
\begin{equation}\label{eq:baccr}
b_\mathrm{acc} = \frac{L_\mathrm{acc}^\mathrm{tot}}{L_\mathrm{bb}^\mathrm{tot}}
\end{equation}
We tested various values of $T_{\mathrm{acc}}$, the results shown along this work are obtained using a shock temperature of $20 000$ $\mathrm{K}$. As discussed for the band fluxes results (Section~\ref{subsec:fluxesresults}), assuming lower $T_{\mathrm{acc}}$ does not vary the fluxes results significantly.

The gas accretion onto the planet is modelled following the prescription in \cite{kley1999}, also \cite{durmannkley}. At each time-step, a fraction $f_\mathrm{acc} \cdot \Delta t \cdot \Omega$ of the gas is removed from the cells enclosed by a sphere of radius $r_{\mathrm{sink}}$ centred at the planet (values for $r_{\mathrm{sink}}$ in Table~\ref{tab:userparameters}). This method mimics the direct accretion onto the planet surface. The values of $r_{\mathrm{sink}}$ have been chosen to guarantee convergence; as shown in \citet{tanigawa}, the estimated accretion rates converge to a stable value if $r_{\mathrm{sink}} \lesssim 0.07$ $R_{\mathrm{Hill}}$. This is independent of the $f_\mathrm{acc}$ value, that is set to $1$. The mass accreted is removed from the computational domain instead of being added to the planet mass (accreted mass is negligible compared to the total planet mass, about $10^{-5}$ times lower).

\subsection{Derivation of expected magnitudes}\label{subsec:model}
The extinction in our model is due to the disc material around the planet. We assume that there are no additional astronomical objects of significant brightness or size between the studied planet and the observer. Extinction in the $V$-band is derived from the column mass density obtained from the simulations. Using the magnitude-flux conversion formula from \citep{guever+2009}, assuming constant gas-to-dust ratio of $100$ and a molecular weight of $2.353$:
\begin{equation}\label{eq:A_V}
A_V [\mathrm{mag}] = \frac{N_H [\mathrm{atoms} \cdot {\mathrm{cm^{-2}}}]}{2.2\cdot10^{21} [\mathrm{atoms} \cdot {\mathrm{cm^{-2}}}\cdot{\mathrm{mag^{-1}}}]}
\end{equation}

The above relation was inferred from observations and it applies to an averaged interstellar medium (ISM) in the Milky Way. While the gas-to-dust ratio used is a good first approximation, variations may be expected, particularly in the vicinity of the planet. Recent surveys in close star-forming regions indicate that this ratio might indeed be different in many discs \citep{miotello+2017}.

From the inferred $A_V$, the extinction coefficients ($A_{\mathrm{band}}$) and optical depths ($\tau_{\mathrm{band}}$) are obtained using the diffuse ISM extinction curves of \cite{cardelli} for $J$-, $H$-, $K$-bands, and \cite{ChiarTielens2006} (which accounts for the silicate feature around $10\mu m$) for $L$-, $M$- and $N$-bands. The expected fluxes are then given by:
\begin{equation}\label{eq:fluxobserved}
F_\mathrm{expected}^{\mathrm{band}} = F_{\mathrm{pl}}^{\mathrm{band}} \cdot e^{- \tau_\mathrm{band}}
\end{equation}
with $F_{\mathrm{pl}}^{\mathrm{band}}$ being the total planet flux in that band. The resulting $F_\mathrm{expected}^{\mathrm{band}}$ is the expected band flux that we would observe for a planet embedded in a protoplanetary disc with ongoing accretion onto the planet.


\section{Model results}\label{sec:results}
In this section we first present the results from the HD simulations (subsection \ref{subsec:simresults}), followed by the results from our model (subsection \ref{subsec:modelresults})

\subsection{Results from the HD simulations}\label{subsec:simresults}
\begin{figure*}
  \centering
  \includegraphics[width=1.0\linewidth]{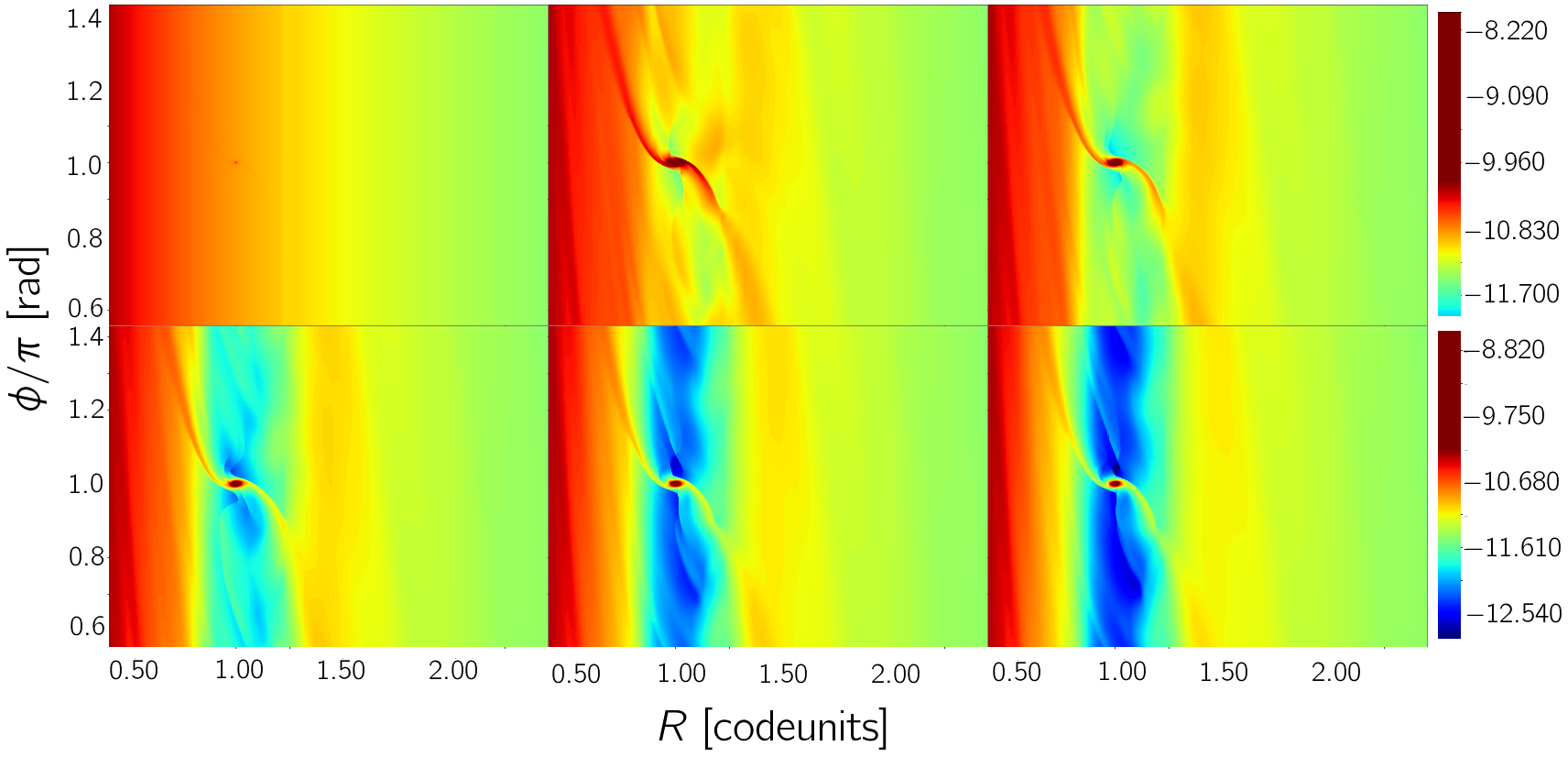}
  \caption[Evolution in time of the planet-disc system.]{Evolution in time of the 2D density map at the mid-plane for the viscous disc with an embedded $1$ $M_\mathrm{J}$ planet. A value of $1$ in radial code units is equivalent to $5.2$ $\mathrm{AU}$ (Jupiter semi-major axis). Density is represented in logarithmic scale, with values in $\mathrm{g} \cdot \mathrm{cm}^{-3}$. From top left to bottom right, each snapshot represents the density $\rho$ of the disc after $1$, $20$, $50$, $100$, $150$ and $200$ orbits.}
  \label{fig:1mcartesianevolution}
\end{figure*}

Jupiter-size planets embedded in a disc generally carve a gap  after several orbits \citep{bryden1999}. In our simulations, this can be seen in 2D density maps of the disc as a function of time (Figure~\ref{fig:1mcartesianevolution}, for the $1$ $M_{\mathrm{J}}$ case). For more massive planets, the gap carving process is faster, as one would expect from planet formation theory.

\begin{figure}
  \centering
  \includegraphics[width=1.0\linewidth]{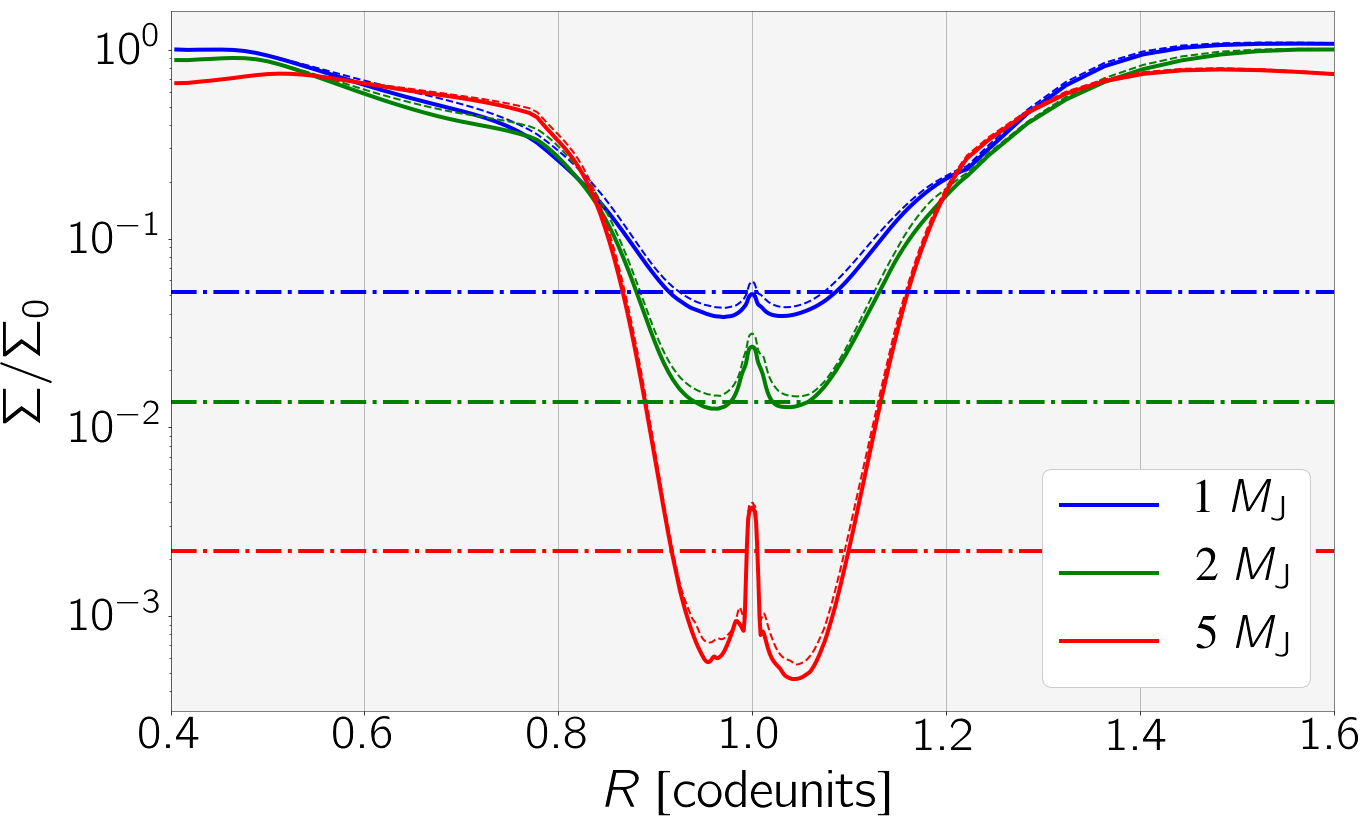}
  \caption[Gap depth for each planet.]{Surface density radial profile for $1$, $2$ and $5$ $M_J$ planets after the simulations reach a steady-state, shown as solid lines. The dash-dot lines are the respective predicted gap-depths, derived from an analytical model for gaps in protoplanetary discs \citep{duffell2015}. The dashed lines represent the surface density after $20$ orbits. A value of $1$ in radial code units is equivalent to $5.2$ $\mathrm{AU}$.}
  \label{fig:gapdepth}
\end{figure}

The gap opened by each planet can be compared to disc models to verify the quality of our simulations. We tested our resulting surface density profiles with an analytical model for gaps in protoplanetary discs, as described in \citet{duffell2015}. An algebraic solution of the gap profiles is presented in that work, together with the derivation of a formula for the gap depth. In Figure~\ref{fig:gapdepth} we show the azimuthally averaged surface density radial profiles relative to the unperturbed surface density ($\Sigma_0$) for the $1$, $2$ and $5$ $M_J$ simulated planets in a viscous disc. The solid lines represent the profiles after a steady-state is reached, and the dashed lines denote the surface density after the first $20$ orbits. The predicted gap depths from the model \citep[][equation 9]{duffell2015} are shown in the figure as horizontal dash-dot lines.

Our results for $1$ and $2$ $M_J$ planets are in very good agreement with the analytical model. The gap for a $5$ $M_J$ planet is relatively deeper than the prediction from the model. Nevertheless, the model by \citet{duffell2015} fails at reproducing the gap profile produced by planets with very high masses, as discussed in that work. Therefore, we can trust the quality of our simulations from the concordance at low planet masses with analytical models.

\subsubsection{Column mass density}\label{subsec:coldensresults}
\begin{figure}
  \centering
  \includegraphics[width=1.0\linewidth]{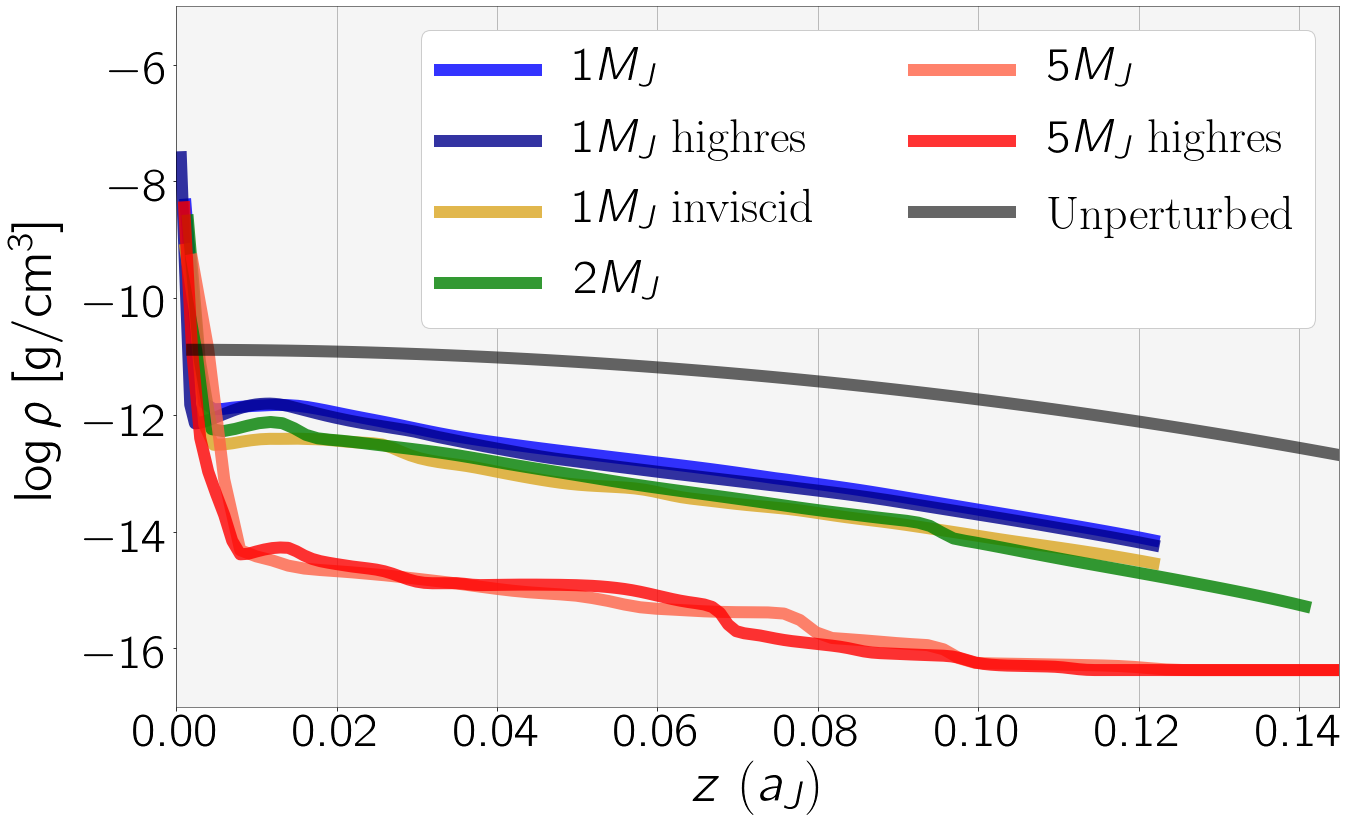}
  \caption[Vertical density evolution for each simulation.]{Vertical density above the planet for each simulation. Horizontal axis represents the height from the mid-plane in Jupiter semi-major axis ($a_J$); vertical axis shows density in logarithmic scale, in $\mathrm{g} \cdot \mathrm{cm}^{-3}$ units.}
  \label{fig:verticaldensityall}
\end{figure}

The column density is obtained by integrating the disc density over the line-of-sight towards the planet. We considered our system to be face-on, since is the most likely geometry for a direct protoplanet detection. The density above the planet for every simulation is shown in Figure~\ref{fig:verticaldensityall}, together with the unperturbed (initial) density. The highest densities correspond to the $1$ $M_\mathrm{J}$ simulation. For more massive planets, the disc densities are lower, since the planet carves a deeper gap. In the inviscid case, the carved gap can not be refilled with adjacent material, resulting in lower densities compared to the viscously evolving case.

\begin{figure}
  \centering
  \includegraphics[width=1.0\linewidth]{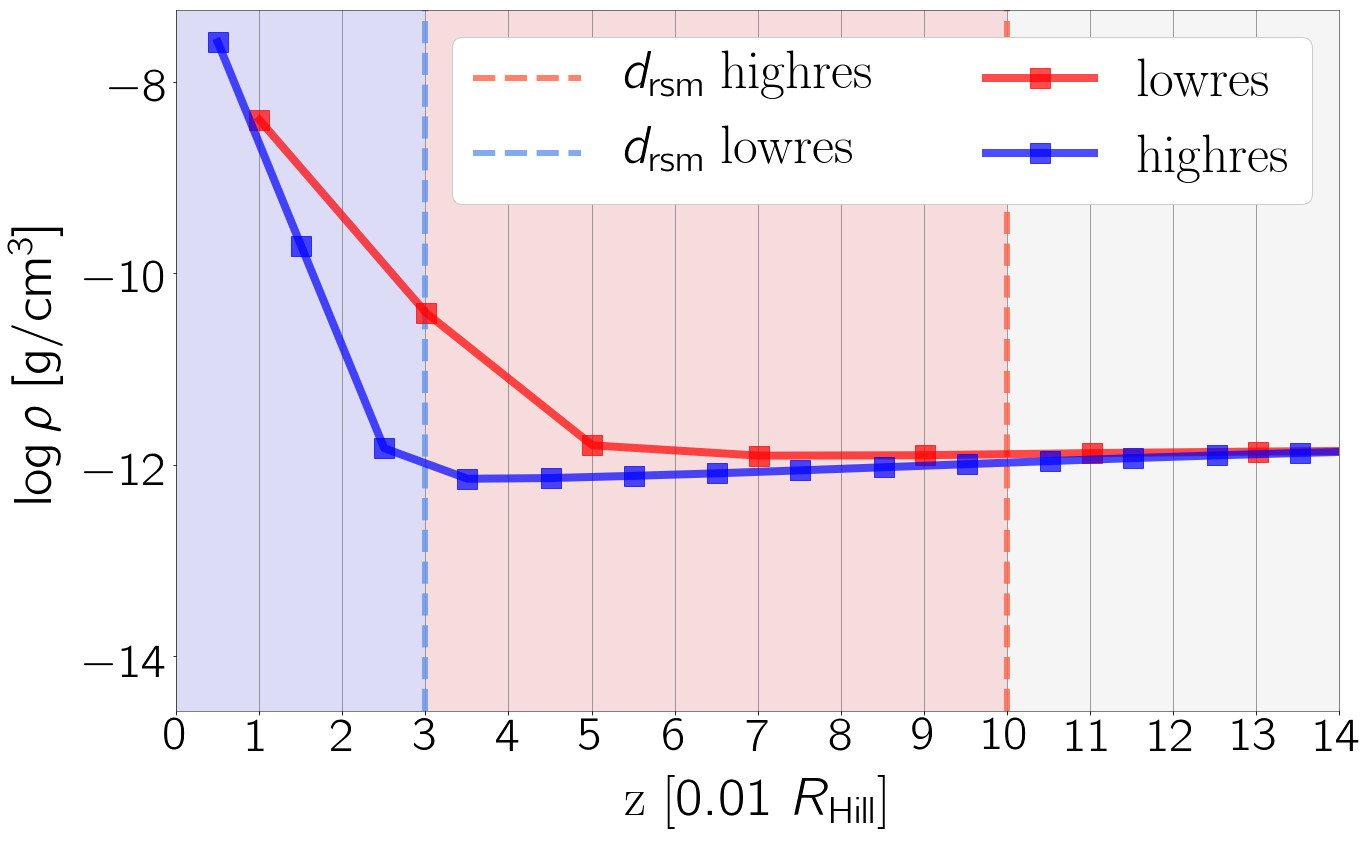}
  \caption[Vertical density profile near the planet.]{Vertical density profile close to the $1$ $M_\mathrm{J}$ planet, comparing the doubled resolution (blue line) to the original case (red). The vertical grid spacing is $0.01$ $R_\mathrm{Hill}$, equivalent to the cell-size for the doubled resolution run, and half the cell-size of the original run. The red and blue dashed lines represent the smoothing radii for the original and the new run respectively.}
  \label{fig:highrescolumnzoom}
\end{figure}

The planet and its atmosphere are not resolved in these simulations, therefore the first cell is assumed to be the planet outer radius. The sharp peak from the vertical density profiles extends over the first 2-3 cells. This is expected to be a combination of several effects, mainly an artefact of the simulations due to the potential smoothing (Eq.~\ref{eq:potentialexpansion}), which affects every cell within $d_\mathrm{rsm}$. However we cannot exclude the possibility that a fraction of the peak might also be the real density stratification of the material at the layers closest to the planet. Additionally, this over-density is also altered by the accretion radius (see \ref{subsec:fluxes}), and limited by the grid resolution. A fully resolved CPD would be necessary to disentangle between the different causes. However, we tested whether the extension of the peak is an artefact due to the potential softening and the mentioned resolution limitations. To this aim we performed 2 additional simulations with doubled resolution over the entire grid. We also decreased both accretion and smoothing radii to $0.03$-$0.05$ $R_\mathrm{Hill}$. The values of the main parameters for the doubled resolution runs are summarised in Table~\ref{tab:userparameters}. This test was done for the discs with $1$ and $5$ $M_\mathrm{J}$ planets. The grid from the last snapshot of the original simulations were readjusted to the new resolution, and we let the system evolve until a new steady-state was reached ($\leq 15$ orbits needed in both cases). The vertical density at the closest cells above the planet for both original and doubled resolution for the $1$ $M_\mathrm{J}$ case are shown in Figure~\ref{fig:highrescolumnzoom}. The vertical grid-lines represent $0.01$ $R_\mathrm{Hill}$, and the coloured lines illustrate the $d_{\mathrm{rsm}}$ in both original and doubled resolution runs. The figure shows that the over-density with doubled resolution spans approximately half the original case. This is in accordance with what we expect if the over-density is due to the smoothing within $d_{\mathrm{rsm}}$. If we were able to completely remove the potential smoothing we would only see a peak inside the planet radius, i.e. within the first cell.

From the results of this test, we consider the column mass density $\sigma$ as the integrated density for all the cells above the smoothing radius. The resulting $\sigma$ for each of the simulated systems are included in Table~\ref{tab:userparameters}. The uncertainty considered is the dispersion of the column mass density for the last $10$ orbits of each simulation. The predicted magnitudes for $1$ and $5$ $M_J$ planets are derived using the $\sigma$ values from the doubled resolution runs, since in these cases the planet-disc interaction is represented more accurately. The values for the original and doubled resolution cases are within their respective uncertainty: for a $1$ $M_J$ we obtained $2.8\pm0.9$ and $2.7\pm1.2$ $\mathrm{g \cdot \mathrm{cm}^{-2}}$, while for the $5$ $M_J$ runs the column mass densities were $0.011\pm0.007$ and $0.01\pm0.01$ $\mathrm{g \cdot \mathrm{cm}^{-2}}$ respectively.

\subsubsection{Mass accretion rates}\label{subsec:maccrresults}
The prescription used to derive the mass accretion rates from the simulations was described in Subsection \ref{subsec:fluxes}. The final value of $\dot{M}_\mathrm{acc}$ for each simulated planet is the optimal value of parameter $c$ when fitting the evolution of $\dot{M}_\mathrm{acc}$ in the last $50$ orbits by an exponential function defined as $f(x) = a \cdot e^{-b \cdot x} + c$. The resulting accretion rates are of the order of $10^{-8}$ $M_{\odot} \cdot \mathrm{yr}^{-1}$, summarised in Table~\ref{tab:bolfluxes}. The highest $\dot{M}_\mathrm{acc}$ is obtained for a system with $2$ $M_\mathrm{J}$ planet. This is somewhat counter-intuitive, as one may expect the least massive planet to have the lowest accretion rate, and the most massive planet to have the highest. There are two effects to consider, the viscosity that refills the gap and the gravitational force of the planet: a more massive planet creates a stronger gravitational potential, a larger CPD and has a larger accretion rate, on the other hand if the disc evolves viscously it can replenish the accreted material with new material, thus keeping a higher $\dot{M}_\mathrm{acc}$ value. Beyond a certain planet mass, the gravitational potential clears large regions quicker, which limits the refill of the gap material thus ultimately reducing the $\dot{M}_\mathrm{acc}$. Our results indicate that the mass of the planet for which this occurs is between $2$ and $5$ $M_\mathrm{J}$. The different accretion rates obtained for the $1$ $M_\mathrm{J}$ planet in a viscous and an inviscid disc can be understood by considering that an inviscid disc cannot replenish the gap opened by a planet, and consequently there is less material to feed the CPD.

Care should be taken when considering these results because of the limitations posed by isothermal simulations, which do not account for accretion heating in the vicinity of the planet. This results in an overestimation of the accretion rate. These simulations provide an upper limit of the expected accretion rates, and this should be taken into account when the model is applied to real systems.

\subsubsection{Gas streamlines}\label{subsec:streamlinesresults}
Gas streamlines provide useful insights on the disc behaviour and the planet-disc interaction. In the close-up top view of the system (Figure~\ref{fig:streamlinesXY}), the gas streamlines at the system mid-plane are plotted as vectors. As expected from the accretion models, a CPD is formed and the gas motion is concentric to the planet. The gas streamlines follow closed trajectories in the regions nearest to the planet, which indicates that the grid does describe the accreting planet with a CPD accurately \citep[as discussed in][]{ormelkuiper}.

\begin{figure}
  \centering
  \includegraphics[width=1.0\linewidth]{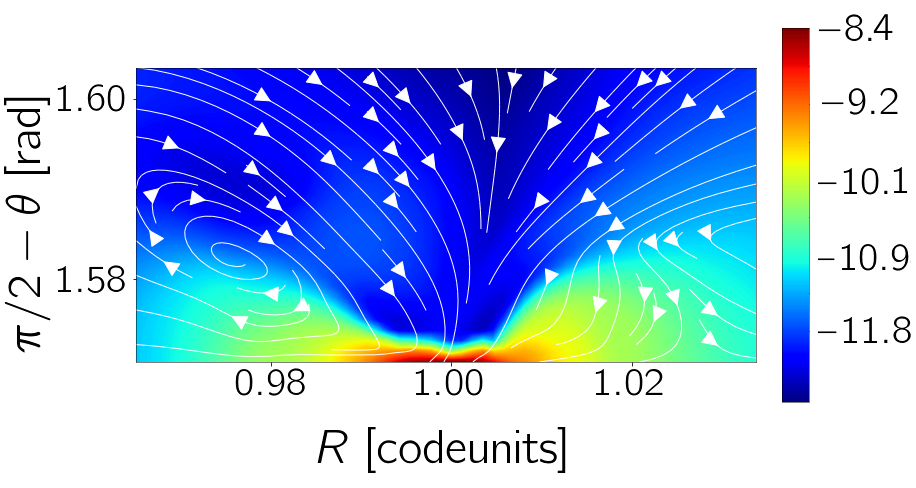}
  \caption[Gas stream lines near the planet location, cut view.]{Edge on view density map with gas streamlines around the planet. An important part of the gas is being accreted vertically. A value of 1 in radial code units is equivalent to $5.2$ $\mathrm{AU}$. The scale represents the logarithm of the density, in $\mathrm{g} \cdot \mathrm{cm}^{-3}$.}
  \label{fig:streamlinesRTheta}
\end{figure}

Accretion onto the planet occurs not only in the orbital plane from the CPD but also vertically \citep[see e.g.][]{szulagyi2016}. In our simulated systems, the gas is indeed falling onto the planet from its pole or with small inclination angle from the vertical. Figure~\ref{fig:streamlinesRTheta} shows that a high fraction of the material is being accreted vertically onto the $1$ $M_{J}$ planet.

\subsection{Predicted planet magnitudes}\label{subsec:modelresults}

\subsubsection{Bolometric and band fluxes}\label{subsec:fluxesresults}
\begin{figure*}
    \begin{center}
    \includegraphics[width=.495\textwidth]{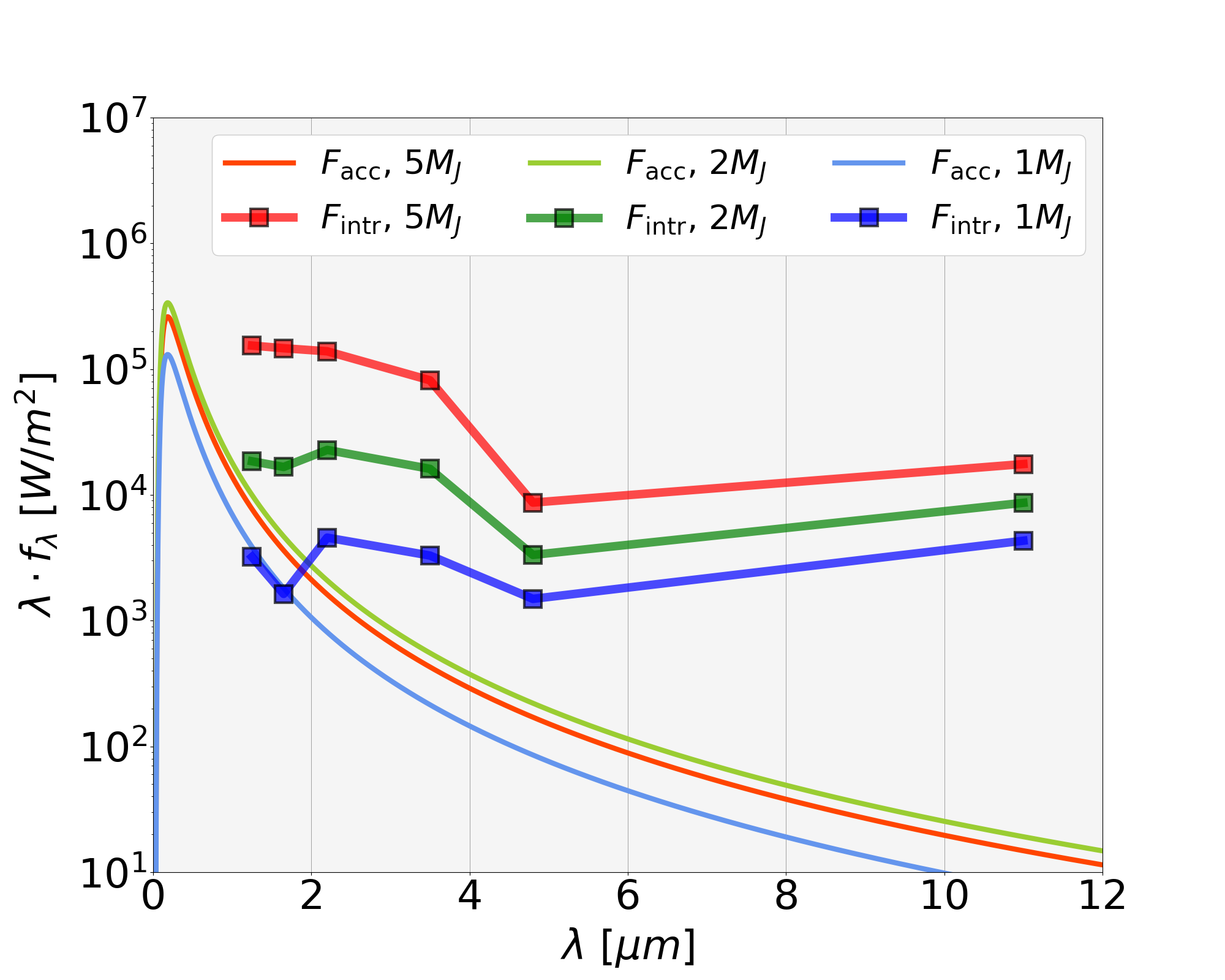}
    \includegraphics[width=.495\textwidth]{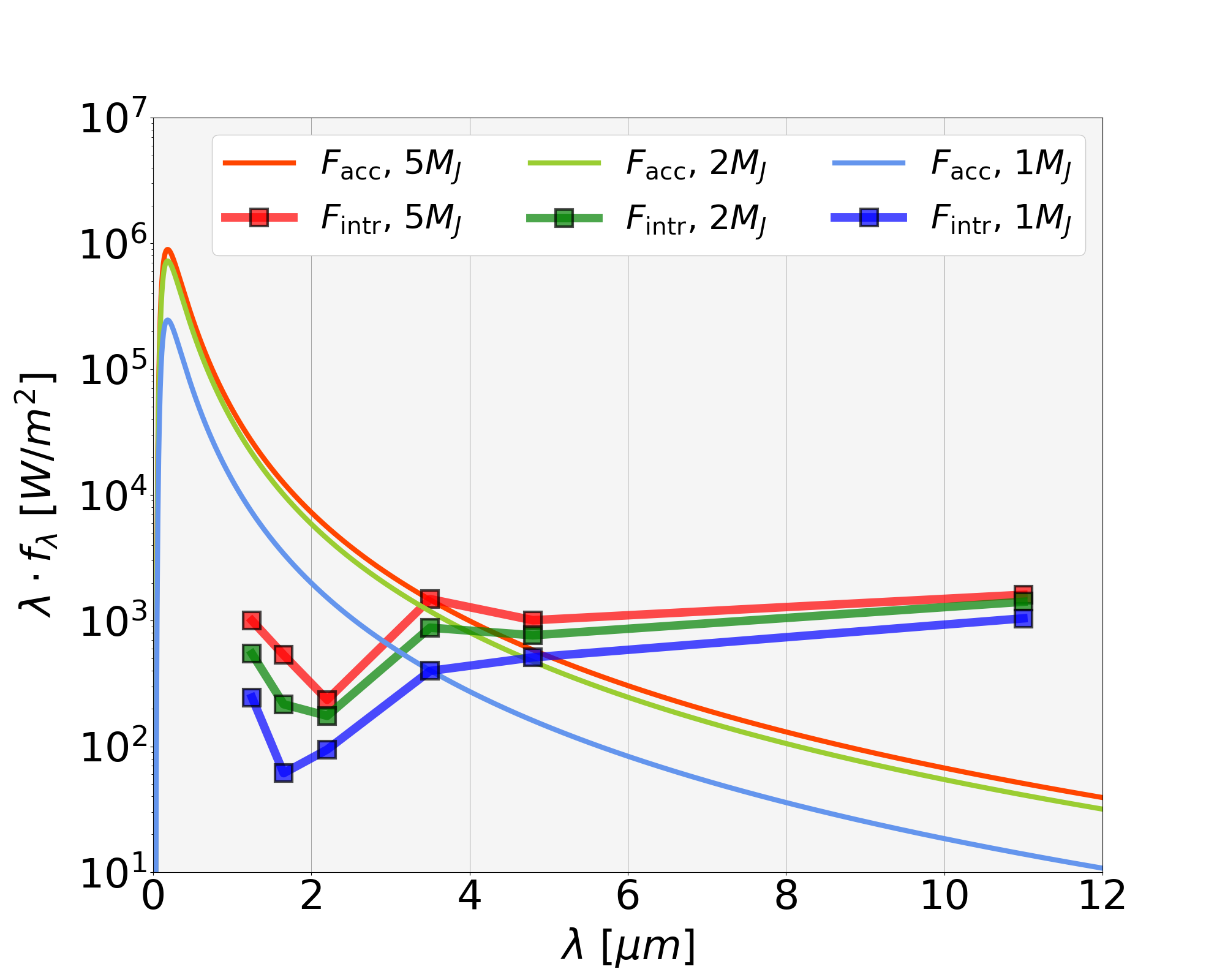}
    \end{center}
  \caption[Accretion flux vs. Intrinsic flux.]{Intrinsic ($F_{\mathrm{intr}}$) and accretion fluxes ($F_{\mathrm{acc}}$) for $1$, $2$ and $5$ $M_J$ planets. The results for planets with a \textit{hot-start} model are shown on the left panel, while \textit{cold-start} models on the right. The intrinsic fluxes are plotted at the central wavelength of each band.}
  \label{fig:fluxesresults}
\end{figure*}

Accretion and intrinsic bolometric fluxes for each planet considered in this work are shown in Table~\ref{tab:bolfluxes}. Mass accretion rates are also included in the table. The difference in $F_\mathrm{acc}$ between \textit{hot} and \textit{cold-start} models arises from the different planet radius $R_{\mathrm{pl}}$ of each model \citep[taken from][]{spiegel2012}, since the $R_{\mathrm{acc}}$ used to compute the accretion flux (Equation~\ref{eq:laccr}) is assumed to be $\equiv 4 R_{\mathrm{pl}}$. The bolometric accretion flux is higher than the planet's intrinsic flux for all cases. Nevertheless, the accretion flux (which peaks at $0.15$ $\mathrm{\mu m}$, with $T_\mathrm{eff} \sim 20000$ $\mathrm{K}$) is in most cases lower in the IR bands considered than the intrinsic planet flux, whose spectrum peaks in the IR ($\sim1$-$10$ $\mu m$). This can be seen in the left panel of Figure~\ref{fig:fluxesresults}. The figure shows the accretion and intrinsic fluxes of planets with $1$, $2$ and $5$ $M_\mathrm{J}$ for \textit{hot} and \textit{cold-start} planet models. Intrinsic fluxes are considerably lower for \textit{cold-start} planets than for the \textit{hot} models. The accretion contribution can be significant, especially in the \textit{cold-start} cases in $J$-, $H$-, $K$- and $L$-bands. Reducing the $T_\mathrm{acc}$ shifts the accretion maximum to longer wavelengths, but it highly decreases its bolometric value, and consequently the overall picture does not vary significantly.
\begin{table}
	\centering
    \caption[Bolometric fluxes results.]{Mass accretion rates and bolometric fluxes for every simulated system. Fluxes are expressed in [$\mathrm{W} \cdot \mathrm{m}^{-2}$].}
    \label{tab:bolfluxes}
	\begin{tabular}{lccccr}
		\hline
$M_\mathrm{pl}$ & $\dot{M}_\mathrm{acc}$[$M_{\odot} \cdot \mathrm{yr}^{-1}$] & $F_\mathrm{intr}^\mathrm{Hot}$ & $F_\mathrm{acc}^\mathrm{Hot}$ & $F_\mathrm{intr}^\mathrm{Cold}$ & $F_\mathrm{acc}^\mathrm{Cold}$ \\
		\hline
		
$1M_\mathrm{J}$ & $5.7e{-8}$ & $2.8e{4}$ & $1.7e{5}$ & $4.8e{3}$ & $3.2e{5}$ \\

$2M_\mathrm{J}$ & $6.8e{-8}$ & $1.2e{5}$ & $4.3e{5}$ & $8.9e{3}$ & $9.3e{5}$ \\

$5M_\mathrm{J}$ & $2.8e{-8}$ & $7.4e{5}$ & $3.3e{5}$ & $1.2e{4}$ & $1.1e{6}$ \\

$1M_\mathrm{J}$ inviscid & $1.0e{-8}$ & $2.8e{4}$ & $3.0e{4}$ & $4.8e{3}$  & $5.6e{4}$ \\

		\hline
	\end{tabular}
\end{table}

These results indicate that radiation from accretion shocks near the planet is an important factor of the planet flux. Nevertheless, it is worth keeping in mind that these results are for accretion rates inferred from isothermal simulations, which are generally overestimated. For more realistic accretion rates at this stage ($\sim {10}^{-10}$ $M_{\odot} \cdot \mathrm{yr}^{-1}$), intrinsic flux dominates at these wavelengths. When scaling our models to larger distances from the host star, accretion rates are highly reduced due to the scaling for lower disc densities. Consequently intrinsic fluxes dominate in planets further out in the disc for both \textit{hot} and \textit{cold-start} scenarios.

\subsubsection{Extinction coefficients and predicted magnitudes}\label{subsec:expectedmagresults}
\begin{figure*}
    \begin{center}
    \includegraphics[width=.495\textwidth]{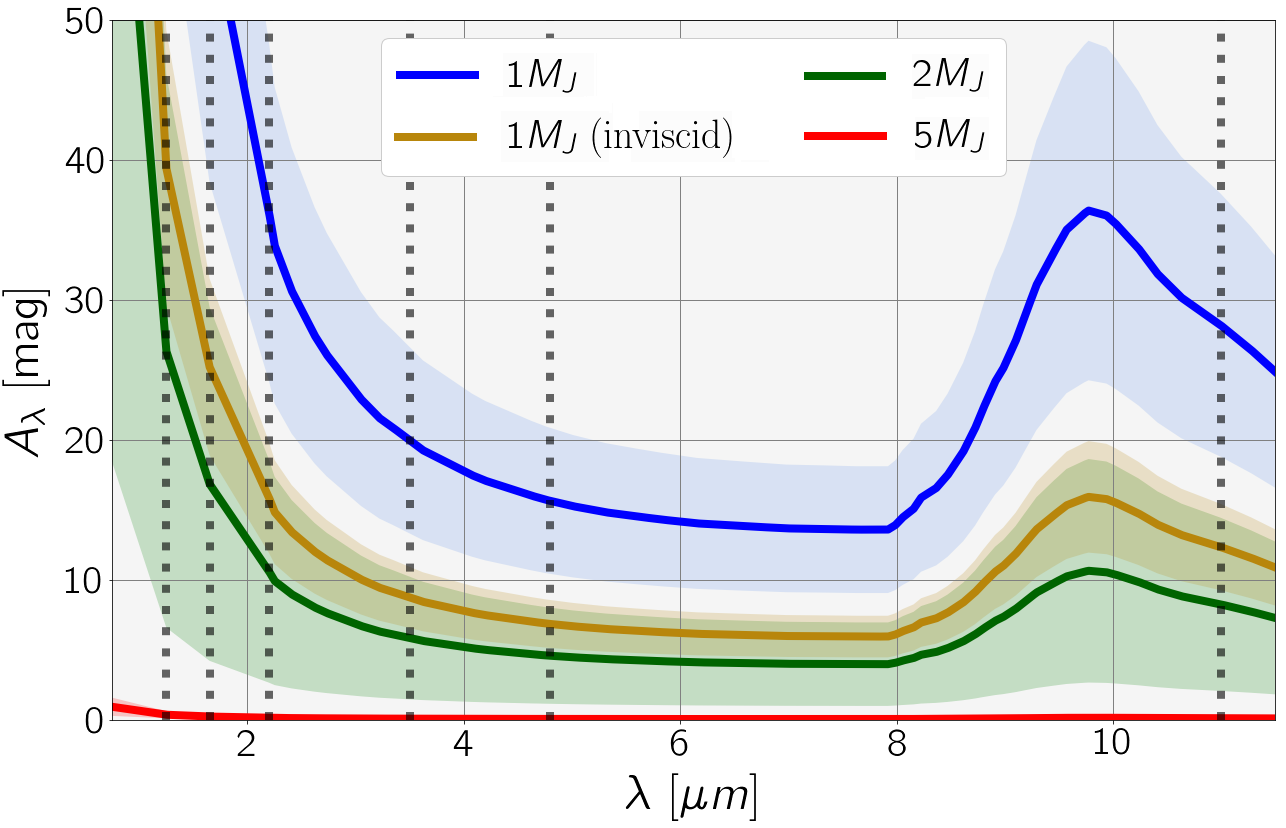}
    \includegraphics[width=.495\textwidth]{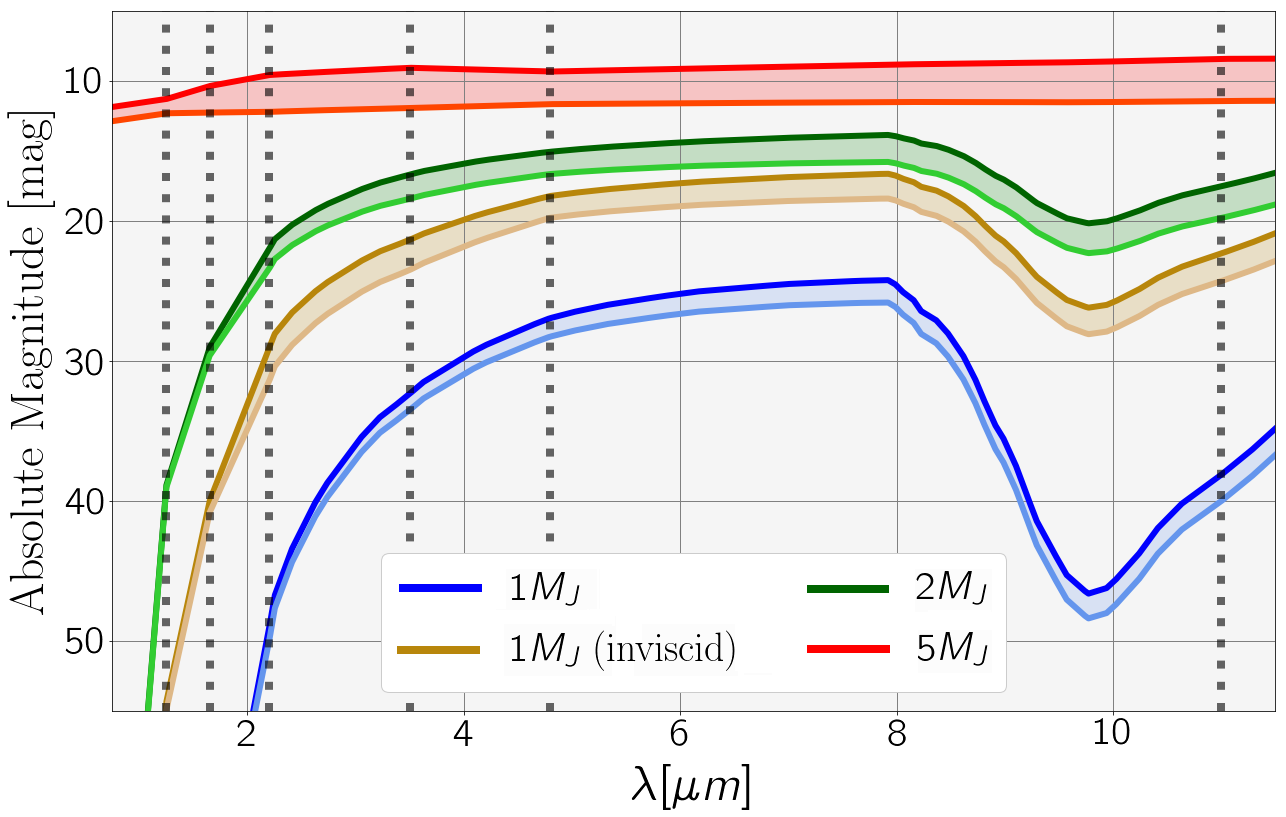}
    \end{center}
    \begin{center}
    \includegraphics[width=.495\textwidth]{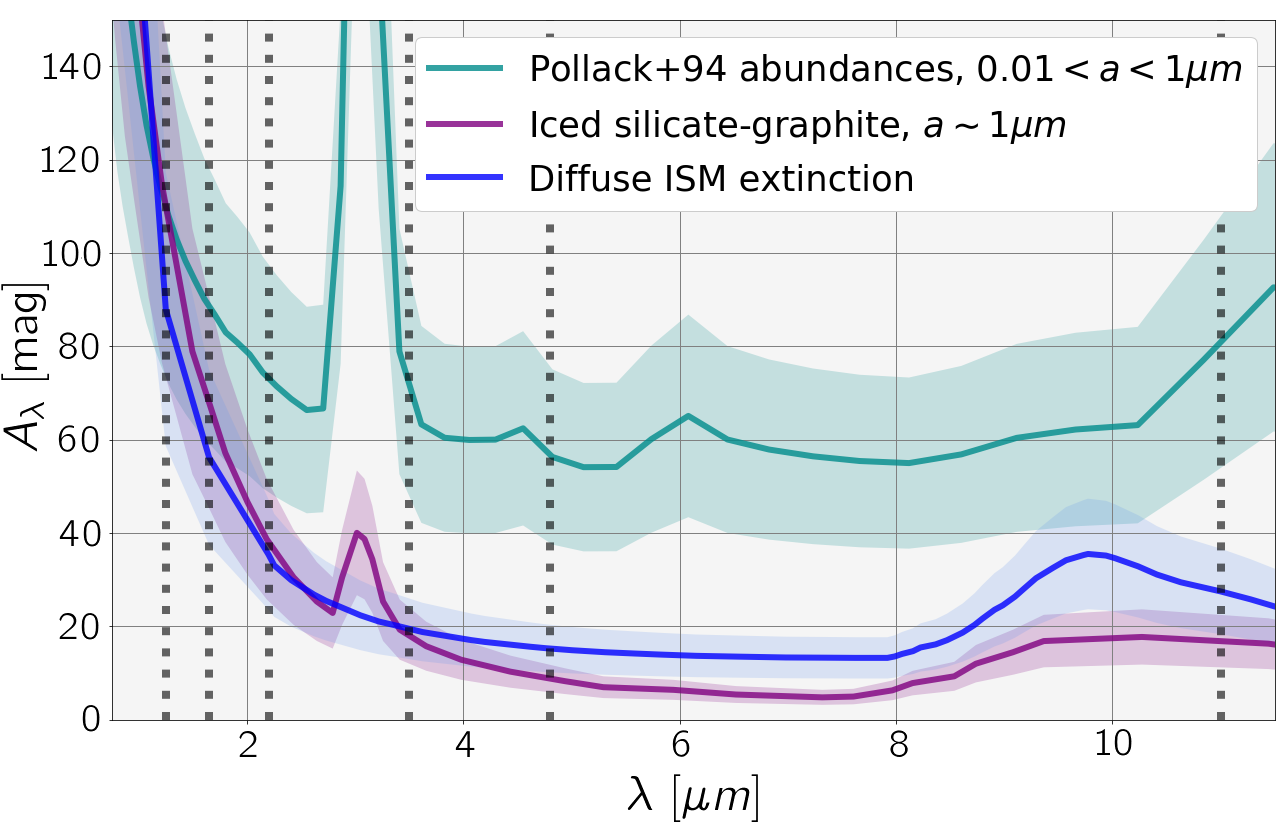}
    \includegraphics[width=.495\textwidth]{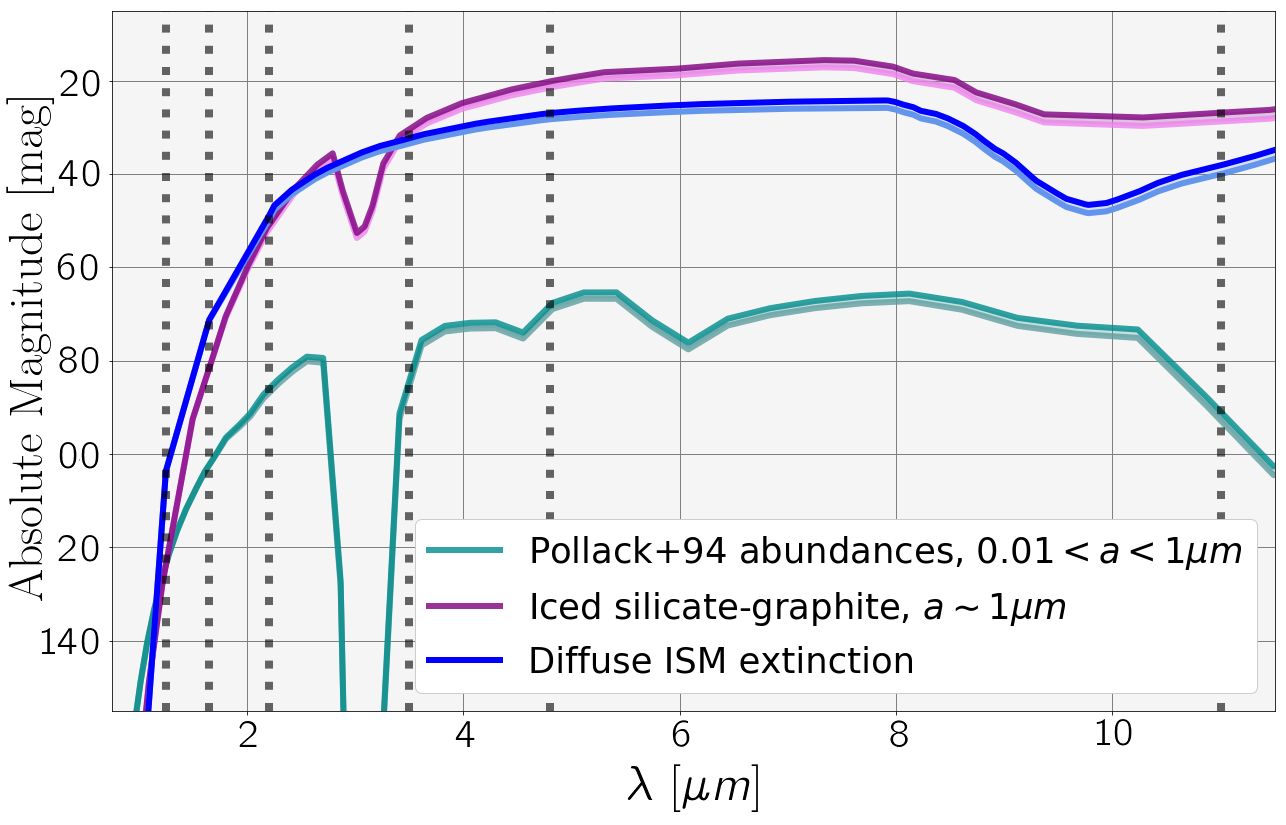}
    \end{center}
  \caption[]{
  Top panels: extinction coefficients with uncertainties (left), and predicted magnitudes for the simulated systems (right), both as a function of wavelength. The predicted planet magnitudes are shown as an area delimited by the \textit{hot} and \textit{cold} planetary model. Bottom panels: extinction and predicted magnitudes of the $1$ $M_J$ viscous case using different dust grain models. The results for the various dust models are normalised at $A_V$. For every panel, the vertical dotted lines represent (from left to right) the central wavelength of $J$-, $H$-, $K$-, $L$-, $M$-, and $N$-bands.}
  \label{fig:magexpected}
\end{figure*}

From the column mass densities we derived extinction coefficients for each planet in $J$- $H$-, $K$-, $L$-, $M$- and $N$-bands. The inferred coefficients are included in Table~\ref{tab:expectedmags5au}, and plotted on the top left panel in Figure~\ref{fig:magexpected}. Extinction coefficients at wavelengths $\lesssim 2$ $\mu m$ are extremely high for $1$ and $2$ $M_J$. The effect of extinction decreases for longer wavelengths, but it raises up again at $\sim8$-$12\mu m$ due to the silicate feature present in the diffuse ISM. For a $5$ $M_J$, extinction coefficients are very low in every IR band. This is due to its ability to open a gap in the disc very effectively, and consequently, disc density around the planet and the inferred column density are very low.

\begin{table*}
	\centering
\caption[Magnitudes for planets at $5.2 \mathrm{AU}$.]{Absolute magnitudes for planets at $5.2$ $\mathrm{AU}$ to the host star, with masses: $1$ $M_\mathrm{J}$ --viscous and inviscid scenarios-- $2$ $M_\mathrm{J}$ and $5$ $M_\mathrm{J}$. $\mathrm{Mag}_\mathrm{pl}$ is the total magnitude of the planet, including accretion flux; $A_{\mathrm{band}}$ is the extinction coefficient in each band, $\mathrm{Mag}_\mathrm{expected}$ is the magnitude of the planet considering extinction due to disc material. All values are in $\mathrm{mag}$.}
\label{tab:expectedmags5au}
	\begin{tabular}{lccccccccccccccr}
		\hline
 \multicolumn{3}{c}{ } & \multicolumn{6}{ c }{\textit{Hot-start} planet} & \multicolumn{1}{c}{ } & \multicolumn{6}{ c }{\textit{Cold-start} planet} \\
 &  &  & J & H & K & L & M & N &  & J & H & K & L & M & N \\
		\hline
 & $\mathrm{Mag}_{\mathrm{pl}}$ &  & $13.74$ & $13.75$ & $12.92$ & 12.40 & 11.33 & 10.00 &  & $13.72$ & $13.81$ & $13.84$ & $13.56$ & $12.65$ & $11.88$ \\ 
$1M_{\mathrm{J}}$ & $A_{\mathrm{band}}$ &  & $91.32$ & $58.41$ & $36.75$ & $19.96$ & $15.63$ & $28.15$ &  & $91.32$ & $58.41$ & $36.75$ & $19.96$ & $15.63$ & $28.15$ \\ 
 & $\mathrm{Mag}_{\mathrm{expected}}$ &  & $105.07$ & $72.16$ & $49.67$ & $32.36$ & $26.96$ & $38.16$ &  & $105.04$ & $72.21$ & $50.59$ & $33.52$ & $28.28$ & $40.04$ \\ 
 &  &  &  &  &  &  &  &  &  &  &  &  &  &  &  \\
 & $\mathrm{Mag}_{\mathrm{pl}}$ &  & $12.58$ & $12.22$ & $11.44$ & $10.87$ &  $10.51$ &  $9.31$ &  & $12.72$ & $12.80$ & $12.84$ &  $12.60$ &  $12.10$ &  $11.58$ \\ 
$2M_\mathrm{J}$ & $A_{\mathrm{band}}$ &  & $26.33$ & $16.84$ & $10.60$ &  $5.84$ &  $4.57$ &  $8.24$ &  & $26.33$ & $16.84$ & $10.60$ &  $5.84$ &  $4.57$ &  $8.24$ \\ 
 & $\mathrm{Mag}_{\mathrm{expected}}$ &  & $38.91$ & $29.06$ & $22.04$ &  $16.71$ &  $15.08$ &  $17.55$ &  & $39.06$ & $29.65$ & $23.44$ &  $18.44$ &  $16.67$ &  $19.81$ \\ 
 &  &  &  &  &  &  &  &  &  &  &  &  &  &  &  \\
 & $\mathrm{Mag}_{\mathrm{pl}}$ &  & $11.09$ & $10.21$ & $9.48$ &  $9.00$ &  $9.29$ &  $8.33$ &  & $12.61$ & $12.69$ & $12.73$ &  $11.88$ &  $11.62$ &  $11.34$ \\ 
$5M_\mathrm{J}$ & $A_{\mathrm{band}}$ &  & $0.36$ & $0.23$ & $0.15$ &  $0.08$ &  $0.06$ &  $0.11$ &  & $0.36$ & $0.23$ & $0.15$ &  $0.08$ &  $0.06$ &  $0.11$ \\ 
 & $\mathrm{Mag}_{\mathrm{expected}}$ &  & $11.45$ & $10.45$ & $9.63$ &  $9.08$ &  $9.35$ &  $8.45$ &  & $12.97$ & $12.92$ & $12.88$ &  $11.96$ &  $11.68$ &  $11.45$ \\ 
 &  &  &  &  &  &  &  &  &  &  &  &  &  &  &  \\
 & $\mathrm{Mag}_{\mathrm{pl}}$ &  & $14.98$ & $14.83$ & $13.28$ &  $12.63$ &  $11.39$ &  $10.02$ &  & $15.56$ & $15.66$ & $15.62$ &  $14.79$ &  $12.95$ &  $12.00$ \\ 
$1M_{\mathrm{inviscid}}$ & $A_{\mathrm{band}}$ &  & $39.40$ & $25.20$ & $15.86$ &  $8.74$ &  $6.84$ &  $12.32$ &  & $39.40$ & $25.20$ & $15.86$ &  $8.74$ &  $6.84$ &  $12.32$ \\ 
 & $\mathrm{Mag}_{\mathrm{expected}}$ &  & $54.38$ & $40.03$ & $29.14$ &  $21.37$ &  $18.24$ &  $22.35$ &  & $54.96$ & $40.86$ & $31.48$ &  $23.52$ &  $19.79$ &  $24.33$ \\ 
		\hline
	\end{tabular}
\end{table*}

The derived magnitudes for each simulated planet as a function of wavelength are shown on the top right panel in Figure~\ref{fig:magexpected}. For each planet, the upper and lower curves represent its \textit{hot} and \textit{cold} models. The results include extinction from the disc material and radiation from the accretion shocks near the planet. The magnitudes for every IR-band and planet at $5.2$ $\mathrm{AU}$ are summarised in Table~\ref{tab:expectedmags5au}. For each simulated planet, the rows in the table show: the absolute magnitude of the planet including the contribution from accretion; the extinction coefficients due to the disc material, and the predicted absolute magnitude including extinction effects.

The predicted magnitudes for a given planet decrease as a function of wavelength, except at $\sim10 \mu m$ due to the silicate feature in dust grain opacities. The curves are more flattened for more massive planets. This is due to the more efficient depletion of material in the planet vicinity, which yields lower column densities, hence lower extinction. For the most massive planet considered ($5$ $M_J$), the gap clearing is extremely effective and extinction in any IR band is negligible. In the inviscid scenario with a $1$ $M_\mathrm{J}$, the gap can not be replenished efficiently compared to the viscously evolving case. This results on a gap with lower density and extinction. At $\lesssim 2$ $\mu m$, $1$-$2$ $M_J$ planets are completely obscured by the disc material (with extinction above 30 and 15 $\mathrm{mag}$ respectively). These results are obtained for an unperturbed surface density of $\Sigma = 127$ $\mathrm{g} \cdot \mathrm{cm}^{-2}$ at $5.2$ $\mathrm{AU}$.

We have used the ISM law to estimate the extinction under the assumption that mostly small grains will be present in the disc atmosphere above the planet. The actual value of extinction depends on the assumption on the dust properties. To investigate the implications of our assumption, we evaluated the impact of using different assumptions on the dust composition and size distributions. The results for the $1$ $M_J$ viscously evolving case are shown in bottom panels of Figure~\ref{fig:magexpected}. Two other dust models were investigated: one model of grains with fractional abundances comparable to the expected in protoplanetary discs mid-plane \citep[][]{pollack+1994} and grain population with number density $n(a)\propto a^{-3.5}$ (where $a$ is the grain size) between $0.01\mu m < a < 1 \mu m$ \citep[][for details]{tazzari+2016}, and a dust coagulation model for ice-coated silicate-graphite aggregates \citep[type II grain mixing, see][]{ormel+2009,ormel+2011}, applicable to dust in protoplanetary discs (extinction shown is for grain sizes $a \sim 1\mu m$). The results of the ice silicate graphite model are in very good agreement with the ISM extinction used (in $M$-, and $N$-bands the diffuse ISM extinction becomes larger). The other model provides similar results in the $J$-, and $H$-bands, however, for longer wavelengths extinctions are $\sim2$-$3$ times larger than for the diffuse ISM. Thus, when considering dust with different properties (e.g. composition, size, level of processing), the resulting predicted planet magnitudes might change due to the opacity variations in IR wavelengths.

On the other hand, the extinction is obtained assuming a gas-to-dust ratio of 100 along the disc. In the atmospheric layers of the disc above the planet, this ratio might be larger due to dust processing and settling. In this regard, our analysis provides a conservative estimate of extinction, and the presented results can be interpreted as the worst case scenario.

\subsubsection{Scaling results}\label{subsec:scalingresults}
The simulations performed in this work are locally isothermal, therefore the results can be scaled to account for different disc densities without altering the dynamics of the disc. We re-scaled column mass densities and accretion rates for different planet positions and disc densities. The normalisation of the surface density is readjusted in order to preserve the surface density profile $\Sigma(R)$ as in Equation~\ref{eq:surfacedensity}. In this way we could extend the results of our model to study different systems. From dimensional analysis, the column mass density $\sigma$ is directly proportional to the surface density, while the accretion rate $\dot{M}_\mathrm{acc}$ is proportional to the density and inversely proportional to orbital time. Surface density $\Sigma$ decreases as $R^{-0.5}$ (Equation~\ref{eq:surfacedensity}), while the density at the planet location $\rho \propto R^{-1.5}$ (Equation~\ref{eq:radialstructure}). These relationships are used to derive the scaling factors for $\sigma$ and $\dot{M}_\mathrm{acc}$ for planets at a distance $\neq 5.2$ $\mathrm{AU}$. In case of re-normalising the surface density (as done for real systems in Section~\ref{sec:applicationresults}) the ratio between the unperturbed new and original surface densities at a fiducial distance multiplies the scaling factors of the column mass density and the mass accretion rate.

Scaling the distance to the central star would change the temperature at the planet location. While this has no \textit{direct} impact on the scaling applied to the $\dot{M}_\mathrm{acc}$ obtained from the simulations (which has no explicit dependence on temperature), it would  also change the disc aspect ratio $ H = h/R$ since discs are generally flared. This would have an \textit{indirect} effect on $\dot{M}_\mathrm{acc}$. In addition, the disc aspect ratio also determines the gap opening planet mass and therefore the depth of the gap. We neglect these effects and remark that this is a limitation of our approach; the performed scaling provides a valuable understanding of how the planet location influences accretion rates and densities, but it does not capture all possible effects.
\begin{figure*}
    \begin{center}
    \includegraphics[width=.495\textwidth]{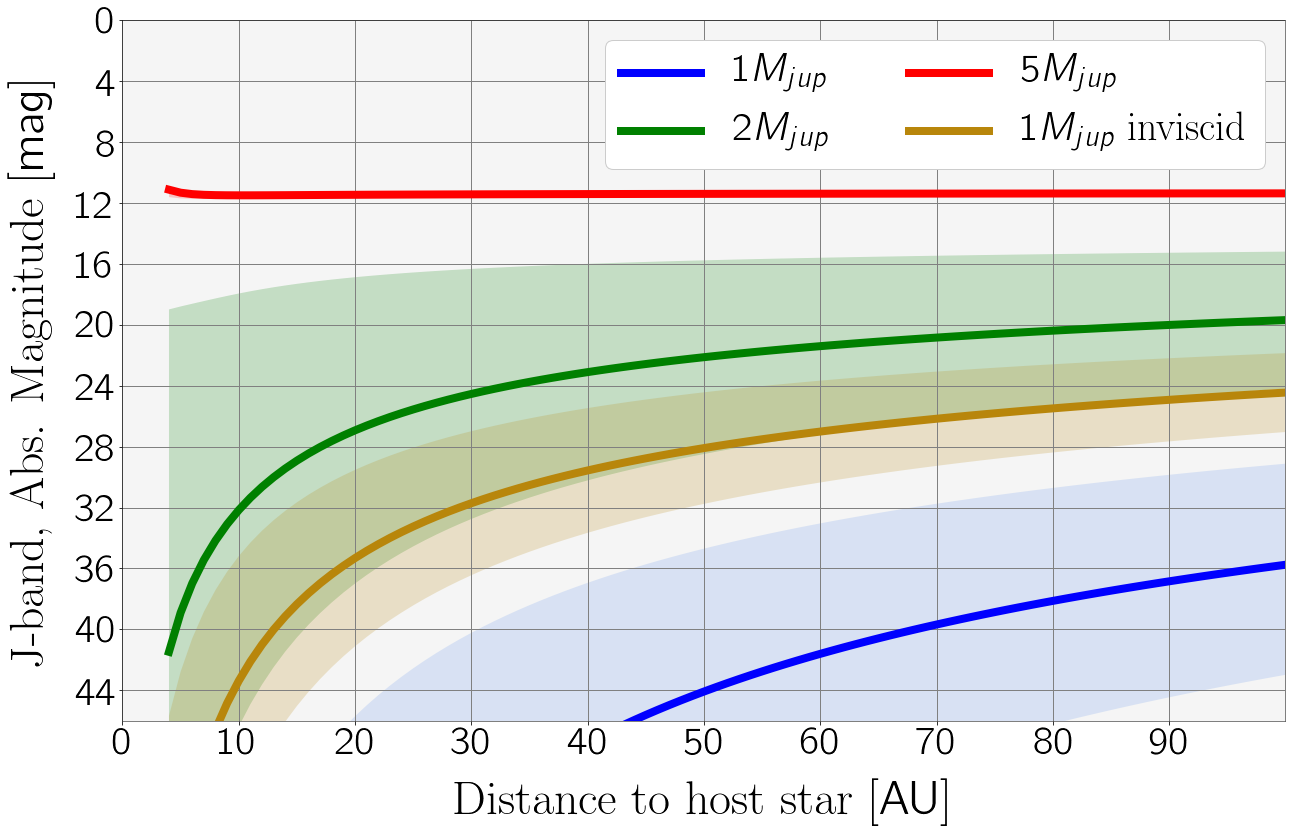}
    \includegraphics[width=.495\textwidth]{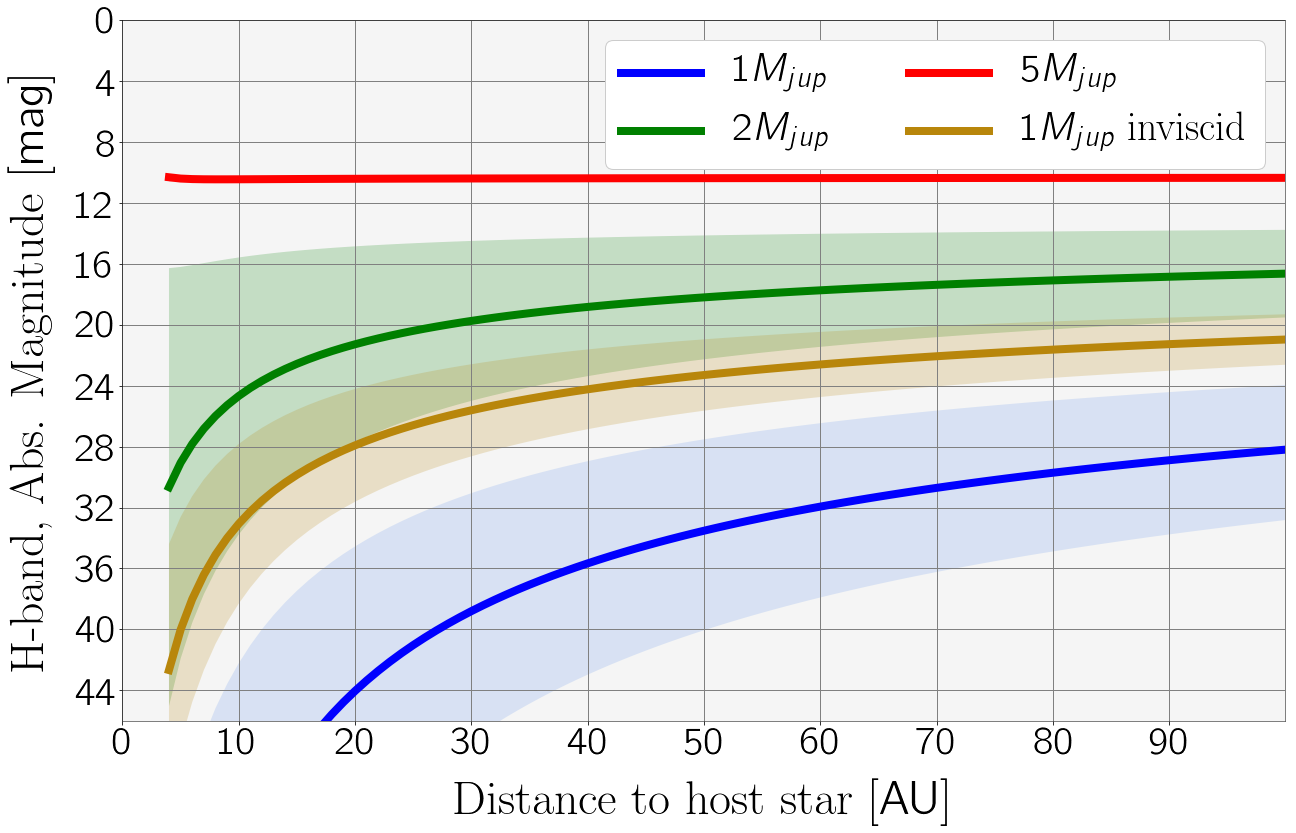}
    \includegraphics[width=.495\textwidth]{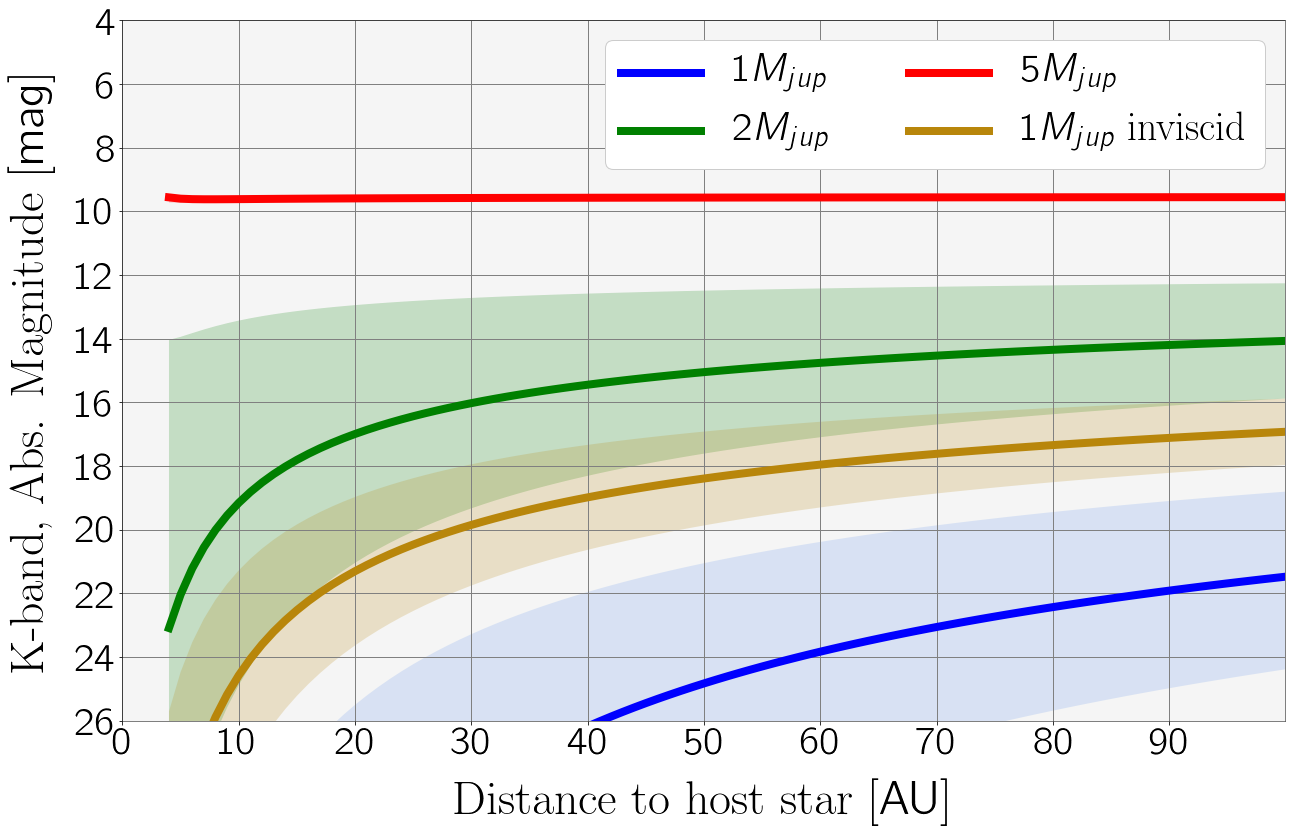}
    \includegraphics[width=.495\textwidth]{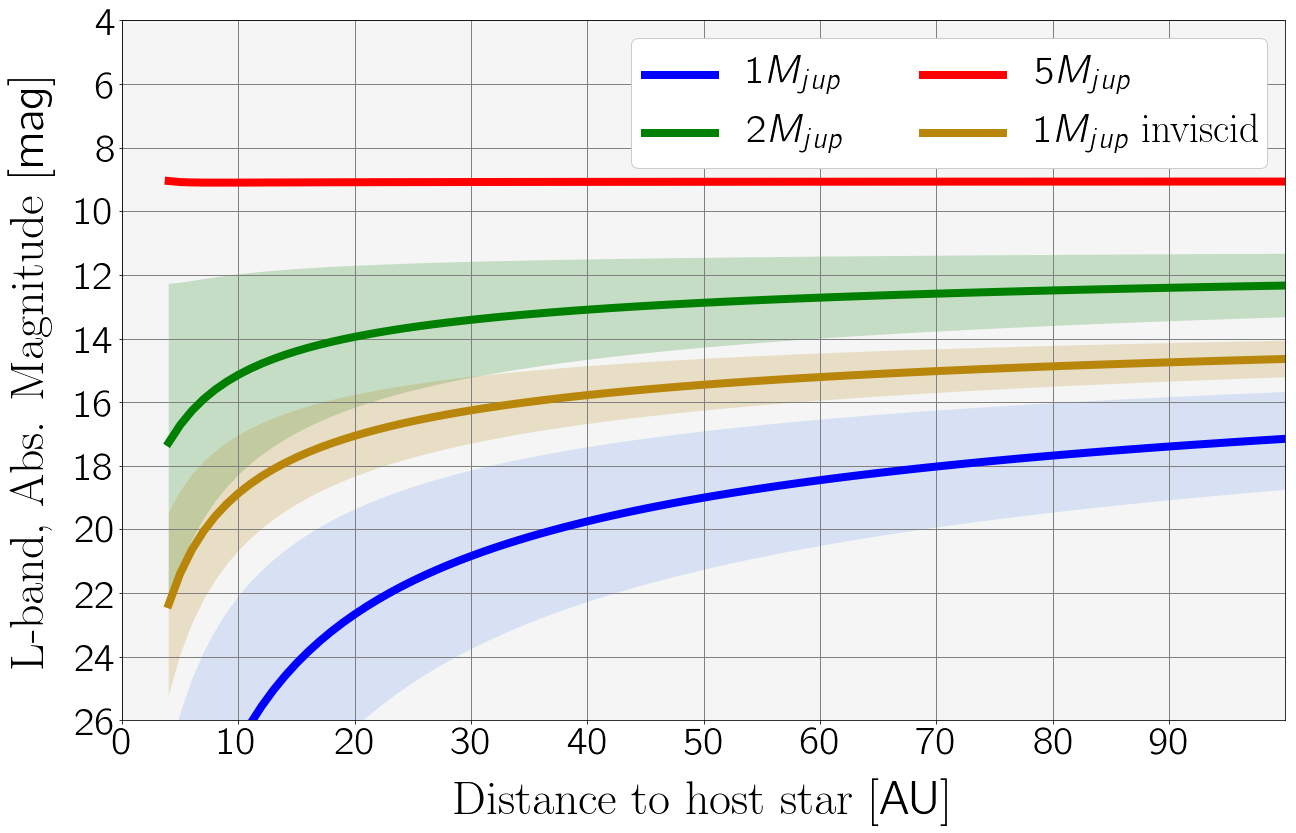}
    \includegraphics[width=.495\textwidth]{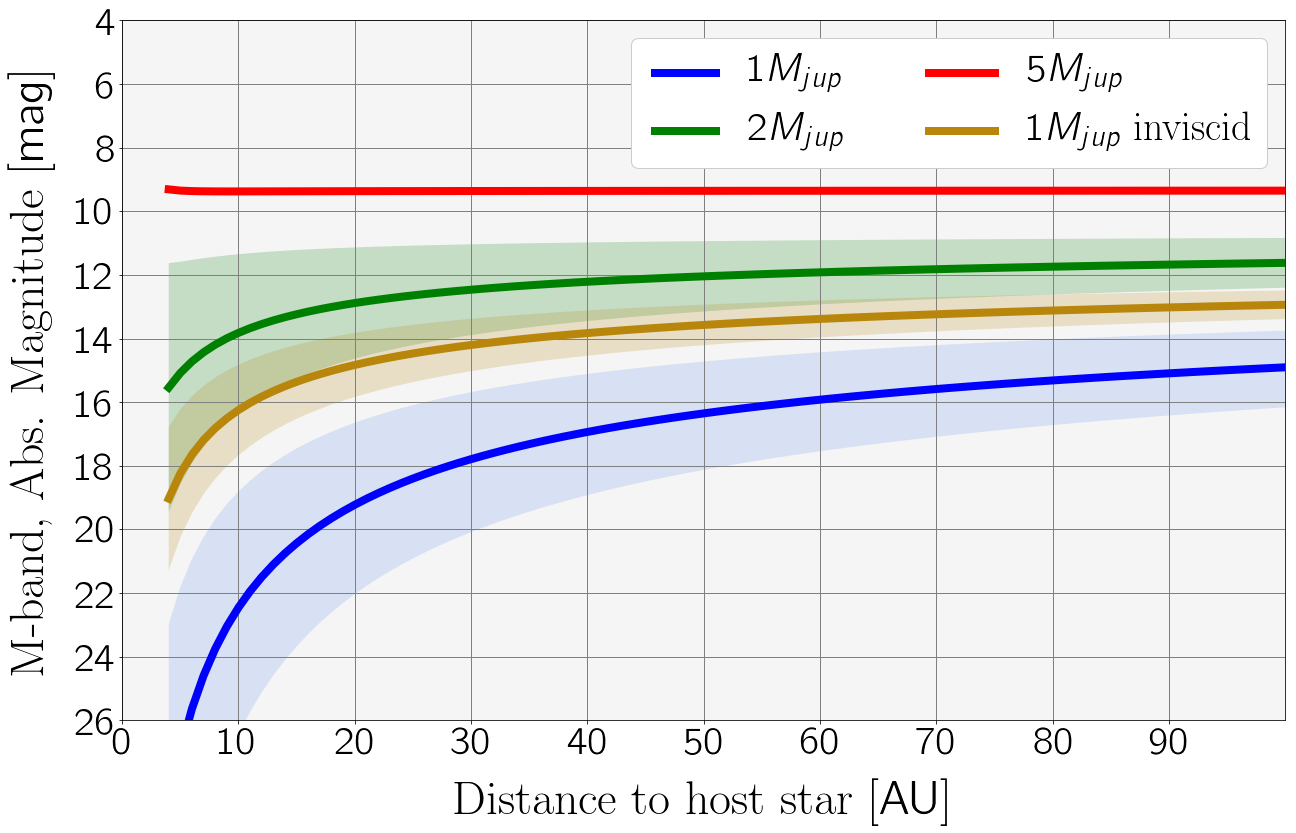}
    \includegraphics[width=.495\textwidth]{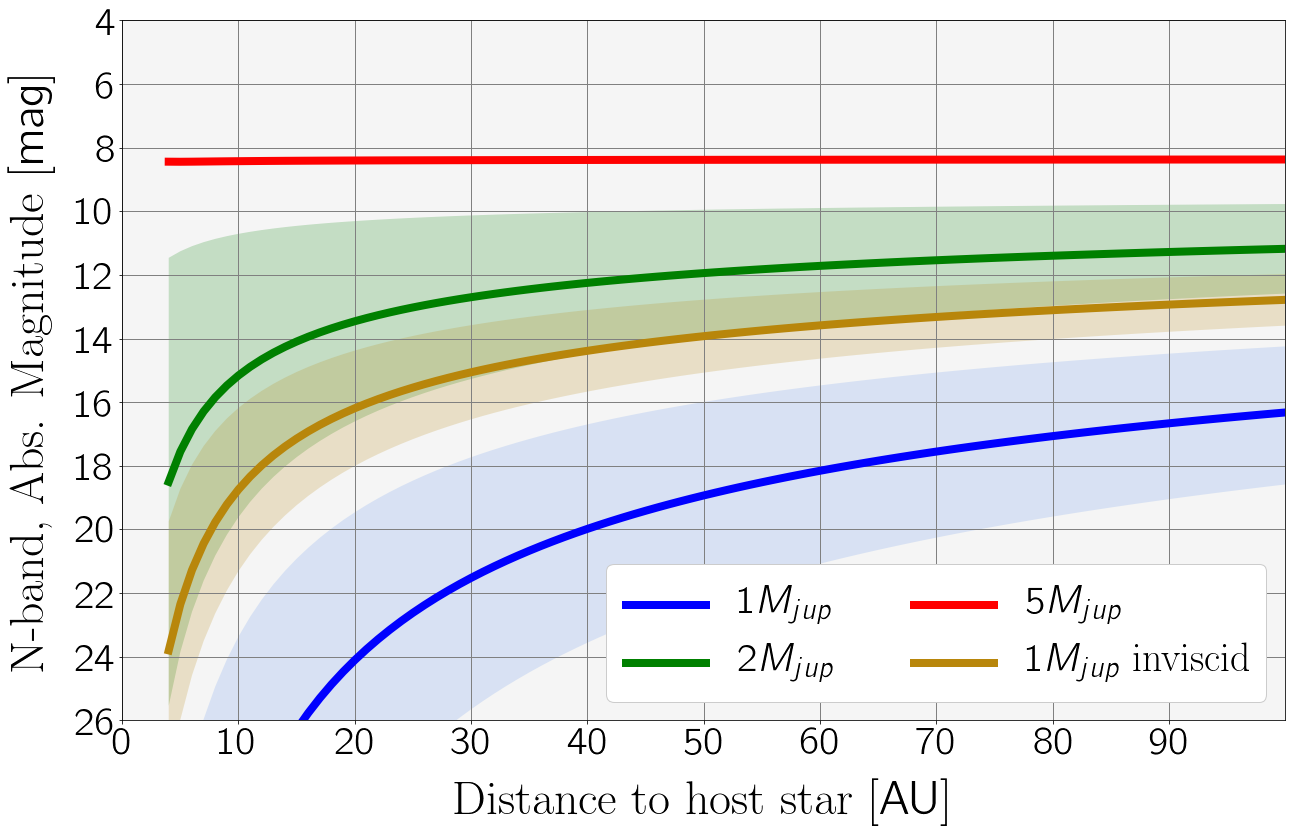}
    \end{center}
    \caption[]{Absolute magnitudes at $J$-, $H$-, $K$-, $L$-, $M$, and $N$-bands for the simulated disc with an embedded planet ($1$, $2$, $5$ $M_J$, or a $1$ $M_J$ planet inviscid case) at different distances to the central star. The results are shown for a \textit{hot-start} model of the formation scenario. The vertical axis for $J$ and $H$-bands covers a wider range in order to include all the planets in the same panel.}
    \label{fig:modelaplall}
\end{figure*}

Tables with extinction coefficients and predicted magnitudes of the simulated systems at $10$, $20$, $50$ and $100 \mathrm{AU}$ for every band are included in the Appendix~\ref{sec:appendixmags} (Tables~\ref{tab:expectedmags10au}, ~\ref{tab:expectedmags20au}, ~\ref{tab:expectedmags50au}, ~\ref{tab:expectedmags100au}). Figure~\ref{fig:modelaplall} shows the expected absolute magnitudes of planets with $1$ (for both viscous and inviscid cases), $2$ and $5$ $M_J$ for different distances to the central star, in $J$-, $H$-, $K$-, $L$-, $M$, and $N$-bands. The coloured area of each planet represents the uncertainty associated to the column density.

Extinction decreases for planets further out in the disc due to lower column densities. This behaviour is driven by the surface density profile. Accretion flux is higher at shorter distances but its effect is minor compared to extinction. Further out in the disc, the planet magnitude is dominated by the intrinsic flux since the accretion drops with distance. The dispersion of the results is considerable, it can not be neglected as a source of indeterminacy in our results. Nevertheless, large uncertainties are linked to very high values of the column density, and in such cases extinction is so strong that the planet would be completely hidden.

At shorter wavelengths ($J$- $H$- and $K$-bands), the magnitude of a $1$ $M_J$ planet is completely dominated by extinction: at the furthermost location considered, $100$ $\mathrm{AU}$, the extinction is $20.4$, $13.1$, and $8.2$ $\mathrm{mag}$ in each of these bands. For a $2$ $M_J$ planet the effect of extinction is lower but still large, e.g. at $100$ $\mathrm{AU}$ the extinction in these bands is $5.9$, $3.8$, and $2.4$ $\mathrm{mag}$.

The magnitude of a $5$ $M_J$ planet in the simulated disc is barely affected by extinction in any band (the highest extinction coefficient, $A_J$, ranges between $0.36$ to $0.08$ between $5.2$ and $100$ $\mathrm{AU}$): since the planet is substantially more massive, it has cleared almost all the material at the gap and its vicinity, thus both column density and the extinction coefficients are exceptionally low compared to the other simulated planets.

In $L$-band, the disc material above the least massive planet causes an extinction of $10$ $\mathrm{mag}$ at $20$ $\mathrm{AU}$, and is reduced to $4.5$ $\mathrm{mag}$ at $100$ $\mathrm{AU}$. A $2$ $M_J$ planet is less affected by extinction, with $A_L$ of $2.9$ and $1.3$ at those distances. In the $M$-band, the extinction coefficients of all the planets considered are the lowest out of all bands considered, nevertheless still considerable except for the $5$ $M_J$ planet. For instance, a $1$ $M_J$ planet at $100$ $\mathrm{AU}$ would be extincted by $3.5$ $\mathrm{mag}$. In the $N$-band, extinction is increased due to the silicate feature in the opacity of ISM dust: the extinction coefficients are almost 2 times larger than the coefficients in the $M$-band.

Our models can as well be used for different values of the stellar mass. Scaling our results for different $M_{\star}$ is especially useful when applying our detectability model to real systems, as discussed in the next section. The ratio $\frac{M_{\star}}{M_\mathrm{pl}}$ cannot change in order to keep the dynamics of the system valid. The planet mass is re-scaled accordingly to keep this ratio constant. Since the planetary models from \cite{spiegel2012} only provide data for planets with $1$, $2$, $5$ or $10$ $M_{J}$ masses, the planet intrinsic magnitudes were interpolated for the new $M_\mathrm{pl}$.


\section{Application to observed systems}\label{sec:applicationresults}
Our results can be applied to real systems to study the detectability of embedded planets, proceeding as explained in \ref{subsec:scalingresults}. In what follows we present the results of our model for the Class II discs of CQ~Tau, PDS~$70$, HL~Tau, TW~Hya and HD~$163296$. For the last three systems, our results are combined with contrast limits from previous IR observations \citep{testi2015, ruane2017, guidi2018}. The improvement of this revision is the inclusion of the extinction due to the disc material, and the emission from the shocks due to planet accretion.

Additionally, to understand how likely would be to detect the simulated planets directly with ALMA, we estimated the $1 M_J$ planet and CPD fluxes at $890 \mu m$ wavelength. The expected planet flux is $\sim 10^{-6}$ mJy, below the ALMA sensitivity limit. For the CPD, simplified to a disc of $1 R_\mathrm{Hill}$ radius centred at the planet and $0.24$ $R_\mathrm{Hill}$ high (i.e. the region around the planet with a disc-shaped overdensity), we obtain a dust mass of $M_{\mathrm{dust}}^{\mathrm{CPD}} \approx 0.003 M_{\oplus}$, which is comparable to the CPD measurements in PDS~$70$b \citep{isella+2019}. Assuming a constant CPD temperature of $121$ $\mathrm{K}$, the continuum emission in ALMA Band 7 would be $0.07$ mJy if emission is assumed optically thin, and $0.23$ mJy if optically thick. Thus, the CPD of the simulated disc could be detected by ALMA observations with enough sensitivity.

\subsection{CQ~Tau}\label{subsec:cqtau}
CQ~Tau is a young star from the Taurus-Auriga region, spectral type A$8$ \citep{trotta2013} and $M_{\star} = 1.5$ $M_{\odot}$ \citep{testi2003}. It has an estimated age of $\sim5$-$10$ $\mathrm{Myr}$ \citep{chapillon2008}, and very low disc mass, of the order of $10^{-3}$-$10^{-4}$ $M_{\odot}$. A fiducial surface density of $\Sigma = 1.6 \mathrm{g \cdot \mathrm{cm}^{-2}}$ at $40$ $\mathrm{AU}$ (a factor $\times0.035$ respect the simulated disc) was used for the re-normalisation to our unperturbed profile, derived from ALMA observations \citep{margi2018}.

\begin{figure} 
  \centering
  \includegraphics[width=1.0\linewidth]{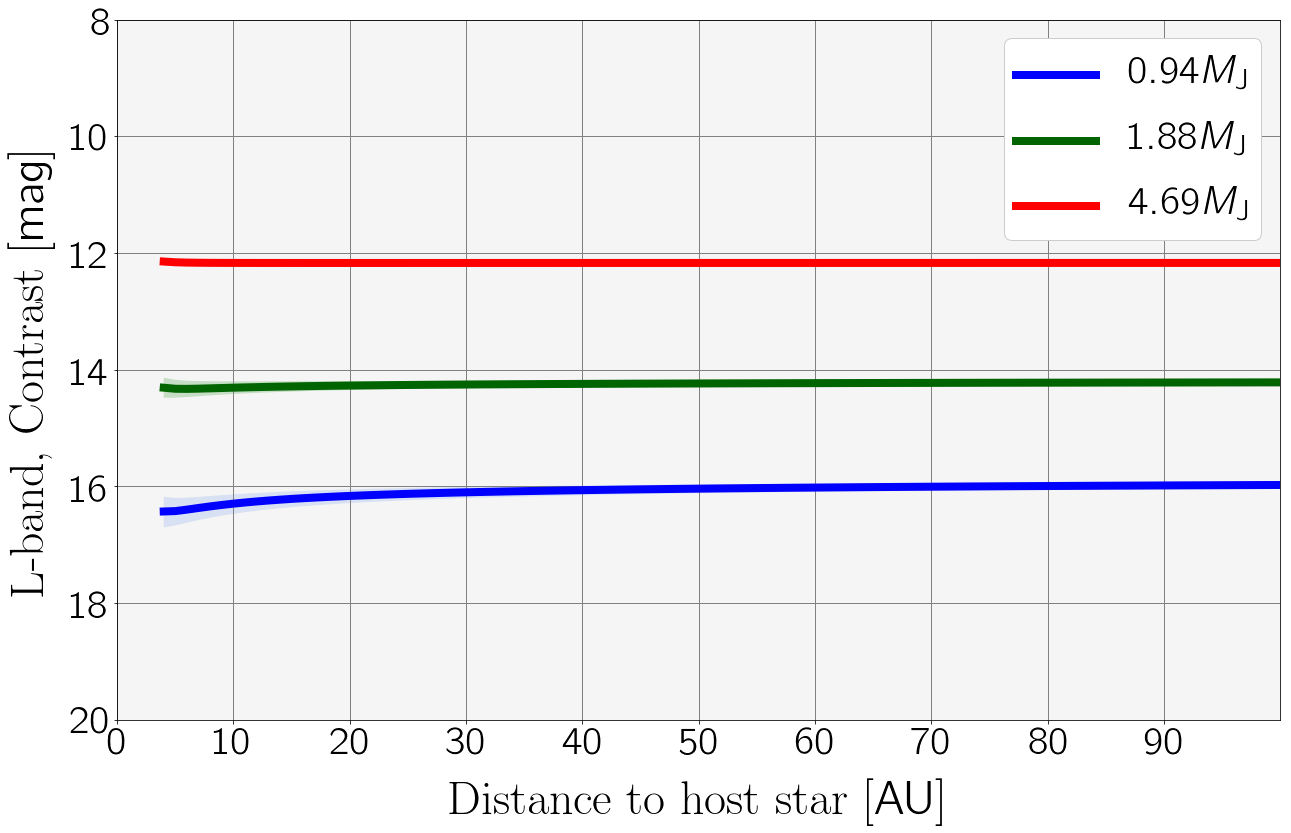}
  \caption[CQ~Tau Contrast plot including extinction effects.]{Application of our model to planets with $0.94$, $1.88$ and $4.69$ $M_J$ masses embedded in the CQ~Tau disc. The coloured lines represent the contrast in $L$-band of each planet as a function of the distance to the central star placed at different distances along the disc.}
  \label{fig:contrastcqtau}
\end{figure}

We applied our models to investigate the effects of disc extinction on potential planets embedded in the disc. Due to the re-scaling with the stellar mass, the contrast curves are derived for planets with $0.94$, $1.88$ and $4.69$ $M_{J}$. A distance of $163.1$ $\mathrm{pc}$ \citep{gaiadr2} was used for the predicted contrast. Figure~\ref{fig:contrastcqtau} shows the contrast of the planets in the $L$-band as a function of distance to the central star; the planet contrast is shown relative to the stellar value. The coloured area for each planet represents the associated dispersion. The results in $J$-, $H$- and $K$-bands are included in Appendix~\ref{sec:appendixallothercontrasts}.

At a fixed distance, the more massive the planet, the less affected by extinction, since the planet is more effective at clearing the gap. Due to the very low surface density inferred from the ALMA observations, extinction in $L$-band is only relevant for the lightest planets. At $20$ $\mathrm{AU}$, a $0.94$ $M_J$ planet would have a contrast of $16.16$ $\mathrm{mag}$ and extinction $A_L = 0.34$ $\mathrm{mag}$ ($A_V = 5.47$ $\mathrm{mag}$). The contrast of a $1.88$ $M_J$ planet is $14.27$ $\mathrm{mag}$ with an extinction of only $0.10$ $\mathrm{mag}$ ($A_V = 1.60$ $\mathrm{mag}$). The most massive planet ($4.69$ $M_J$) is barely affected by extinction, its contrast is equivalent to the case of a completely depleted disc, $12.17$ $\mathrm{mag}$.

\subsection{PDS70}\label{subsec:pds}
PDS~$70$ is a member of the Upper Centaurus-Lupus subgroup \citep[at $\sim 113$ $\mathrm{pc}$,][]{gaiadr2}, with a central star of $5.4$ $\mathrm{Myr}$ and mass $0.76$ $M_{\odot}$ \citep{mueller+2018}. It is surrounded by a transition disc with estimated total disc mass of $1\cdot 10^{-3}$ $M_{\odot}$. A first companion (PDS~$70$b) was found combining observations with VLT/SPHERE, VLT/NaCo and Gemini/NICI at various epochs, detected as a point-source in $H$-, $K$- and $L$-bands at a projected averaged separation of $194.7$ $\mathrm{mas}$ \citep{keppler+2018}. In $J$-band, PDS~$70$b could only be marginally detected when collapsing the $J$- and $H$-band channels. Due to the high uncertainties, $J$-band magnitude was not given. Atmospheric modelling of the planet was used to constrain its properties \citep{mueller+2018}, with an estimated mass range from $2$ to $17 M_J$.

Recent $H_{\alpha}$ line observations using VLT/MUSE confirmed a $8 \sigma$ detection from a second companion (PDS~$70$c) at $240$ $\mathrm{mas}$ \citep{haffert+2019}. Dust continuum emission (likely from its CPD) has been also observed \citep{isella+2019}. This second source is very close to an extended disc feature, consequently its photometry should be done with caution. In \cite{mesa+2019}, the planetary nature of this companion has been confirmed, and absolute magnitudes in $J$-, $H$-, and $K$-bands could be inferred for two SPHERE epochs. The spectrum in the $J$-band is very faint, and indistinguishable from the adjacent disc feature, thus the $J$-band magnitude should be regarded as upper limit. The NaCo $L$-band map detected emission that is partly covered by the disc, therefore its $L$-band magnitude should also be taken as an upper limit. Using various atmospheric models, \cite{mesa+2019} constrained the mass PDS~$70$c to be between $1.9$ and $4.4$ $M_J$.
\begin{figure}
    \begin{center}
    \includegraphics[width=.495\textwidth]{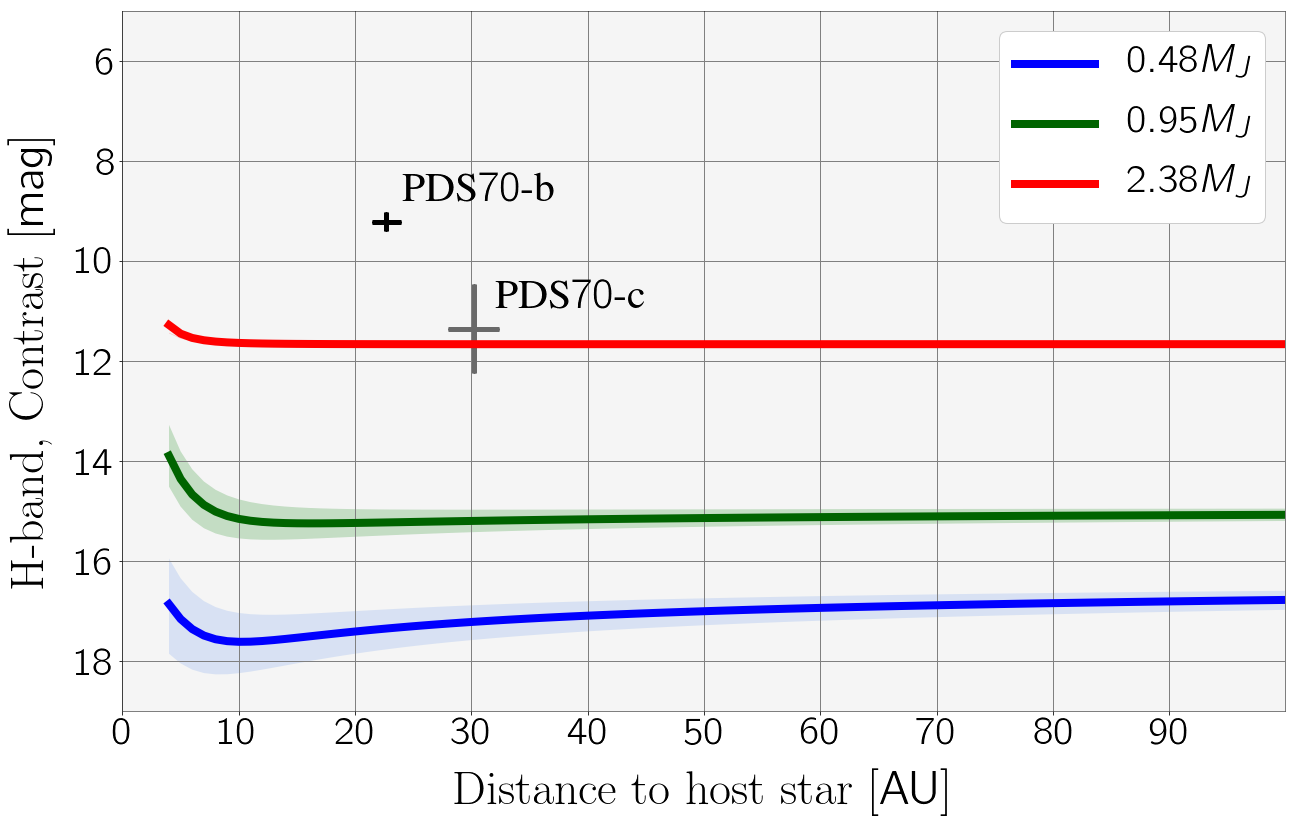}
    \includegraphics[width=.495\textwidth]{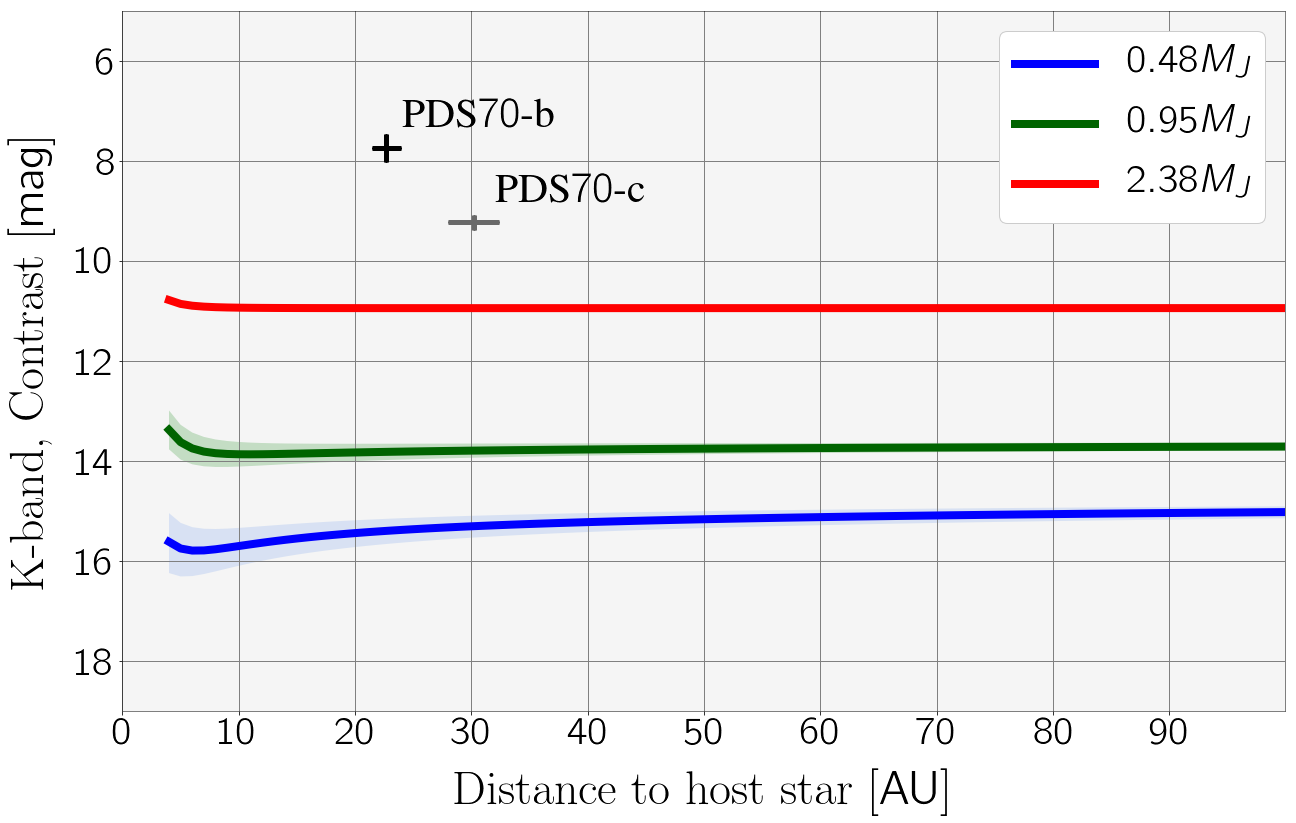}
    \includegraphics[width=.495\textwidth]{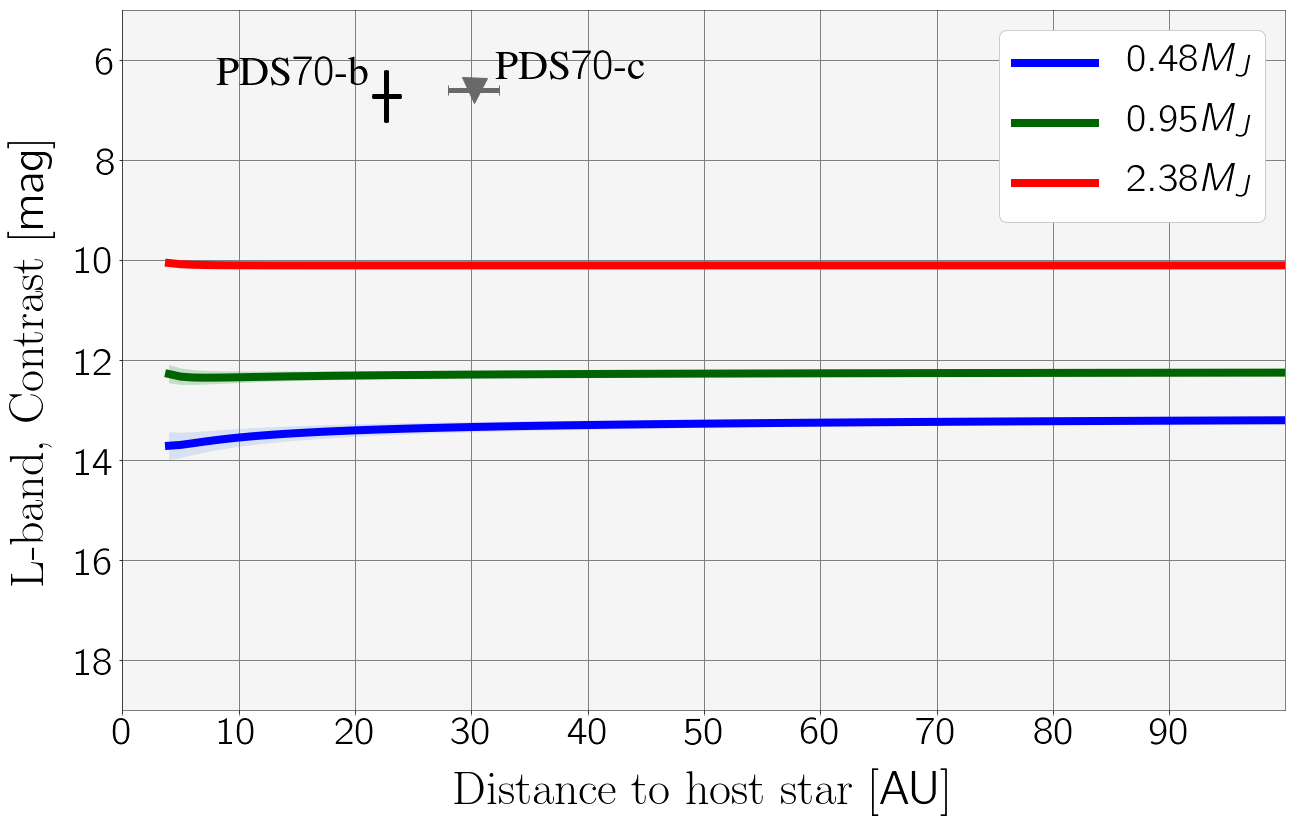}
    \end{center}
\caption[PDS~$70$ Contrast plot including extinction effects.]{Application of our model to planets with $0.48$, $0.95$ and $2.38$ $M_J$ embedded in the PDS~$70$ disc. The contrast curves shown for $H$-, $K$- and $L$-bands were obtained considering stellar magnitudes of $H = 8.8$ $\mathrm{mag}$, $K = 8.5$ $\mathrm{mag}$ and $L = 7.9$ $\mathrm{mag}$ \citep{cutri2003, cutri2014}. The two planetary companions \citep{keppler+2018,mesa+2019,haffert+2019} are shown as black and grey crosses, with the corresponding uncertainties.}
  \label{fig:contrastpds70}
\end{figure}

Our models were re-scaled using a fiducial surface density of $\Sigma = 12.5$ $\mathrm{g \cdot \mathrm{cm}^{-2}}$ at $1$ $\mathrm{AU}$ \citep[taking the unperturbed surface density model with depletion factor $\delta_{disc} = 1$ and gas-to-dust ratio of 100, from][]{keppler+2018}.  This corresponds to a surface density scale factor of $\times0.043$ with respect to the simulated disc. From the re-scaling, we obtained contrast curves of planets embedded in the PDS~$70$ disc with $0.48$, $0.95$ and $2.38$ $M_J$ (Figure~\ref{fig:contrastpds70}). The results show the effect of a disc with very low surface density: extinction has an incidence in $J$- and $H$-bands for $0.95$ and $0.48$ $M_J$ planets located within $\lesssim40$ $\mathrm{AU}$. In the $L$-band extinction has only a minor effect on the lightest planet model at distances below $20$ $\mathrm{AU}$. From the assumed surface density profile, none of the planetary companions would be affected by extinction due to material from the protoplanetary disc in the IR bands.

The observed contrast of the primary companion in three bands is considerably higher than the value for the most massive planet of our models, thus setting a mass lower limit of $2.38$ $M_J$ for PDS~$70$b. The second companion lays on top of the $2.38$ $M_J$ model in $H$-band, and above it in $K$-band. The redness of this source can explain the difference in the bands contrast. This reddening might be due to material from its own CPD or from the contiguous disc feature. Our models are in agreement with the previous mass ranges estimated for the two companions; further observations and modelling of the disc and their atmospheres are needed to better constrain their masses.

The estimated accretion rates of the companions are of the order of $\sim 10^{-11}$ $M_{\sun} \cdot {\mathrm{yr}}^{-1}$ \citep{haffert+2019}, thus radiation from accretion shocks near both planets are negligible. From our results, accretion flux would only have an incidence in the modelled IR planet fluxes at distances $\sim5$ $\mathrm{AU}$, since accretion rates are expected to be higher due to the scaling. This can be appreciated in the contrast curve of the three planet models in $H$-band: planets' contrasts decrease at these distances. The effect of the accretion shock's radiation becomes negligible at $\gtrsim10$ $\mathrm{AU}$.

\subsection{HL~Tau}\label{subsec:hltau}
HL~Tau is one of the most extensively studied protoplanetary discs, with several rings and gaps detected in the dust continuum \citep{alma2015}. It is a young stellar object of $\leq1$ $\mathrm{Myr}$ at around $140$ $\mathrm{pc}$ to us \citep{kenyon2008}, with an estimated stellar mass of $\sim 0.7$ $M_{\odot}$ \citep{kenyonhartmann1995, close1997}. Observations were carried out using the LBTI L/M IR Camera \citep[LMIRcam,][]{skrutskie2010, leisenring2012}, using only one of the two primary mirrors of the LBT telescope. No point-sources were detected. For the normalisation of the surface density, we took the inferred gas surface density from CARMA observations \citep{kwon+2011, kwon+2015} at a fiducial distance of $40$ $\mathrm{AU}$, $\Sigma = 34$ $\mathrm{g \cdot \mathrm{cm}^{-2}}$ (a factor $\times0.74$ compared to the simulated disc).
\begin{figure} 
  \centering
  \includegraphics[width=1.0\linewidth]{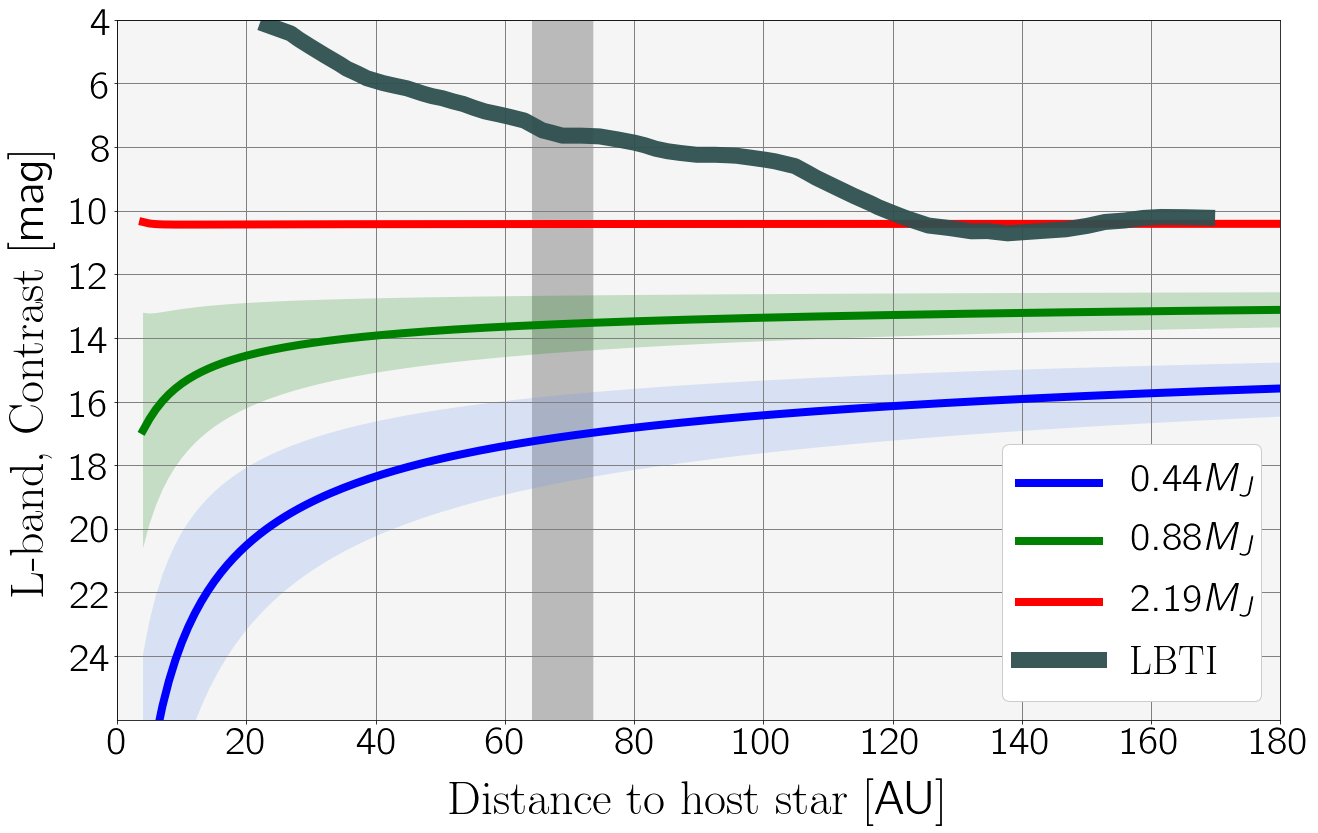}
  \caption[HL~Tau Contrast plot in $L$-band including extinction effects.]{Contract curves in $L$-band for planets embedded in HL~Tau, including the $5\sigma$ detection limit of the observation from \cite{testi2015}. The observations were performed using LBTI/LMIRcam. The contrast curves are for planet masses of $0.44$, $0.88$ and $2.19$ $M_J$. The considered apparent magnitude of the central star was $L = 6.23$ $\mathrm{mag}$ \citep{testi2015}. The coloured regions  accounts for the uncertainty in the planet contrast. The grey vertical area is delimited by the D$5$ and D$6$ rings detected in dust continuum \citep{alma2015}.}
  \label{fig:contrasthltau}
\end{figure}

In Figure~\ref{fig:contrasthltau} we show the contrast limit of the LBTI observation in $L$-band as a function of the angular separation to the central star, together with the derived contrast of the re-scaled models for planets with $0.44$, $0.88$ and $2.19$ $M_J$. In Appendix~\ref{sec:appendixallothercontrasts}, contrast curves in $J$-, $H$- and $K$-bands are included for completeness. In $L$-band, a high extinction is predicted for $0.44$ and $0.88$ $M_J$ along the entire disc, especially at distances $\lesssim60$ $\mathrm{AU}$; at that distance, $A_L$ values are $4.27$ $\mathrm{mag}$ ($A_V = 68.29$ $\mathrm{mag}$) and $1.25$ $\mathrm{mag}$ ($A_V = 19.98$ $\mathrm{mag}$) for these planets respectively. For planets outer in the disc, the extinction contribution is smaller but still significant: $3.02$ $\mathrm{mag}$ ($A_V = 48.29$ $\mathrm{mag}$) for $0.44$ $M_J$, and $0.88$ $\mathrm{mag}$ ($A_V = 14.13$ $\mathrm{mag}$) for $0.88$ $M_J$ at $120$ $\mathrm{AU}$. These planets are not massive enough to clear the gap efficiently. On the other hand, for the most massive planet ($2.19$ $M_J$), extinction is negligible at any distance.

Six gaps were observed in the ALMA continuum observation; following the example as in \cite{testi2015}, within the gap delimited by D5 and D6 rings (marked as grey vertical line) the contrast limit of the instrument does not allow us to constrain the mass of the companion that could be responsible for the gap. Nevertheless, from the inferred contrast curves, extinction would have an incidence in a hypothetical point-source detection only for planet masses $\lesssim 0.88$ $M_{J}$.

\subsection{TW~Hya}\label{subsec:twhya}
\begin{figure} 
  \centering
  \includegraphics[width=1.0\linewidth]{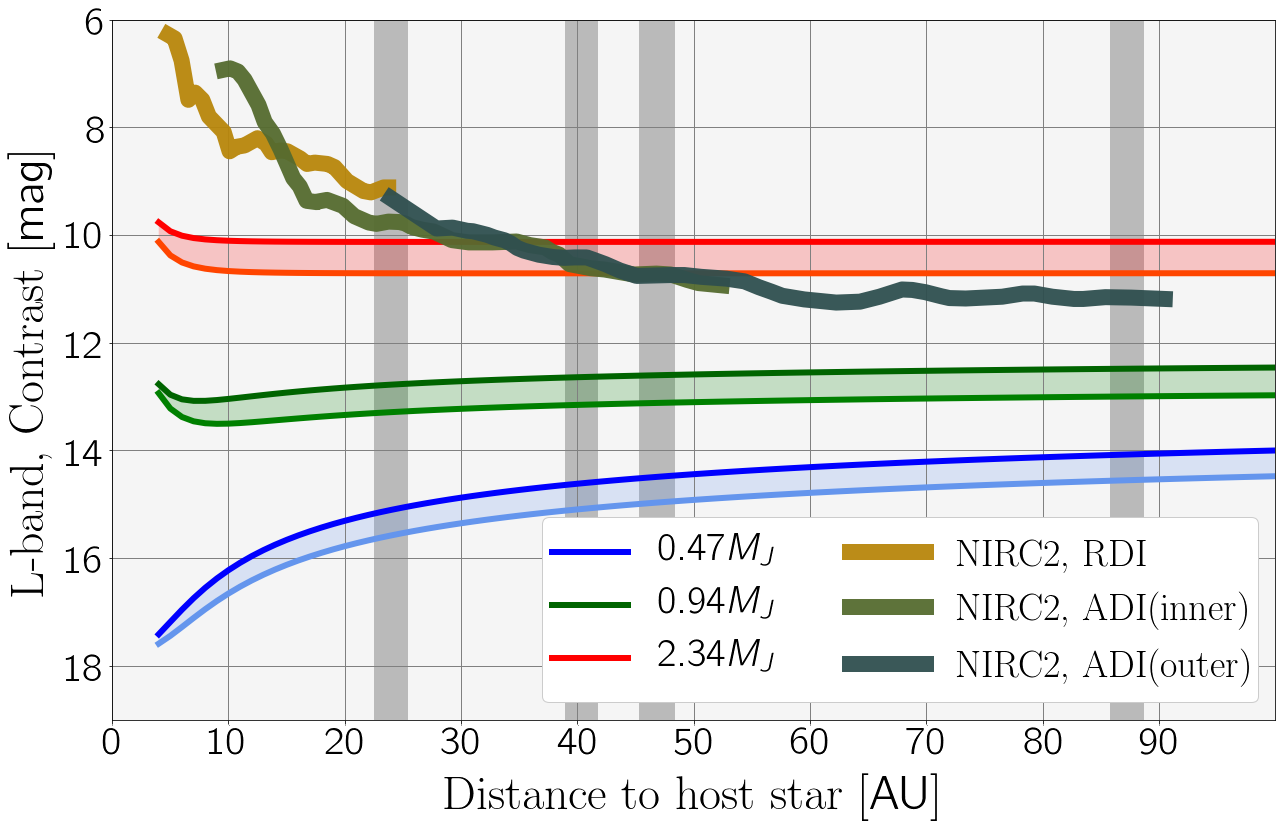}
  \caption[TW~Hya Contrast plot in $L$-band including extinction effects.]{Contract curves in $L$-band for planets embedded in TW~Hya, including $95\%$ significance detection limits of Keck/NIR$2$ observations \citep{ruane2017}. The contrast limits are shown for angular differential imaging (ADI) and reference star differential imaging (RDI). The contrast curves shown are for planet masses of $0.47$, $0.94$ and $2.34$ $M_J$. The apparent magnitude of the central star is $L = 7.01$ $\mathrm{mag}$, taken from the W1 band in the WISE catalogue \citep{wright2010}. The coloured regions for each planet model are delimited by the estimated ages \citep[$7$-$10$ $\mathrm{Myr}$,][]{ruane2017}. The grey vertical lines account for the gaps observed in \cite{andrews2016} and \cite{vanboekel2017}.}
  \label{fig:contrasttwhya}
\end{figure}

Observations using the Keck/NIRC2 vortex coronagraph were performed by \cite{ruane2017} searching for point-sources in the TW~Hya disc. This system is the closest known protoplanetary disc to us \citep[$60.2$ $\mathrm{AU}$,][]{gaiadr2}, with a central star of $0.7$-$0.8$ $M_{\odot}$ \citep{andrews2012, HerczegHillenbrand2014} relatively old \citep[$7$-$10$ $\mathrm{Myr}$,][]{ruane2017}, and with an estimated total disc mass of $0.05$ $M_{\odot}$ \citep{bergin2013}. The surface gas density has been modelled from ALMA line emission observations \citep{kama2016, trapman2017}. From their unperturbed models, we used a fiducial surface density of $10.9$ $\mathrm{g \cdot \mathrm{cm}^{-2}}$ at $40$ $\mathrm{AU}$ for the re-scaling  (a surface density factor $\times0.24$ of the simulated disc). The instrument allows for IR high-contrast imaging in $L$-band, using angular differential imaging (ADI) and reference star differential imaging (RDI). In Figure~\ref{fig:contrasttwhya}, the detection limits for ADI and RDI are shown together with the expected contrast of planets with $0.47$, $0.94$ and $2.34$ $M_J$. In this observation, RDI allows for detections of point-sources at distances as low as $\sim5$ $\mathrm{AU}$. At this distance the accretion flux overcomes the intrinsic flux for a planet of $\gtrsim2.34$ $M_J$, thus the contrast decreases compared to the non-accreting case. The contrast curves of these planets in $J$-, $H$- and $K$-bands are shown in Appendix~\ref{sec:appendixallothercontrasts}.

Different observations of TW~Hya confirmed several gaps in the disc. \cite{andrews2016} detected three dark annuli at $24$, $41$ and $47$ $\mathrm{AU}$ distances to the host star (distances corrected with newest Gaia parallax). An unresolved gap in the inner disc was also seen from $870$ $\mu m $ continuum emission using ALMA. Scattered light using SPHERE detected three gaps in the polarised intensity distribution, at $\lesssim 7$, $23$, and $88$ $\mathrm{AU}$. In the figure we show the gaps in the outer disc ($\sim23$, $40$, $46$ and $87$ $\mathrm{AU}$). No point-sources were detected in the Keck/NIRC2 observations. \cite{ruane2017} set upper limits for planets located at these gaps ($1.6$-$2.3$ $M_J$, $1.1$-$1.6$ $M_J$, $1.1$-$1.5$ $M_J$, and $1.0$-$1.2$ $M_J$ from inner to outer distances). Analogously, we can infer upper limits of the planets interpolating our results, since the contrast curves lay between our models. Using the models for $7$ $\mathrm{Myr}$ planets, the upper limits for these gaps would be $2.5$ $M_J$, $2.1$ $M_J$, $2.0$ $M_J$ and $1.7$ $M_J$. Considering an age of $10$ $\mathrm{Myr}$, the upper limits are marginally higher: $2.8$ $M_J$, $2.4$ $M_J$, $2.3$ $M_J$ and $2.1$ $M_J$. Taking into account the extinction due to the disc above the planet increases the estimated upper limits compared to the previous work, which did not consider this effect. This indicates the importance of extinction when looking for protoplanets with direct imaging methods.

\subsection{HD~163296}\label{subsec:hd163296}
In \cite{guidi2018}, the HD~$163296$ disc was studied in the $L$-band using the same instrument (Keck/NIRC2 vortex coronagraph). The scattered polarised emission in the $J$-band was also studied with the Gemini Planet Imager in \cite{monnier+2017}, detecting a ring with an offset that can be explained by an inclined flared disc. This system has a central star of $2.3$ $M_{\odot}$ \citep{natta2004} and estimated age of $\sim 5$ $\mathrm{Myr}$ \citep{montesinos2009}. Observations in the dust continuum using ALMA \citep{isella2016b} confirmed the existence of three gaps at distances of $\sim 50$, $\sim 81$ and $\sim 136$ $\mathrm{AU}$ (corrected with the new Gaia distance of $101.5$ $\mathrm{pc}$). Kinematical analysis of gas observations suggested the presence of two planets at the second and third gaps \citep{teague2018}. In \cite{pinte2018}, HD models showed that a third planet is expected further out. The estimated masses of the three potential planets are $1$ $M_J$ (at $83$ $\mathrm{AU}$), $1.3$ $M_J$ (at $127$ $\mathrm{AU}$) and $\approx 2$ $M_J$ at ($\approx260$ $\mathrm{AU}$). The new DSHARP/ALMA observations confirmed an additional gap at $\sim 10$ $\mathrm{AU}$ \citep{isella+2018}; assuming that this gap is caused by a planet, \cite{zhang+2018} estimated a planet mass between $0.2$ and $1.5$ $M_J$ from 2D HD simulations.
\begin{figure} 
  \centering
  \includegraphics[width=1.0\linewidth]{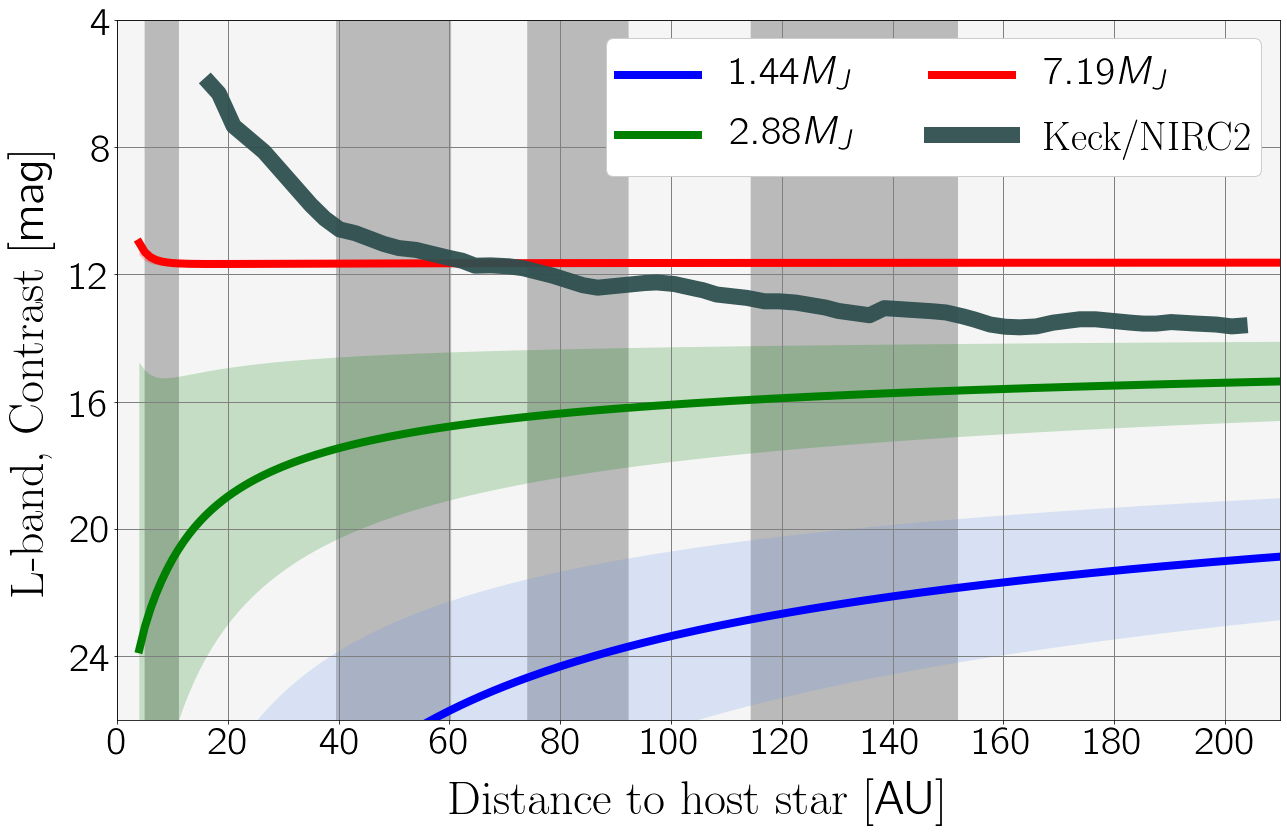}
  \caption[HD~$163296$ Contrast plot in $L$-band including extinction effects.]{Contract curves in $L$-band for planets embedded in HD~$163296$, including the $5\sigma$ detection limits of the observation from \cite{guidi2018}. The observations were performed using Keck/NIRC2. The contrast of planets with $1.44$, $2.88$ and $7.19$ $M_J$ are shown. The apparent magnitude of the central star is $L = 3.7$ $\mathrm{mag}$, inferred from the W1 band in the WISE catalogue \citep{wright2010}. The grey vertical lines account for the gaps observed in \cite{isella2016b, isella+2018} .}
  \label{fig:contrasthd163296}
\end{figure}

The $L$-band high-contrast imaging \citep{guidi2018} detected a point-like source at a distance of $67.7$ $\mathrm{AU}$ with $4.7\sigma$ significance. None of the observations in $L$- or $J$-band found any point-sources at the gaps observed in the continuum. Our models allow to set upper limits for planets at the location of the gaps. We used a fiducial surface density of $\Sigma = 82.8$ $\mathrm{g \cdot \mathrm{cm}^{-2}}$ at $40$ $\mathrm{AU}$ \citep[from][]{isella2016b}, corresponding to a factor $\times1.8$ of the simulated disc, to obtain contrast curves for $1.44$, $2.88$ and $7.19$ $M_J$ ($L$-band in Figure~\ref{fig:contrasthd163296}, and $J$-, $H$- and $K$-bands in Figures~\ref{fig:hd163296all} in the Appendix~\ref{sec:appendixallothercontrasts}). The innermost gap in Figure~\ref{fig:contrasthd163296} is within the masked region in the Keck/NIR2 observations, thus a mass upper limit can not be inferred. At the second gap, the model for our most massive planet lays slightly below the detection limit of the observation. A rough extrapolation would yield an upper-limit of $7.6$ $M_J$, slightly below the range provided by \cite{guidi2018} ($8$-$15$ $M_J$ in that work). For the third and fourth gaps, we obtain upper limits of $6.7$ $M_J$ and $5.5$ $M_J$ from interpolating our models. These values are slightly higher than the upper limits inferred in \cite{guidi2018} ($4.5$-$6.5$ $M_J$, and $2.5$-$4$ $M_J$ respectively). Taking into account extinction on the contrast of the planets increase the inferred upper limits of the non-detected planets. In every gap, extinction does have an important effect for planets with masses lower than the inferred upper limits. Compared to the estimates of \cite{teague2018} and \cite{pinte2018} from indirect analysis, our inferred upper limits are significantly higher; consequently a direct detection of these companions would only be possible improving the detection limit to much higher contrast.


\section{Conclusions}\label{sec:conclusions}
In this work we studied the effect of extinction for direct imaging of young planets embedded in protoplanetary discs. A set of HD simulations were performed to reproduce planet-disc interaction at high resolution for several planet masses. Column densities and extinction coefficients were derived in order to model the planet predicted magnitudes in $J$-, $H$-, $K$-, $L$-, $M$-, $N$-bands. Exploiting properties of locally isothermal discs, we applied the models to planets embedded in CQ~Tau, PDS~$70$ and HL~Tau protoplanetary discs, and inferred upper-limits for planets at the gaps observed in TW~Hya and HD~$163296$. The most important results of this work are:

\begin{itemize}
    \item For the simulated planets at $5.2$ $\mathrm{AU}$, the $5$ $M_\mathrm{J}$ clears its surrounding material very effectively. The resulting column density is extremely low, and, as consequence, extinction is not significant in any band. The $1$ and $2$ $M_\mathrm{J}$ planets are completely hidden by the disc at $\lesssim 2$ $\mu m$ wavelengths (with respective extinction coefficients of $> 30$, $> 15$ $\mathrm{mag}$), while at wavelengths between $5$ to $8$ $\mu m$ their corresponding coefficients are reduced, below $15$ and $4$ $\mathrm{mag}$. In the $N$-band, extinction is higher compared to $L$- and $N$-bands due to the silicate feature in the assumed ISM dust opacities.
    \item Jupiter-like planets embedded in discs with very low unperturbed surface densities (of the order of $\lesssim1$ $\mathrm{g \cdot \mathrm{cm}^{-2}}$) have very low extinction coefficients in IR at any distance considered. In CQ~Tau, only planets with $\lesssim 2$ $M_J$ are affected by extinction in $J$- and $H$-bands at distances $\lesssim 20$ $\mathrm{AU}$. In PDS~$70$, extinction has an incidence only for the least massive planet model at distances <$50$ $\mathrm{AU}$, more significant at shorter wavelengths.
    \item In more dense discs like HL~Tau and HD~$163296$, direct detection of companions is unlikely in $J$-, $H$-, $K$-, and $N$-bands due to the extinction effects. Only the most massive planet from our models would be detectable, since the extinction is negligible.
    \item We inferred upper limits of the  gaps in TW~Hya and HD~$163296$, slightly higher than previous work due to the effect of extinction. This points out the importance of extinction from the disc material in high-contrast imaging of protoplanetary discs.
    \item Radiation from accretion shocks onto the planet has been considered in our models. It can have an important effect on the total planet emission for accretion rates of the order of $\sim 10^{-8}$ $M_{\odot}/{\mathrm{yr}}$; these high rates occur at distances $\lesssim 10$ $\mathrm{AU}$ in our models.
\end{itemize}

The scarcity of detections so far might suggest different scenarios: giant planet formation further out in the disc is rare, or perhaps planets formed at these early stages are still not massive enough ($\lesssim 2$ $M_J$) to be detected with current instrumentation. 


\section*{Acknowledgements}
This work was partly supported by the Italian Ministero dell\'\,Istruzione, Universit\`a e Ricerca through the grant Progetti Premiali 2012 -- iALMA (CUP C52I13000140001), by the Deutsche Forschungs-gemeinschaft (DFG, German Research Foundation) - Ref no. FOR 2634/1 TE 1024/1-1, and by the DFG cluster of excellence Origin and Structure of the Universe (\href{http://www.universe-cluster.de}{www.universe-cluster.de}). This work is part of the research programme VENI with project number 016.Veni.192.233, which is (partly) financed by the Dutch Research Council (NWO). The simulations were partly run on the computing facilities of the Computational Center for Particle and Astrophysics (C2PAP) of the Excellence Cluster Universe. BE acknowledges funding by the Deutsche Forschungs-gemeinschaft (DFG, German Research Foundation) under Germany's Excellence Strategy -- EXC-2094 -- 390783311. GP and BE acknowledge support from the DFG Research Unit "Transition Disks" (FOR 2634/1, ER 685/8-1).

\bibliographystyle{mnras}
\bibliography{detectability-ref}

\appendix

\section{Magnitudes for planets at 10, 20, 50 and 100 AU}\label{sec:appendixmags}
The predicted magnitudes of the modelled planets at various distances to the central star are included in Tables~\ref{tab:expectedmags10au}, \ref{tab:expectedmags20au}, \ref{tab:expectedmags50au} and \ref{tab:expectedmags100au} for completeness.

\begin{table*}
	\centering
\caption[Magnitudes for planets at $10$ $\mathrm{AU}$.]{Expected magnitudes for planets at $10$ $\mathrm{AU}$ to the host star, with masses: $1$ $M_\mathrm{J}$ --viscous and inviscid scenarios-- $2$ $M_\mathrm{J}$ and $5$ $M_\mathrm{J}$. $\mathrm{Mag}_\mathrm{pl}$ is the total magnitude of the planet, including accretion flux; $A_{\mathrm{band}}$ is the extinction coefficient in each band, $\mathrm{Mag}_\mathrm{expected}$ is the predicted magnitude of the planet considering extinction due to disc material.}
\label{tab:expectedmags10au}
	\begin{tabular}{lccccccccccccccr}
		\hline
 \multicolumn{3}{c}{ } & \multicolumn{6}{ c }{\textit{Hot-start} planet} & \multicolumn{1}{c}{ } & \multicolumn{6}{ c }{\textit{Cold-start} planet} \\
 &  &  & J & H & K & L & M & N &  & J & H & K & L & M & N \\
		\hline
 & $\mathrm{Mag}_{\mathrm{pl}}$ &  & $15.13$ & $14.95$ & $13.31$ &  $12.65$ &  $11.40$ &  $10.02$ &  & $15.92$ & $16.02$ & $15.95$ &  $14.94$ &  $12.97$ &  $12.01$ \\ 
$1M_{\mathrm{J}}$ & $A_{\mathrm{band}}$ &  & $64.58$ & $41.30$ & $25.99$ &  $14.11$ &  $11.05$ &  $19.91$ &  & $64.58$ & $41.30$ & $25.99$ &  $14.11$ &  $11.05$ &  $19.91$ \\ 
 & $\mathrm{Mag}_{\mathrm{expected}}$ &  & $79.70$ & $56.25$ & $39.30$ &  $26.76$ &  $22.45$ &  $29.93$ &  & $80.49$ & $57.32$ & $41.94$ &  $29.05$ &  $24.03$ &  $31.91$ \\ 
 &  &  &  &  &  &  &  &  &  &  &  &  &  &  &  \\
 & $\mathrm{Mag}_{\mathrm{pl}}$ &  & $13.55$ & $12.76$ & $11.66$ &  $11.01$ &  $10.59$ &  $9.34$ &  & $14.93$ & $15.01$ & $15.00$ &  $14.12$ &  $12.64$ &  $11.82$ \\ 
$2M_\mathrm{J}$ & $A_{\mathrm{band}}$ &  & $18.62$ & $11.91$ & $7.49$ &  $4.13$ &  $3.23$ &  $5.82$ &  & $18.62$ & $11.91$ & $7.49$ &  $4.13$ &  $3.23$ &  $5.82$ \\ 
 & $\mathrm{Mag}_{\mathrm{expected}}$ &  & $32.17$ & $24.67$ & $19.16$ &  $15.14$ &  $13.82$ &  $15.16$ &  & $33.55$ & $26.92$ & $22.50$ &  $18.25$ &  $15.88$ &  $17.64$ \\ 
 &  &  &  &  &  &  &  &  &  &  &  &  &  &  &  \\
 & $\mathrm{Mag}_{\mathrm{pl}}$ &  & $11.25$ & $10.28$ & $9.52$ &  $9.04$ &  $9.33$ &  $8.35$ &  & $14.80$ & $14.84$ & $14.89$ &  $13.52$ &  $12.39$ &  $11.76$ \\ 
$5M_\mathrm{J}$ & $A_{\mathrm{band}}$ &  & $0.26$ & $0.16$ & $0.10$ &  $0.06$ &  $0.04$ &  $0.08$ &  & $0.26$ & $0.16$ & $0.10$ &  $0.06$ &  $0.04$ &  $0.08$ \\ 
 & $\mathrm{Mag}_{\mathrm{expected}}$ &  & $11.51$ & $10.44$ & $9.62$ &  $9.10$ &  $9.38$ &  $8.43$ &  & $15.05$ & $15.00$ & $14.99$ &  $13.58$ &  $12.44$ &  $11.84$ \\ 
 &  &  &  &  &  &  &  &  &  &  &  &  &  &  &  \\
 & $\mathrm{Mag}_{\mathrm{pl}}$ &  & $15.53$ & $15.25$ & $13.36$ &  $12.68$ &  $11.41$ &  $10.03$ &  & $17.51$ & $17.69$ & $17.27$ &  $15.33$ &  $13.02$ &  $12.02$ \\ 
$1M_{\mathrm{inviscid}}$ & $A_{\mathrm{band}}$ &  & $27.86$ & $17.82$ & $11.21$ &  $6.18$ &  $4.84$ &  $8.71$ &  & $27.86$ & $17.82$ & $11.21$ &  $6.18$ &  $4.84$ &  $8.71$ \\ 
 & $\mathrm{Mag}_{\mathrm{expected}}$ &  & $43.39$ & $33.07$ & $24.58$ &  $18.86$ &  $16.25$ &  $18.74$ &  & $45.37$ & $35.51$ & $28.49$ &  $21.51$ &  $17.86$ &  $20.74$ \\ 

	\hline
	\end{tabular}
\end{table*}

\begin{table*}
	\centering
\caption[Magnitudes for planets at $20$ $\mathrm{AU}$.]{Absolute magnitudes for planets at $20$ $\mathrm{AU}$ to the host star, with masses: $1$ $M_\mathrm{J}$ --viscous and inviscid scenarios-- $2$ $M_\mathrm{J}$ and $5$ $M_\mathrm{J}$. $\mathrm{Mag}_\mathrm{pl}$ is the total magnitude of the planet, including accretion flux; $A_{\mathrm{band}}$ is the extinction coefficient in each band, $\mathrm{Mag}_\mathrm{expected}$ is the predicted magnitude of the planet considering extinction due to disc material.}
\label{tab:expectedmags20au}
	\begin{tabular}{lccccccccccccccr}
		\hline
 \multicolumn{3}{c}{ } & \multicolumn{6}{ c }{\textit{Hot-start} planet} & \multicolumn{1}{c}{ } & \multicolumn{6}{ c }{\textit{Cold-start} planet} \\
 &  &  & J & H & K & L & M & N &  & J & H & K & L & M & N \\
		\hline
 & $\mathrm{Mag}_{\mathrm{pl}}$ &  & $15.55$ & $15.27$ & $13.37$ &  $12.68$ &  $11.41$ &  $10.03$ &  & $17.77$ & $17.98$ & $17.45$ &  $15.36$ &  $13.02$ &  $12.02$ \\ 
$1M_{\mathrm{J}}$ & $A_{\mathrm{band}}$ &  & $45.66$ & $29.20$ & $18.38$ &  $9.98$ &  $7.82$ &  $14.08$ &  & $45.66$ & $29.20$ & $18.38$ &  $9.98$ &  $7.82$ &  $14.08$ \\ 
 & $\mathrm{Mag}_{\mathrm{expected}}$ &  & $61.22$ & $44.47$ & $31.74$ &  $22.66$ &  $19.22$ &  $24.10$ &  & $63.43$ & $47.18$ & $35.83$ &  $25.34$ &  $20.84$ &  $26.10$ \\ 
 &  &  &  &  &  &  &  &  &  &  &  &  &  &  &  \\
 & $\mathrm{Mag}_{\mathrm{pl}}$ &  & $13.76$ & $12.85$ & $11.69$ &  $11.03$ &  $10.60$ &  $9.34$ &  & $16.86$ & $16.91$ & $16.70$ &  $14.64$ &  $12.74$ &  $11.85$ \\ 
$2M_\mathrm{J}$ & $A_{\mathrm{band}}$ &  & $13.17$ & $8.42$ & $5.30$ &  $2.92$ &  $2.29$ &  $4.12$ &  & $13.17$ & $8.42$ & $5.30$ &  $2.92$ &  $2.29$ &  $4.12$ \\ 
 & $\mathrm{Mag}_{\mathrm{expected}}$ &  & $26.93$ & $21.28$ & $16.99$ &  $13.95$ &  $12.89$ &  $13.46$ &  & $30.03$ & $25.33$ & $22.00$ &  $17.56$ &  $15.02$ &  $15.97$ \\ 
 &  &  &  &  &  &  &  &  &  &  &  &  &  &  &  \\
 & $\mathrm{Mag}_{\mathrm{pl}}$ &  & $11.28$ & $10.29$ & $9.52$ &  $9.05$ &  $9.34$ &  $8.35$ &  & $16.61$ & $16.50$ & $16.55$ &  $14.16$ &  $12.55$ &  $11.83$ \\ 
$5M_\mathrm{J}$ & $A_{\mathrm{band}}$ &  & $0.18$ & $0.12$ & $0.07$ &  $0.04$ &  $0.03$ &  $0.06$ &  & $0.18$ & $0.12$ & $0.07$ &  $0.04$ &  $0.03$ &  $0.06$ \\ 
 & $\mathrm{Mag}_{\mathrm{expected}}$ &  & $11.46$ & $10.40$ & $9.59$ &  $9.09$ &  $9.37$ &  $8.41$ &  & $16.79$ & $16.61$ & $16.62$ &  $14.20$ &  $12.58$ &  $11.89$ \\ 
 &  &  &  &  &  &  &  &  &  &  &  &  &  &  &  \\
 & $\mathrm{Mag}_{\mathrm{pl}}$ &  & $15.62$ & $15.32$ & $13.38$ &  $12.69$ &  $11.41$ &  $10.03$ &  & $18.59$ & $18.94$ & $17.92$ &  $15.43$ &  $13.03$ &  $12.03$ \\ 
$1M_{\mathrm{inviscid}}$ & $A_{\mathrm{band}}$ &  & $19.70$ & $12.60$ & $7.93$ &  $4.37$ &  $3.42$ &  $6.16$ &  & $19.70$ & $12.60$ & $7.93$ &  $4.37$ &  $3.42$ &  $6.16$ \\ 
 & $\mathrm{Mag}_{\mathrm{expected}}$ &  & $35.32$ & $27.92$ & $21.30$ &  $17.06$ &  $14.83$ &  $16.19$ &  & $38.29$ & $31.54$ & $25.84$ &  $19.79$ &  $16.45$ &  $18.19$ \\ 
		\hline
	\end{tabular}
\end{table*}

\begin{table*}
	\centering
\caption[Expected magnitudes for planets at $50$ $\mathrm{AU}$.]{Expected magnitudes for planets at $50$ $\mathrm{AU}$ to the host star, with masses: $1$ $M_\mathrm{J}$ --viscous and inviscid scenarios-- $2$ $M_\mathrm{J}$ and $5$ $M_\mathrm{J}$. $\mathrm{Mag}_\mathrm{pl}$ is the total magnitude of the planet, including accretion flux; $A_{\mathrm{band}}$ is the extinction coefficient in each band, $\mathrm{Mag}_\mathrm{expected}$ is the predicted magnitude of the planet considering extinction due to disc material.}
\label{tab:expectedmags50au}
	\begin{tabular}{lccccccccccccccr}
		\hline
 \multicolumn{3}{c}{ } & \multicolumn{6}{ c }{\textit{Hot-start} planet} & \multicolumn{1}{c}{ } & \multicolumn{6}{ c }{\textit{Cold-start} planet} \\
 &  &  & J & H & K & L & M & N &  & J & H & K & L & M & N \\
		\hline
 & $\mathrm{Mag}_{\mathrm{pl}}$ &  & $15.63$ & $15.32$ & $13.38$ &  $12.69$ &  $11.41$ &  $10.03$ &  & $18.77$ & $19.18$ & $18.00$ &  $15.43$ &  $13.03$ &  $12.03$ \\ 
$1M_{\mathrm{J}}$ & $A_{\mathrm{band}}$ &  & $28.88$ & $18.47$ & $11.62$ &  $6.31$ &  $4.94$ &  $8.90$ &  & $28.88$ & $18.47$ & $11.62$ &  $6.31$ &  $4.94$ &  $8.90$ \\ 
 & $\mathrm{Mag}_{\mathrm{expected}}$ &  & $44.51$ & $33.79$ & $25.00$ &  $19.00$ &  $16.35$ &  $18.93$ &  & $47.65$ & $37.65$ & $29.62$ &  $21.75$ &  $17.97$ &  $20.93$ \\ 
 &  &  &  &  &  &  &  &  &  &  &  &  &  &  &  \\
 & $\mathrm{Mag}_{\mathrm{pl}}$ &  & $13.80$ & $12.87$ & $11.70$ &  $11.03$ &  $10.60$ &  $9.34$ &  & $18.03$ & $18.00$ & $17.46$ &  $14.73$ &  $12.75$ &  $11.86$ \\ 
$2M_\mathrm{J}$ & $A_{\mathrm{band}}$ &  & $8.33$ & $5.33$ & $3.35$ &  $1.85$ &  $1.45$ &  $2.60$ &  & $8.33$ & $5.33$ & $3.35$ &  $1.85$ &  $1.45$ &  $2.60$ \\ 
 & $\mathrm{Mag}_{\mathrm{expected}}$ &  & $22.12$ & $18.19$ & $15.05$ &  $12.88$ &  $12.05$ &  $11.94$ &  & $26.35$ & $23.33$ & $20.81$ &  $16.58$ &  $14.20$ &  $14.46$ \\ 
 &  &  &  &  &  &  &  &  &  &  &  &  &  &  &  \\
 & $\mathrm{Mag}_{\mathrm{pl}}$ &  & $11.28$ & $10.29$ & $9.52$ &  $9.05$ &  $9.34$ &  $8.35$ &  & $17.53$ & $17.21$ & $17.27$ &  $14.28$ &  $12.57$ &  $11.84$ \\ 
$5M_\mathrm{J}$ & $A_{\mathrm{band}}$ &  & $0.11$ & $0.07$ & $0.05$ &  $0.03$ &  $0.02$ &  $0.04$ &  & $0.11$ & $0.07$ & $0.05$ &  $0.03$ &  $0.02$ &  $0.04$ \\ 
 & $\mathrm{Mag}_{\mathrm{expected}}$ &  & $11.39$ & $10.36$ & $9.57$ &  $9.08$ &  $9.36$ &  $8.39$ &  & $17.65$ & $17.28$ & $17.31$ &  $14.31$ &  $12.59$ &  $11.88$ \\ 
 &  &  &  &  &  &  &  &  &  &  &  &  &  &  &  \\
 & $\mathrm{Mag}_{\mathrm{pl}}$ &  & $15.64$ & $15.33$ & $13.38$ &  $12.69$ &  $11.41$ &  $10.03$ &  & $18.87$ & $19.32$ & $18.04$ &  $15.44$ &  $13.03$ &  $12.03$ \\ 
$1M_{\mathrm{inviscid}}$ & $A_{\mathrm{band}}$ &  & $12.46$ & $7.97$ & $5.01$ &  $2.76$ &  $2.16$ &  $3.90$ &  & $12.46$ & $7.97$ & $5.01$ &  $2.76$ &  $2.16$ &  $3.90$ \\ 
 & $\mathrm{Mag}_{\mathrm{expected}}$ &  & $28.10$ & $23.30$ & $18.39$ &  $15.45$ &  $13.57$ &  $13.93$ &  & $31.32$ & $27.29$ & $23.05$ &  $18.20$ &  $15.19$ &  $15.93$ \\ 
		\hline
	\end{tabular}
\end{table*}

\begin{table*}
	\centering
\caption[Expected magnitudes for planets at $100$ $\mathrm{AU}$.]{Expected magnitudes for planets at $100$ $\mathrm{AU}$ to the host star, with masses: $1$ $M_\mathrm{J}$ --viscous and inviscid scenarios-- $2$ $M_\mathrm{J}$ and $5$ $M_\mathrm{J}$. $\mathrm{Mag}_\mathrm{pl}$ is the total magnitude of the planet, including accretion flux; $A_{\mathrm{band}}$ is the extinction coefficient in each band, $\mathrm{Mag}_\mathrm{expected}$ is the predicted magnitude of the planet considering extinction due to disc material.}
\label{tab:expectedmags100au}
	\begin{tabular}{lccccccccccccccr}
		\hline
 \multicolumn{3}{c}{ } & \multicolumn{6}{ c }{\textit{Hot-start} planet} & \multicolumn{1}{c}{ } & \multicolumn{6}{ c }{\textit{Cold-start} planet} \\
 &  &  & J & H & K & L & M & N &  & J & H & K & L & M & N \\
		\hline
 & $\mathrm{Mag}_{\mathrm{pl}}$ &  & $15.63$ & $15.33$ & $13.38$ &  $12.69$ &  $11.41$ &  $10.03$ &  & $18.87$ & $19.33$ & $18.04$ &  $15.44$ &  $13.03$ &  $12.03$ \\ 
$1M_{\mathrm{J}}$ & $A_{\mathrm{band}}$ &  & $20.42$ & $13.06$ & $8.22$ &  $4.46$ &  $3.50$ &  $6.30$ &  & $20.42$ & $13.06$ & $8.22$ &  $4.46$ &  $3.50$ &  $6.30$ \\ 
 & $\mathrm{Mag}_{\mathrm{expected}}$ &  & $36.06$ & $28.39$ & $21.60$ &  $17.15$ &  $14.90$ &  $16.32$ &  & $39.29$ & $32.39$ & $26.26$ &  $19.90$ &  $16.52$ &  $18.32$ \\ 
 &  &  &  &  &  &  &  &  &  &  &  &  &  &  &  \\
 & $\mathrm{Mag}_{\mathrm{pl}}$ &  & $13.80$ & $12.87$ & $11.70$ &  $11.03$ &  $10.60$ &  $9.34$ &  & $18.16$ & $18.12$ & $17.53$ &  $14.74$ &  $12.75$ &  $11.86$ \\ 
$2M_\mathrm{J}$ & $A_{\mathrm{band}}$ &  & $5.89$ & $3.77$ & $2.37$ &  $1.31$ &  $1.02$ &  $1.84$ &  & $5.89$ & $3.77$ & $2.37$ &  $1.31$ &  $1.02$ &  $1.84$ \\ 
 & $\mathrm{Mag}_{\mathrm{expected}}$ &  & $19.69$ & $16.64$ & $14.07$ &  $12.34$ &  $11.62$ &  $11.18$ &  & $24.05$ & $21.89$ & $19.90$ &  $16.04$ &  $13.77$ &  $13.70$ \\ 
 &  &  &  &  &  &  &  &  &  &  &  &  &  &  &  \\
 & $\mathrm{Mag}_{\mathrm{pl}}$ &  & $11.28$ & $10.29$ & $9.52$ &  $9.05$ &  $9.34$ &  $8.35$  &  & $17.63$ & $17.27$ & $17.33$ &  $14.29$ &  $12.57$ &  $11.84$ \\ 
$5M_\mathrm{J}$ & $A_{\mathrm{band}}$ &  & $0.08$ & $0.05$ & $0.03$ &  $0.02$ &  $0.01$ &  $0.03$ &  & $0.08$ & $0.05$ & $0.03$ &  $0.02$ &  $0.01$ &  $0.03$ \\ 
 & $\mathrm{Mag}_{\mathrm{expected}}$ &  & $11.36$ & $10.34$ & $9.55$ &  $9.07$ &  $9.35$ &  $8.38$ &  & $17.71$ & $17.32$ & $17.36$ &  $14.31$ &  $12.58$ &  $11.87$ \\ 
 &  &  &  &  &  &  &  &  &  &  &  &  &  &  &  \\
 & $\mathrm{Mag}_{\mathrm{pl}}$ &  & $15.64$ & $15.33$ & $13.38$ &  $12.69$ &  $11.41$ &  $10.03$ &  & $18.89$ & $19.34$ & $18.05$ &  $15.44$ &  $13.03$ &  $12.03$ \\ 
$1M_{\mathrm{inviscid}}$ & $A_{\mathrm{band}}$ &  & $8.81$ & $5.63$ & $3.55$ &  $1.95$ &  $1.53$ &  $2.76$ &  & $8.81$ & $5.63$ & $3.55$ &  $1.95$ &  $1.53$ &  $2.76$ \\ 
 & $\mathrm{Mag}_{\mathrm{expected}}$ &  & $24.45$ & $20.96$ & $16.92$ &  $14.64$ &  $12.94$ &  $12.79$ &  & $27.70$ & $24.98$ & $21.59$ &  $17.39$ &  $14.56$ &  $14.79$ \\ 
		\hline
	\end{tabular}
\end{table*}

\section{Contrast of planets embedded in CQ~Tau, HL~Tau, TW~Hya and HD~163296}\label{sec:appendixallothercontrasts}
\begin{figure}
    \begin{center}
    \includegraphics[width=.495\textwidth]{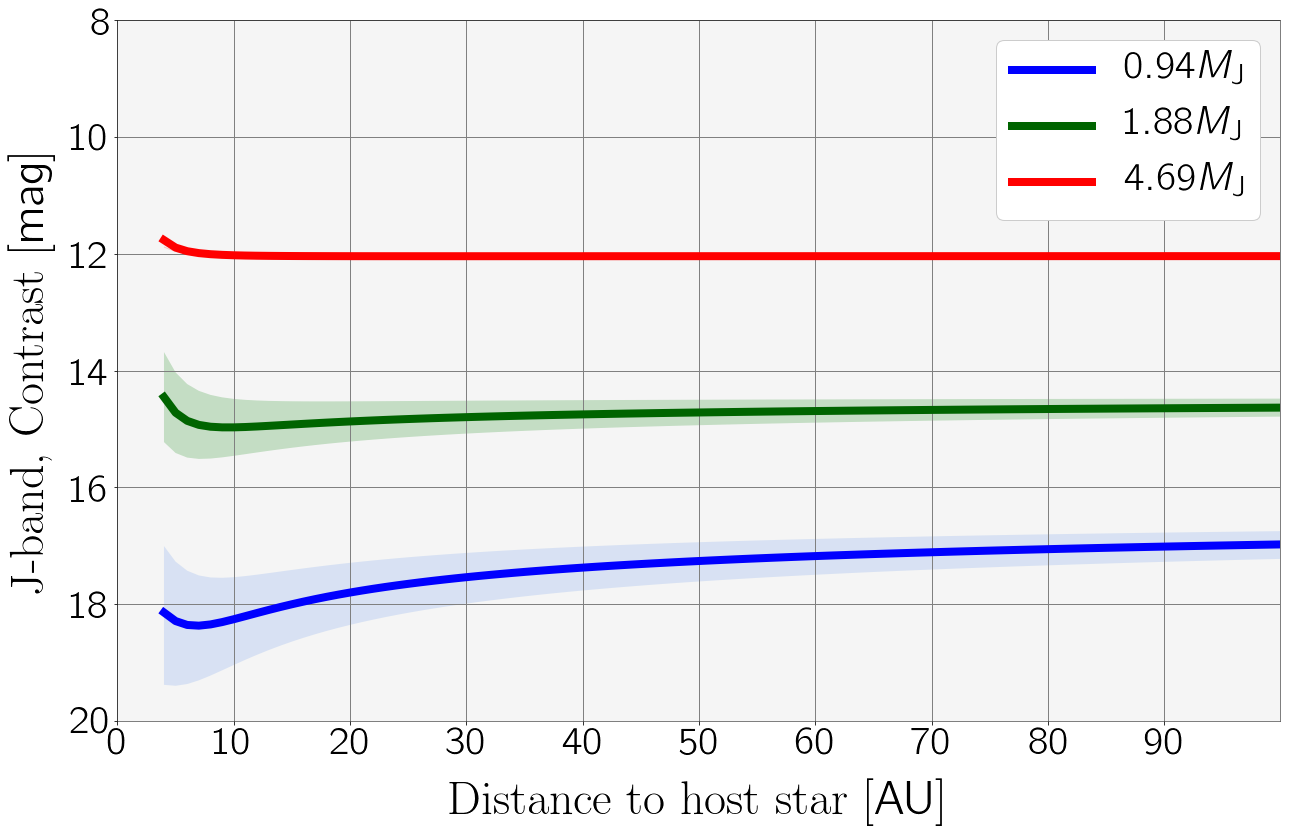}
    \includegraphics[width=.495\textwidth]{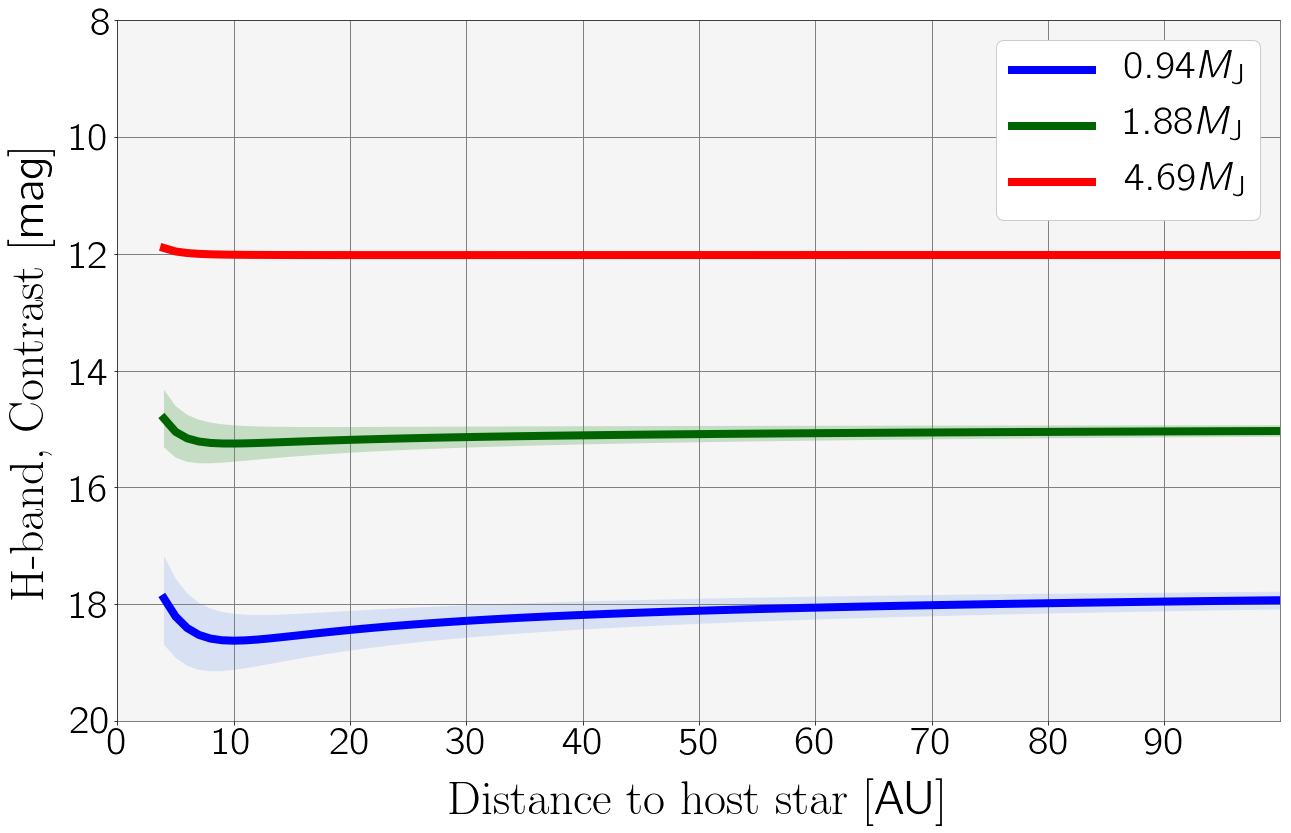}
    \includegraphics[width=.495\textwidth]{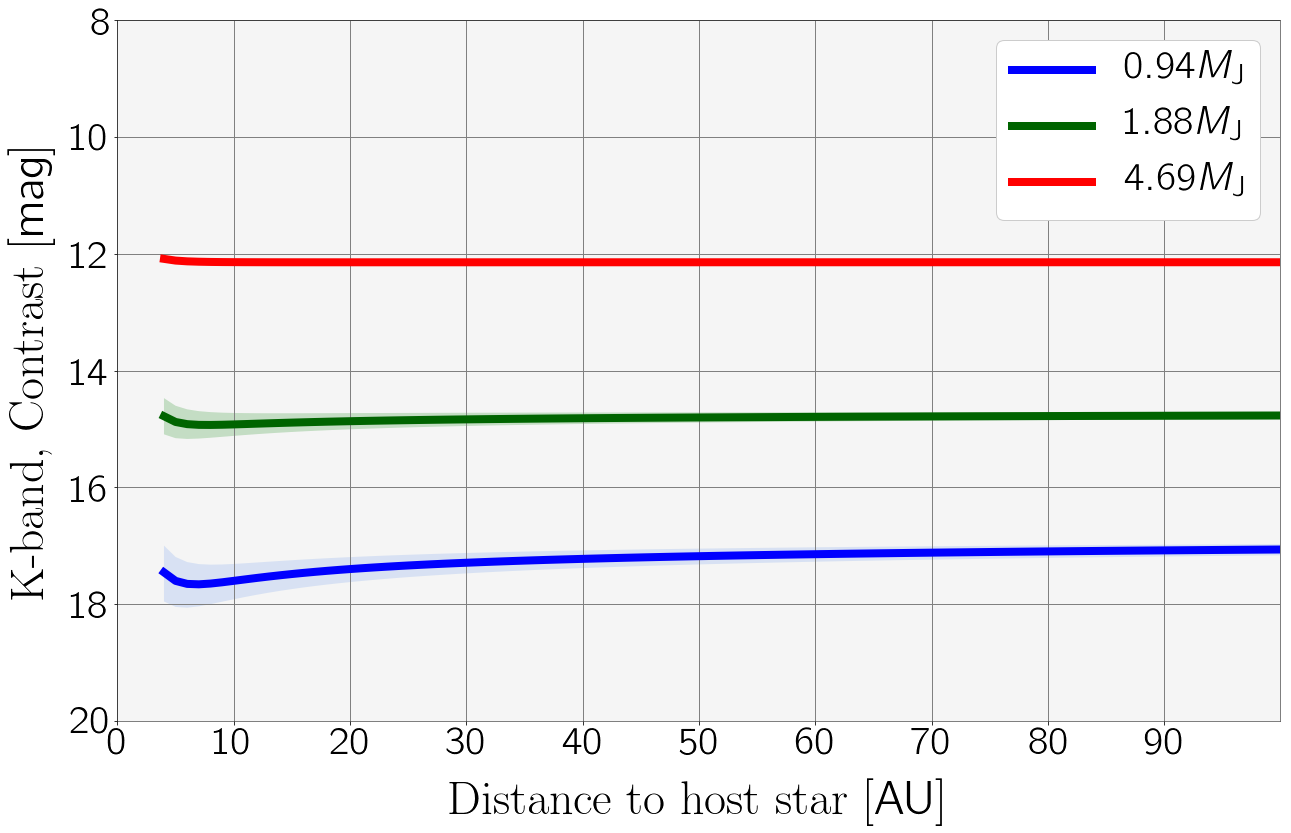}
    \end{center}
    \caption[CQ~Tau Contrast plot including extinction effects.]{Contrast in $J$-, $H$- and $K$-bands of planets embedded in the CQ~Tau disc as a function of distance to the central star . Results are fot planets with $0.94$, $1.88$ and $4.69$ $M_J$.}
    \label{fig:cqtauall}
\end{figure}

\begin{figure}
    \begin{center}
    \includegraphics[width=.495\textwidth]{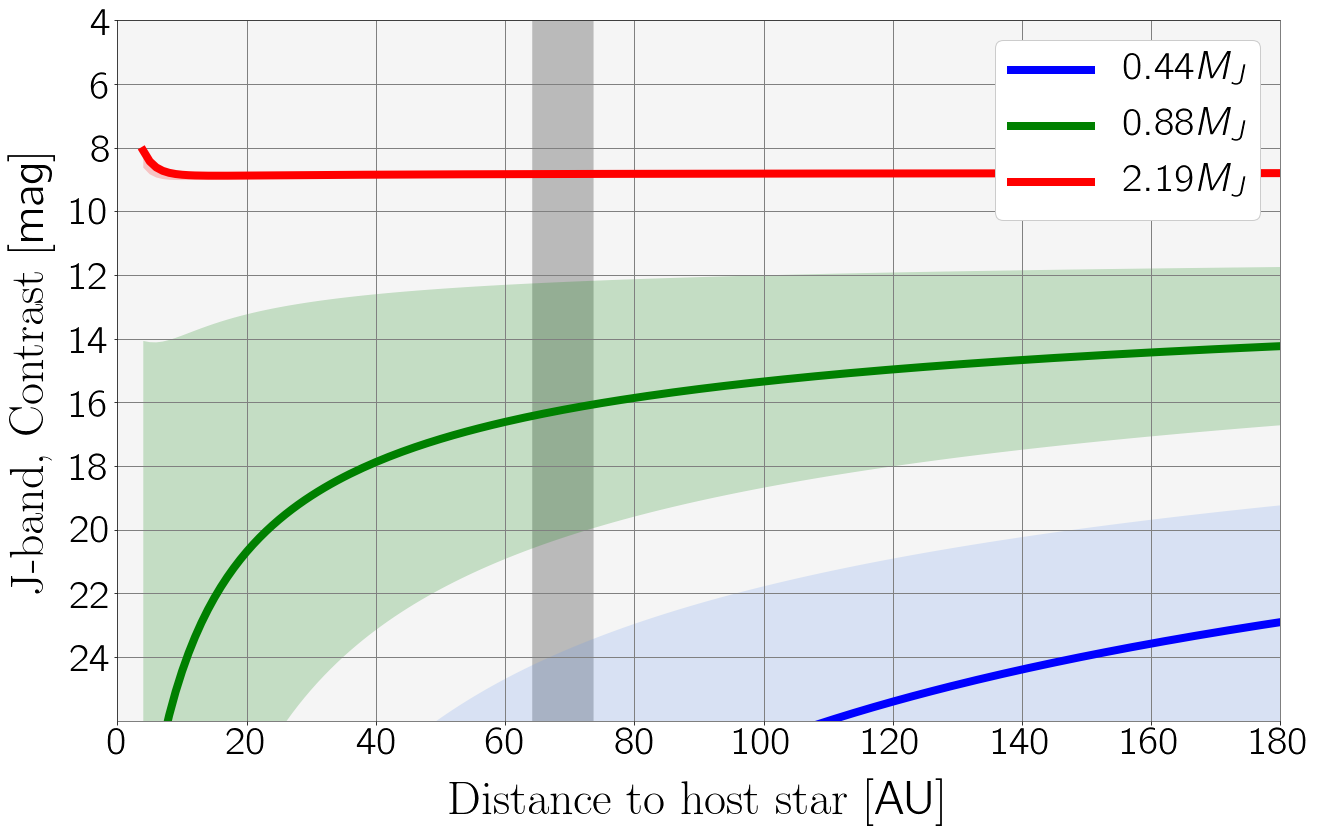}
    \includegraphics[width=.495\textwidth]{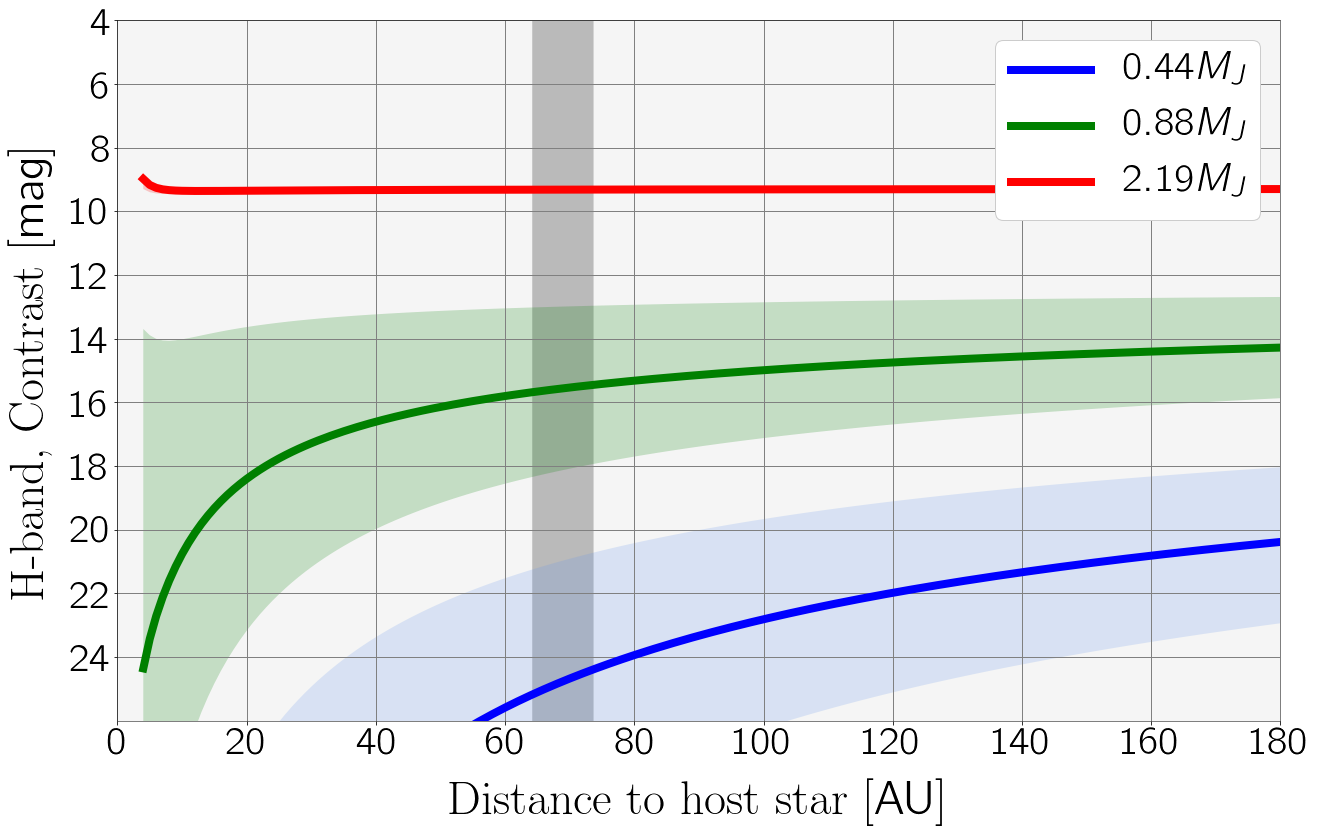}
    \includegraphics[width=.495\textwidth]{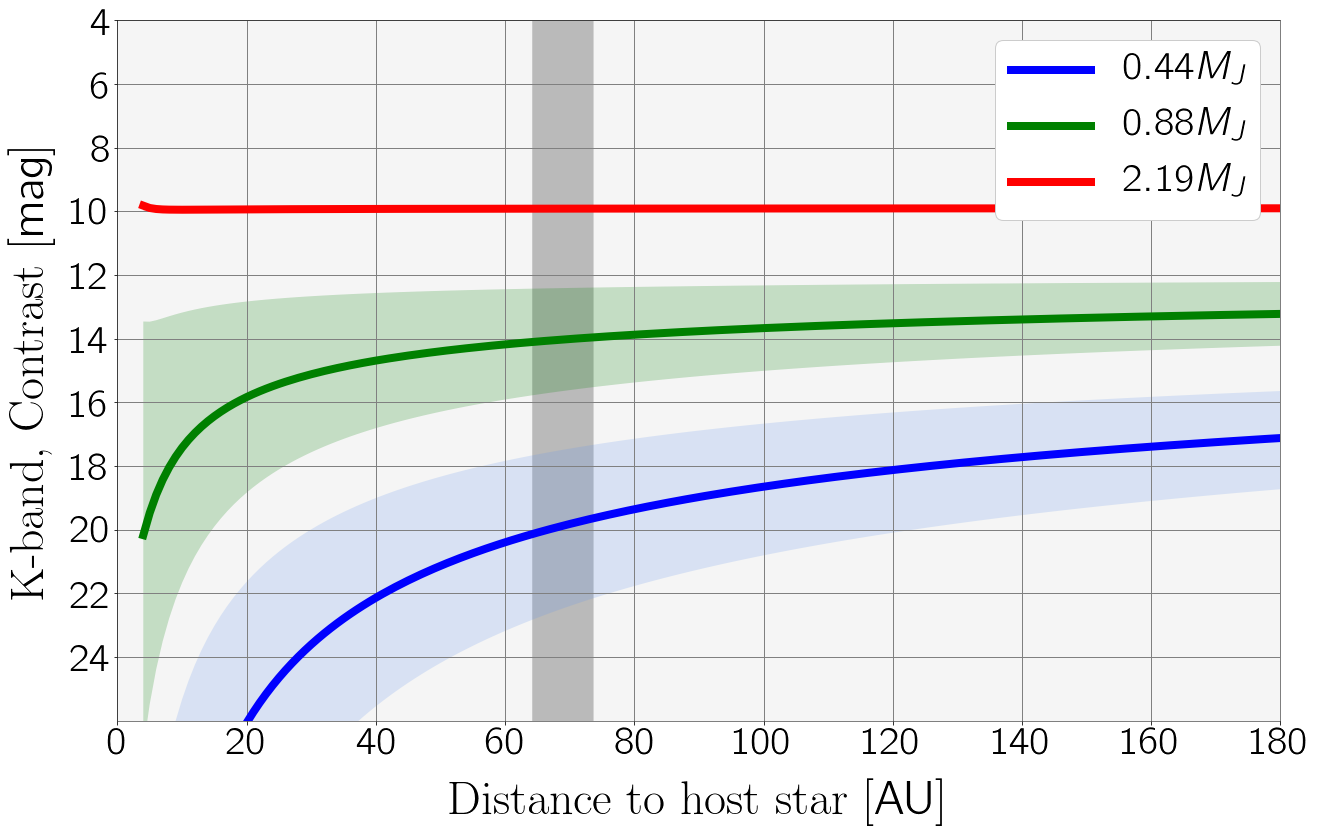}
\end{center}
    \caption[HL~Tau Contrast plot including extinction effects.]{Application of our model to planets with $0.44$, $0.88$ and $2.19$ $M_J$ masses embedded in the HL~Tau  disc. Contrast is shown for different distances to the host star in $J$-, $H$- and $K$-bands.}
    \label{fig:hltauall}
\end{figure}

\begin{figure}
    \begin{center}
    \includegraphics[width=.495\textwidth]{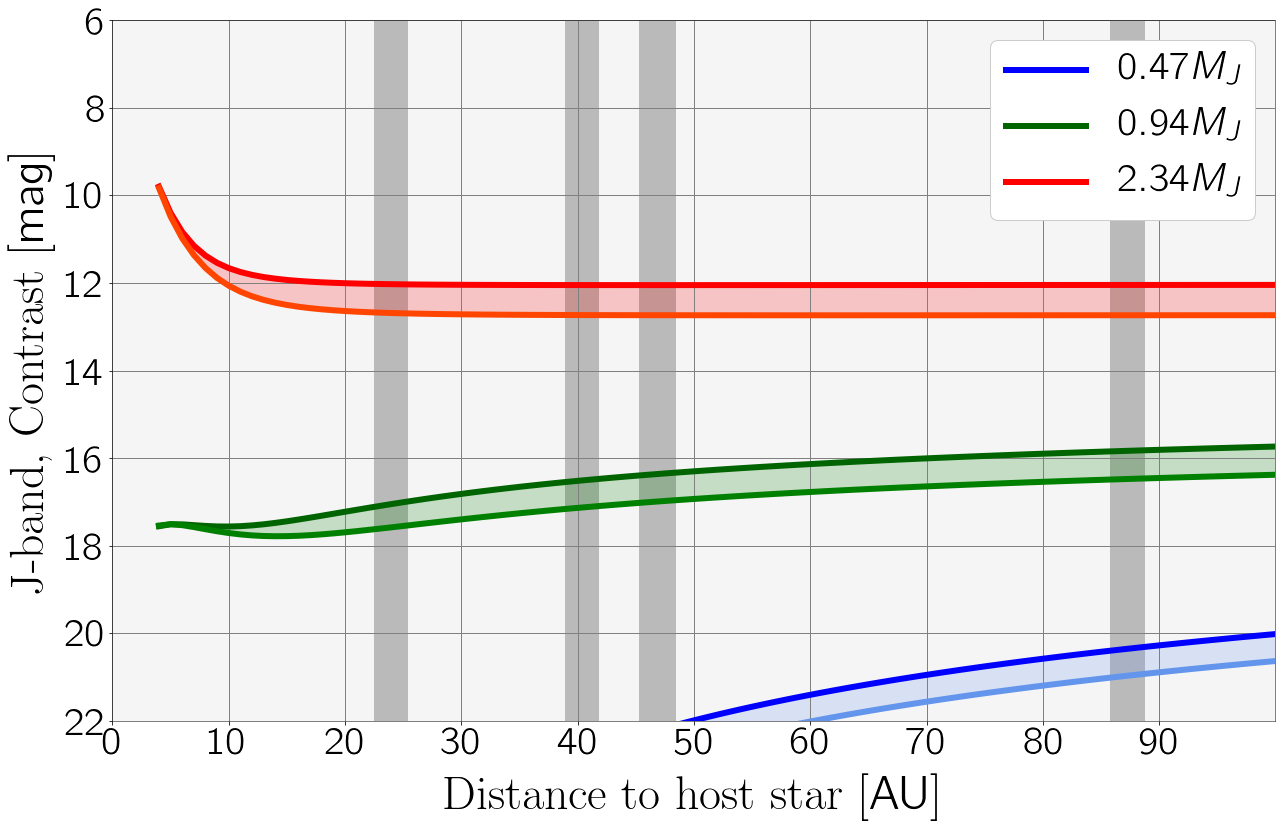}
    \includegraphics[width=.495\textwidth]{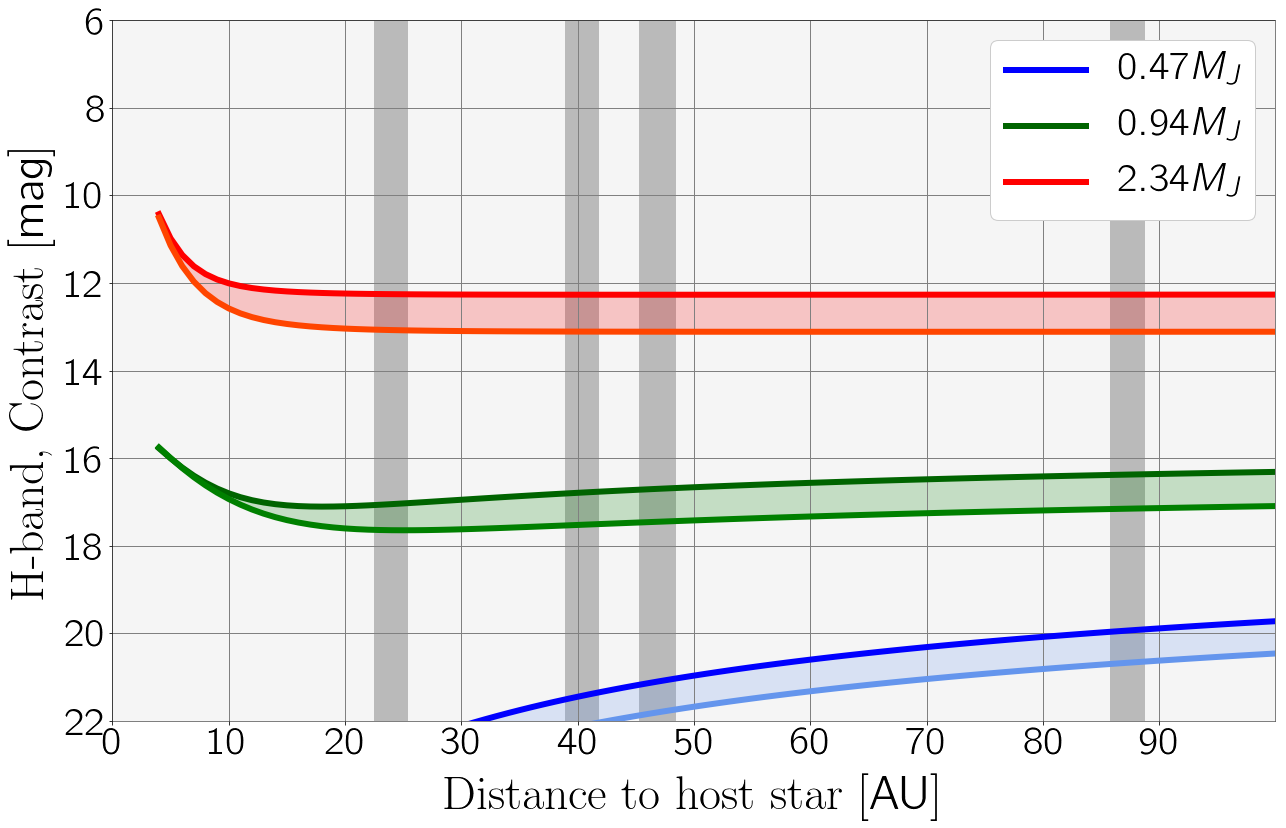}
    \includegraphics[width=.495\textwidth]{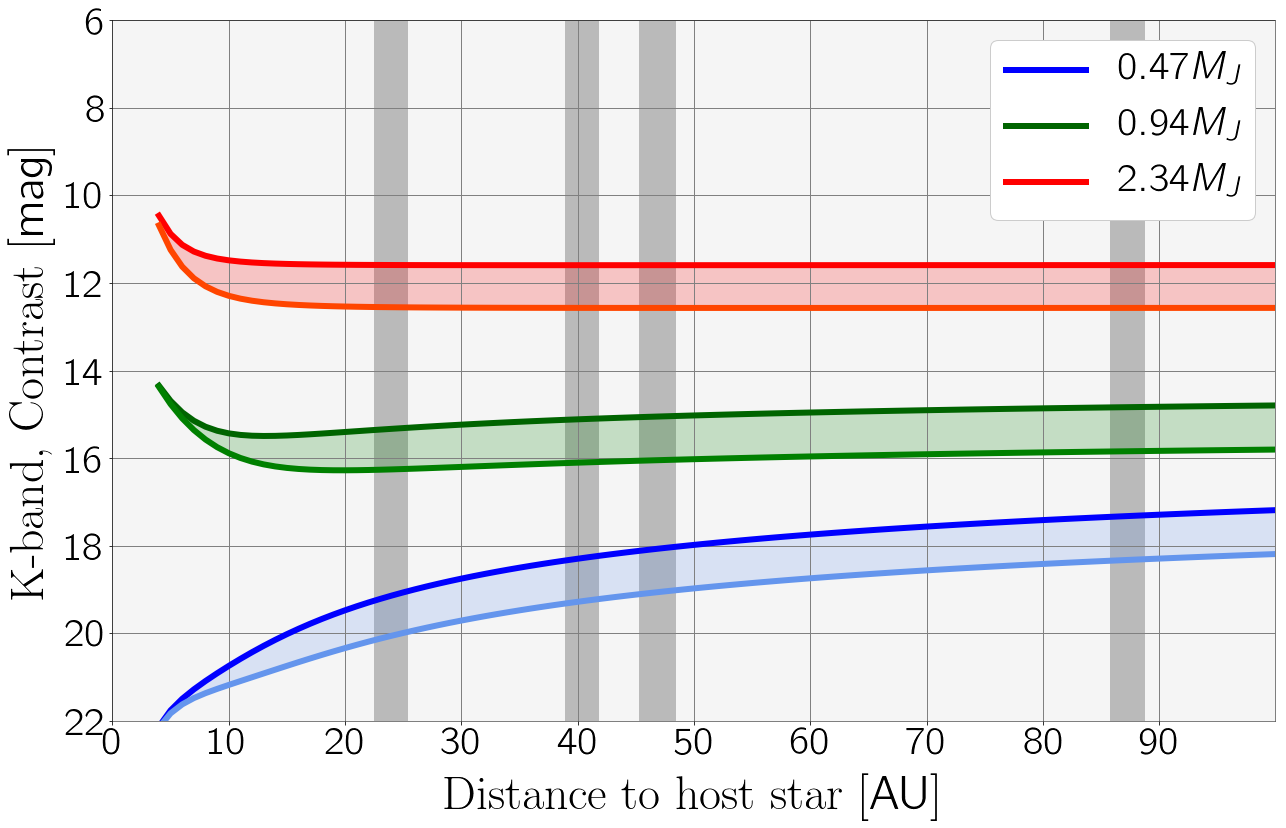}
\end{center}
    \caption[TW~Hya Contrast plot including extinction effects.]{Application of our model to planets with $0.47$, $0.94$ and $2.34$ $M_J$ masses embedded in the TW~Hya disc. Contrast is shown for different distances to the host star in $J$-, $H$- and $K$-bands.}
    \label{fig:twhyaall}
\end{figure}

\begin{figure}
    \begin{center}
    \includegraphics[width=.495\textwidth]{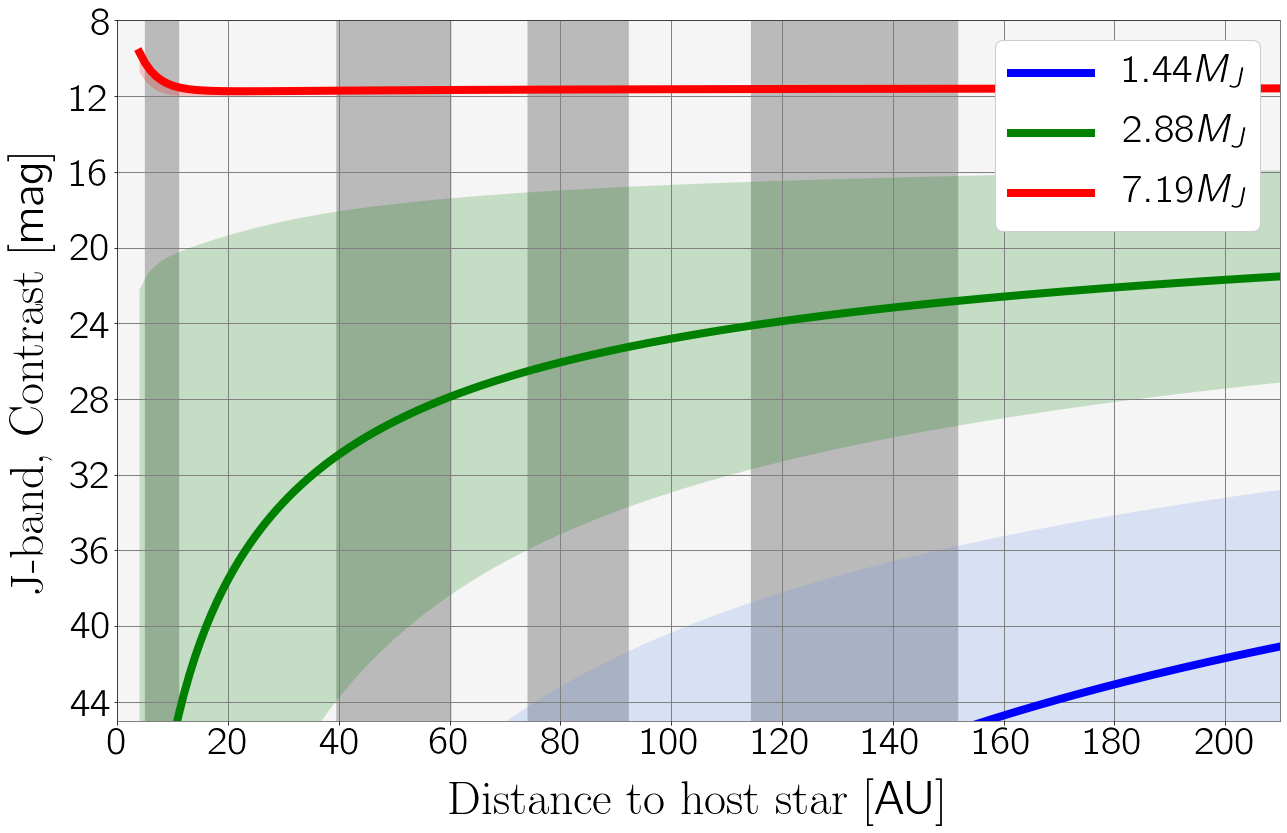}
    \includegraphics[width=.495\textwidth]{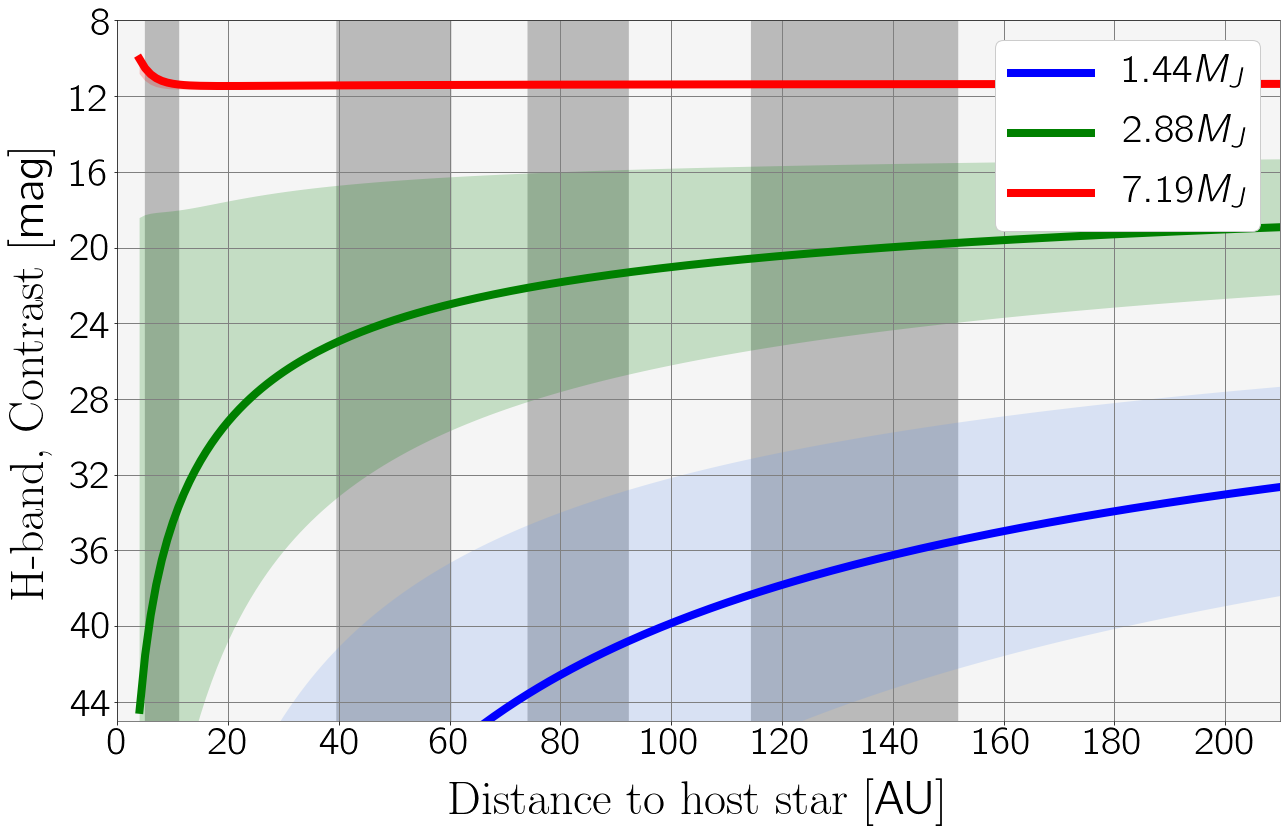}
    \includegraphics[width=.495\textwidth]{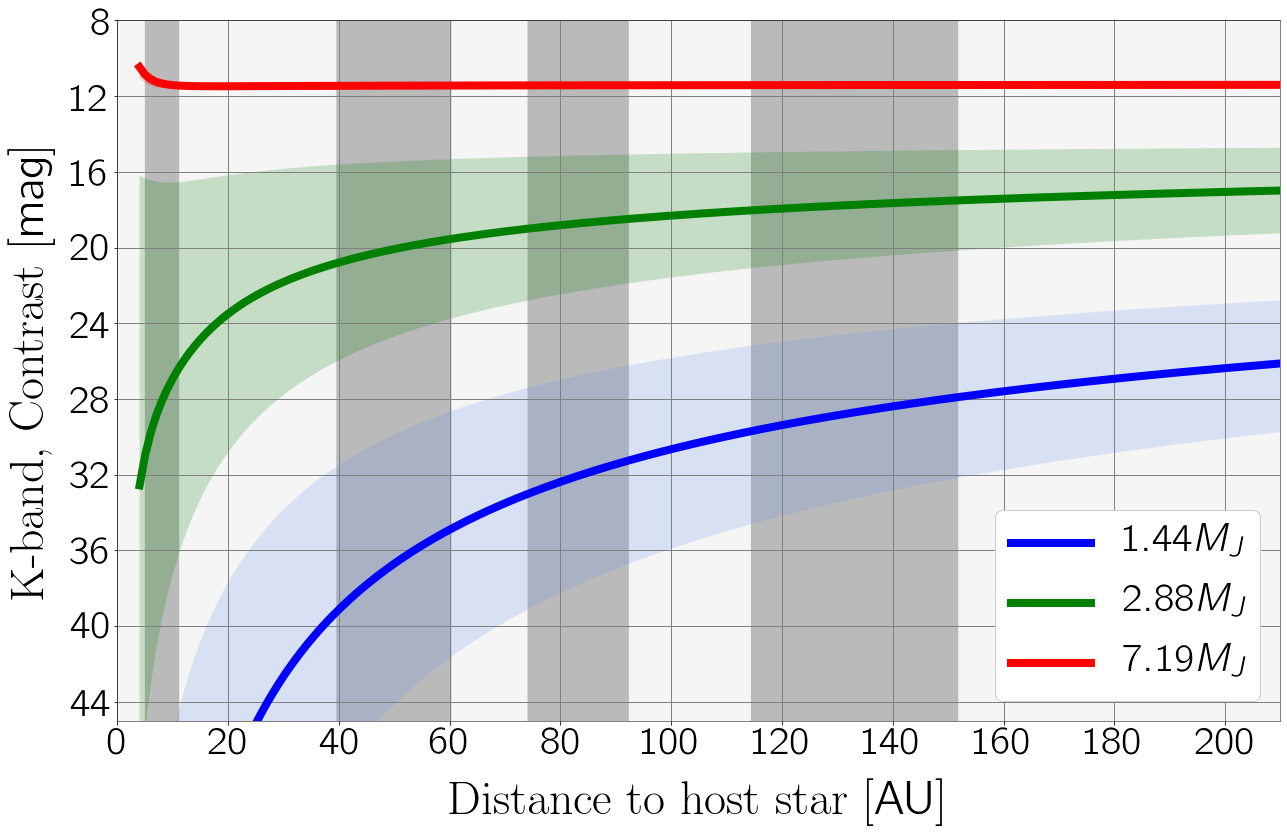}
\end{center}
    \caption[HD~$163296$ Contrast plot including extinction effects.]{Application of our model to planets with $1.44$, $2.88$ and $7.19$ $M_J$ masses embedded in the HD~$163296$ disc. Contrast is shown for different distances to the host star in $J$-, $H$- and $K$-bands.}
    \label{fig:hd163296all}
\end{figure}

The remaining contrast curves of planets as a function of the distance to the host star in $J$-, $H$- and $K$-bands of all the studied systems: $0.94$, $1.88$ and $4.69$ $M_J$ planets in CQ~Tau disc (Figure~\ref{fig:cqtauall}), $0.44$, $0.88$ and $2.19$ $M_J$ planets in HL~Tau (Figure~\ref{fig:hltauall}), $0.47$, $0.94$ and $2.34$ $M_J$ planets in TW~Hya (Figure~\ref{fig:twhyaall}), and $1.44$, $2.88$ and $7.19$ $M_J$ planets in HD~$163296$ (Figure~\ref{fig:hd163296all}).

\bsp
\label{lastpage}
\end{document}